\documentclass[aps,prd,reprint,superscriptaddress,nofootinbib]{revtex4-1}
\usepackage[all]{xy}
\usepackage{amsmath,amsthm,amssymb}
\usepackage[dvips]{graphicx}
\usepackage{comment}
\usepackage{array}
\usepackage{bm}
\allowdisplaybreaks[1]

\usepackage{color}
\usepackage{hyperref}
\usepackage{tcolorbox}
\usepackage{mathrsfs}
\usepackage{float}
\newcommand{\vinf}{v_{\infty}}

\newcommand{\El}{{\sf El}}

\newcommand{\chiinf}{\chi_{\infty}}

\newcommand{\Rdot}{\dot{R}}
\newcommand{\Rdotinf}{\dot{R}_\infty}
\newcommand{\J}[1]{\left[#1\right]}

\newcommand{\vf}{\varphi}

\newcommand{\SP}[1]{\left(#1\right)}
\newcommand{\bP}[1]{\big(#1\big)}

\newcommand{\BB}[1]{\Big[#1\Big]}

\newcommand{\tet}[2]{e_{\bm #1}^{#2}}				%tetrad
\begin{document}

\title{Time-domain metric reconstruction for hyperbolic scattering}

\def\Soton{Mathematical Sciences, University of Southampton, 
Southampton SO17 1BJ, United Kingdom}

\author{Oliver Long}
\affiliation{\Soton}

\author{Leor Barack}
\affiliation{\Soton}

\date{\today}
\begin{abstract}
Self-force methods can be applied in calculations of the scatter angle in two-body hyperbolic encounters, working order by order in the mass ratio (assumed small) but with no recourse to a weak-field approximation. This, in turn, can inform ongoing efforts to construct an accurate model of the general-relativistic binary dynamics via an effective-one-body description and other semi-analytical approaches. Existing self-force methods are to a large extent specialised to bound, inspiral orbits. Here we develop a technique for (numerical) self-force calculations that can efficiently tackle scatter orbits. The method is based on a time-domain reconstruction of the metric perturbation from a scalar-like Hertz potential that satisfies the Teukolsky equation, an idea pursued so far only for bound orbits.  The crucial ingredient in this formulation are certain jump conditions that (each multipole mode of) the Hertz potential must satisfy along the orbit, in a 1+1-dimensional multipole reduction of the problem.  We obtain a closed-form expression for these jumps, for an arbitrary geodesic orbit in Schwarzschild spacetime, and present a full numerical implementation for a scatter orbit. 
In this paper we focus on method development, and go only as far as calculating the Hertz potential; a calculation of the self-force and its physical effects on the scatter orbit will be the subject of forthcoming work.
\end{abstract}

\maketitle

\setcounter{tocdepth}{1}
%\tableofcontents

%%%%%%%%%%%%%%%%%%%%%%%%%%%%%%%%%%%%
\section{Introduction}
%%%%%%%%%%%%%%%%%%%%%%%%%%%%%%%%%%%%

The post-Minkowskian (PM) theory of two-body dynamics in General Relativity has seen a rapid progress in recent years, thanks in part to the introduction of radically new approaches to the problem. These include the effective-one-body method \cite{Damour:2016gwp,Damour:2017zjx,Damour:2019lcq,Bini:2020rzn}, effective-field-theory treatments \cite{Porto:2016zng,Kalin:2019rwq,Kalin:2020fhe,Liu:2021zxr}, and the use of dictionaries that translate between quantum scattering amplitudes and classical gravitational dynamics (``double copy'') \cite{Bern:2019nnu,Bern:2019crd,Bern:2020buy,Bern:2020uwk}. A recent milestone is the derivation of the conservative two-body dynamics through 4PM order [$O(G^4)$] using scattering-amplitude methods \cite{Bern:2021dqo}, and there is also progress on the description of radiative effects \cite{Damour:2020tta,divecchia2021eikonal,Bern:2021dqo}.

An alternative route to high-order PM calculations is provided by black-hole perturbation theory, i.e., methods that rely on an expansion in the mass ratio $\eta$ of the binary, without a weak-field approximation. The remarkable effectiveness of such an avenue of approach was first noted by Damour in \cite{Damour:2019lcq}. At least for structureless point particles, the 2PM conservative 2-body dynamics can be inferred in full simply from the scatter angle of {\em geodesic} orbits on a Schwarzschild background (as a function of, say, the orbit's energy and impact parameter). Knowledge of the $O(\eta)$ back-reaction correction to the scatter angle---the so-called first-order self-force correction---determines the full conservative dynamics through 4PM order (at any mass ratio). A second-order self-force  calculation would achieve the same to as high an order as 6PM. Furthermore, since at each order in $\eta$ the self-force results are ``exact'' (they ``contain all PM terms''), such results can provide a useful benchmark against which to assess the performance of the PM series in the strong-field regime.  

Thus there is a motivation for self-force calculations in scatter-orbit scenarios. Unfortunately, existing calculation methods and codes are to a large extent tailored to tackle bound-orbit or inspiral systems, which are relevant to astrophysical extreme mass-ratio setups, and whose study remains the main driver of such calculations. These codes can not be immediately applied to scatter-type orbits. For example, the most advanced self-force code \cite{vandeMeent:2017bcc} is based on a procedure of metric reconstruction from discrete frequency-mode solutions of the perturbation equations, which crucially relies on the assumption that the orbit is quasi-periodic. (Part of the issue is that 
the so-called ``method of extended homogeneous solutions'' \cite{Barack:2008ms}, which enables the time-domain reconstruction of the metric perturbation near the particle, works a priori only for bounded orbits.)  While these difficulties are unlikely insurmountable, existing frequency-domain methods would require much further development before they can be applied to scatter-type orbits; see \cite{Hopper:2017qus,Hopper:2017iyq} for initial work. 

Time-domain treatments offer an alternative route to unbound-orbit calculations, bypassing some of the difficulties. Ref.\ \cite{Barack_2019} recently presented a first such calculation, based on a numerical integration of the Lorenz-gauge metric perturbation equations on a Schwarzschild background, formulated as an initial-value problem in 1+1 dimensions. This work did not consider scatter-type orbits, but the special case of a particle falling from rest at infinity, eventually (radiation ignored) getting trapped on an unstable circular orbit around the Schwarzschild black hole. The method could be further developed to tackle scatter orbits, but significant hurdles remain. In particular, the Lorenz-gauge formulation admits certain nonphysical, linearly growing gauge modes that develop generically in numerical evolutions and are hard to control \cite{Dolan}. These were tamed in \cite{Barack_2019} by exploiting the asymptotic periodicity of the special ``zoom-whirl'' setup, but it remains unclear how to handle the problem in the case of hyperbolic-like scattering, where similar tricks cannot be used. Furthermore, the Lorenz-gauge formulation involves a rather unwieldy coupled set of partial differential equations, which impacts on computational precision and cost. 

In this paper we develop an alternative time-domain method and illustrate its implementation for scatter orbits. The method is based on metric reconstruction from a scalar-like Hertz potential, which satisfies the Teukolsky equation. The equation is solved as an evolution problem in the time domain, and the metric perturbation is then reconstructed from the solution (and additional, trivial perturbation pieces) in a gauge suitable for self-force calculations \cite{PMB} (the ``no-string'' radiation gauge, to be reviewed in Sec.\ \ref{Sec:review}). This procedure circumvents the main pitfalls of the Lorenz-gauge method: One has to solve a single, simple hyperbolic equation, and one encounters no problematic gauge modes. 
%(still need to compute even-parity dipole mode in some regular CoM-centered gauge.) 

The central idea behind our method is not new: it was introduced by one of us with Giudice in Ref.\ \cite{Barack:2017oir}, where, however, it was fully developed and implemented only for circular (geodesic) orbits. Here we formulate the method for arbitrary (geodesic) orbits, and implement it numerically for scatter orbits. We go as far as computing the Hertz potential along the scatter orbit, in order to demonstrate the applicability of our method and explore its performance. We do not proceed here to calculate the self-force and its effects on the scatter angle; this we hope to accomplish soon in subsequent work.  

We begin in Sec.\ \ref{Sec:review} with a review of metric reconstruction for a point-particle source in a no-string gauge, specializing to a Schwarzschild background and casting the procedure in a form suitable for a time-domain implementation. In Sec.\ \ref{Sec:HertzFormulation} we then formulate an initial-value problem for the (no-string) Hertz potential. In the no-string construction, the spacetime outside the central black hole is split into two vacuum domains, $r>R(t)$ and $r<R(t)$, where $r$ and $t$ are Schwarzschild coordinates and $r=R(t)$ along the particle's trajectory.  The crucial ingredient in our formulation are jump conditions that the Hertz potential and its derivatives must satisfy on the (time-dependent) two-sphere $r=R(t)$. These conditions are derived in Sec.\ \ref{Sec:Jumps} for an arbitrary timelike geodesic trajectory. This is the main new result of the formulation part of this work. 

In Sec.\ \ref{sec:vacuum} we present a new code for numerical integration of the Bardeen-Press-Teukolsky (BPT) equation in 1+1 dimensions on a Schwarzschild background. The code employs a finite--difference scheme on a characteristic grid based on Eddington--Finkelstein coordinates---a simple tried-and-tested architecture that has worked reliably in many past calculations of the Lorenz-gauge and scalar-field self-forces. We demonstrate, however, how a naive implementation of this standard scheme fails when applied to the Teukolsky equation with spin parameter $s=\pm 2$, due to divergences that develop at late time (an exponential divergence for $s=+2$ and a $\sim t^4$ divergence for $s=-2$). We attribute these divergences to certain growing modes of the Teukolsky equation. These modes violate the physical boundary conditions, but since boundary conditions are not actively imposed in our characteristic scheme, they are allowed to grow. The problem persists even in vacuum evolutions. We explain why the issue is not encountered in existing time-domain Teukolsky codes based on hyperboloidal slicing with compactification \cite{Racz:2011qu,Zenginoglu:2012us,Harms2013}. 

Here, restricting to the Schwarzschild case, we opt for a simpler solution. We circumvent the problem of growing modes by transforming to a new field variable (using a time-domain version of the Chandrasekhar transformation), which, in the vacuum case, satisfies the Regge--Wheeler (RW) equation, for which the problem does not occur. In Sec.\ \ref{sec:RWformulation} we reformulate our initial-value problem in terms of the new variable, and, in particular, derive the necessary jump conditions for it on $r=R(t)$.

In Sec.\ \ref{sec:hyperevol} we finally present a full numerical implementation of our method, for a scatter orbit. We evolve the field equation for the RW-like variable, and from it compute (multipole mode by multipole mode) the no-string Hertz potential along the scatter trajectory. We thus numerically construct the necessary input for a calculation of the self-force along the orbit.

We conclude in Sec.\ \ref{Sec:conclusions} by reviewing the extra steps needed to carry our the calculation of the self-force from the Hertz potential. We also discuss the prospects of extending our method to the case of a Kerr background.   

Throughout this work we use units in which $G=1=c$, and adopt the metric signature $({-}{+}{+}{+})$. For quantities that arise in the Newman-Penrose formalism we follow the sign conventions of Ref.\ \cite{Merlin:2016boc}, as summarized in  Appendix A therein; for ease of reference we review the relevant details here, in Appendix \ref{App:convention}, specialised to the Schwarzschild case.

%%%%%%%%%%%%%%%%%%%%%%%%%%%%%%%%%%%%%%%%%%%%%%%%%%%%%%%%%%%%%%%%%%%%%%%%
\section{Review of metric reconstruction in a no-string radiation gauge}
\label{Sec:review}
%%%%%%%%%%%%%%%%%%%%%%%%%%%%%%%%%%%%%%%%%%%%%%%%%%%%%%%%%%%%%%%%%%%%%%%%

In this section we review essential results concerning (i) the reconstruction of vacuum metric perturbations from curvature scalars; (ii) the failure of a naive metric reconstruction in the presence of sources; and (iii) the no-string reconstruction scheme for point-particle sources. From a certain point we will specialize to a Schwarzschild background, introducing a decomposition of the various fields into multipole modes, but refraining from a further frequency-mode decomposition and instead remaining in the time domain. Our purpose here is to remind readers of the relevant theory, introduce notation, and set up the relevant technical background for the rest of the analysis. 

We adopt the Kinnersley null tetrad $e^\alpha_{\bm a}=\{\ell^\alpha,n^\alpha,m^\alpha,\bar m^\alpha\}$ [see Eq.\ (\ref{eq:kerrtetrad})], where boldface Roman indices run over $1,\ldots, 4$ and denote tetrad components: $A_{\bm a}:=e^\alpha_{\bm a}A_{\alpha}$.   
The legs $e^\alpha_{\bm a}$ are all mutually orthogonal, except $\ell^\alpha n_\alpha=-1$ and $m^\alpha \bar m_{\alpha}=1$. An overbar denotes complex conjugation. (Covariant) directional derivatives along the tetrad legs are denoted ${\bm D}_\ell = \ell^\alpha\nabla_\alpha$, ${\bm D}_n = n^\alpha\nabla_\alpha$, ${\bm D}_m = m^\alpha\nabla_\alpha$ and ${\bm D}_{\bar m} = \bar m^\alpha\nabla_\alpha$ (corresponding to the more customary but less transparent $\bm D$, $\bm \Delta$, $\bm \delta$ and $\bar {\bm\delta}$, respectively).

\subsection{Vacuum case }

The reconstruction of vacuum metric perturbations from curvature scalars was first prescribed in \cite{CKK1}, but we follow here the concise presentation by Wald \cite{waldrec}. In what follows hatted sans-serif symbols (${\sf\hat E}, {\sf\hat T}, \ldots$) represent linear differential operators on tensors. 

%Suppose $h_{\alpha\beta}$ is a perturbation of the Kerr metric $g_{\alpha\beta}$, satisfying the vacuum Einstein's equations linearised in $h_{\alpha\beta}$ about $g_{\alpha\beta}$, which we write as
%in  such that $G_{\alpha\beta}(g+h)=0$, where $G_{\alpha\beta}$ is the Einstein tensor. 
 
Suppose $h_{\alpha\beta}$ is a solution of the vacuum Einstein's equation linearised about the Kerr metric:
\begin{equation}\label{EFEvacuum}
{\sf\hat E} h:= \delta G(h) =0.
\end{equation} 
Here $\delta G_{\mu\nu}$ is the linearised Einstein tensor, thought of as a differential operator ${\sf\hat E}$ acting on $h_{\alpha\beta}$, and we have omitted tensorial indices for brevity. To this perturbation there correspond Weyl curvature scalars $\Psi_0=:\Psi_+$ and $\varrho^{-4}\Psi_4=:\Psi_-$ [see Eq.\ (\ref{eq:psi}); $\varrho=-1/r$ for Schwarzschild]. $\Psi_\pm$ are derived from $h_{\alpha\beta}$ via
\begin{equation}\label{T}
{\sf\hat T}_{\pm} h =\Psi_\pm,
\end{equation} 
where the operators $\sf\hat T_{\pm}$ are given explicitly in Eq.\ (\ref{hatT}). Let $\sf\hat S_{\pm}$ be the operators that take the linearised Einstein's equation into the Teukolsky equations with spins $\pm 2$, respectively:
\begin{equation}\label{S}
{\sf\hat S}_{\pm} {\sf\hat E}h ={\sf\hat O}_{\pm}\Psi_\pm ,
\end{equation} 
where ${\sf\hat O}_{\pm}$ is the Teukolsky operator given in Eq.\ (\ref{Ocompact}), and $\sf\hat S_{\pm}$ can be read off the source side of the Teukolsky equation (\ref{eq:kerrteuk}); these operators are given explicitly in Eqs.\ (\ref{eq:kerrsource}) and (\ref{eq:kerrsource-2}). From (\ref{T}) and (\ref{S}) there follows the operator identity
\begin{equation}\label{operator identity}
{\sf\hat S}_{\pm} {\sf\hat E} ={\sf\hat O}_{\pm}{\sf\hat T}_{\pm}  .
\end{equation}

Now let $\Phi_{\pm}$ be (any) solution of the {\em adjoint}\footnote{
For a linear operator $\sf\hat L$ taking an $n$-rank tensor field $\phi$ to an $m$-rank tensor field $\psi$, the adjoint $\sf\hat L^\dagger$ takes $\psi$ to $\phi$ and satisfies 
$({\sf\hat L}^\dagger\psi)\phi=\psi({\sf\hat L}\phi)$
%=s^\alpha_{\ ;\alpha}$, 
(up to a divergence of an arbitrary vector field). 
%where $s^\alpha$ is an arbitrary vector field. 
} 
vacuum Teukolsky equation,
\begin{equation}\label{Teukolski Hertz}
{\sf\hat O}^\dagger_{\pm}\Phi_{\pm}(={\sf\hat O}_{\mp}\Phi_{\pm}) =0 . 
\end{equation}
Noting $\sf\hat E$ is self-adjoint (${\sf\hat E}={\sf\hat E}^\dagger$), we then have
\begin{equation}
{\sf\hat E} {\sf\hat S}^\dagger_{\pm} \Phi_{\pm} = 
%{\sf\hat E}^\dagger {\sf\hat S}^\dagger_{\pm} \Phi_{\pm}=
({\sf\hat S}_{\pm}{\sf\hat E})^\dagger \Phi_{\pm}=
({\sf\hat O}_{\pm}{\sf\hat T}_{\pm})^\dagger \Phi_{\pm} =
{\sf\hat T}_{\pm}^\dagger {\sf\hat O}_{\pm}^\dagger \Phi_{\pm} = 0 ,
\end{equation}
where in the second equality we have used (\ref{operator identity}). Thus 
%\begin{equation}\label{h_complex}
$h^{\pm}:={\sf\hat S}^\dagger_{\pm} \Phi_{\pm}$
%\end{equation}
are (complex-valued) solutions of the vacuum Einstein's equation. 
%To obtain a real solution, we may choose either the real or imaginary parts of $h^{\pm}$. Choosing the former, we obtain the 
A real-valued reconstructed solution is given by
\begin{equation}\label{h_rec}
h^{\rm rec}_{\pm}:={\rm Re\, }{\sf\hat S}^\dagger_{\pm} \Phi_{\pm}.
\end{equation}
%The reconstruction operators ${\sf\hat S}^\dagger_{\pm}$ are given explicitly in (xx).
The explicit form of the reconstruction operator ${\sf\hat S}^\dagger_{\pm}$ is given in Eqs.\ (\ref{eq:kerrh+}). It returns $h^{\rm rec}_{+}$ in an {\em ingoing} radiation gauge (IRG), and $h^{\rm rec}_-$ in an {\em outgoing} radiation gauge (ORG):
\begin{eqnarray}
h^{{\rm rec}+}_{{\bf 1}\beta} = 0\quad\text{(IRG)},\quad\quad
h^{{\rm rec}-}_{{\bf 2}\beta} = 0 \quad\text{(ORG)},
\end{eqnarray}
with both perturbations being traceless.

For $h_{+}^{\rm rec}$ and $h_{-}^{\rm rec}$ to each reproduce the original perturbation $h$, we must have ${\sf\hat T}_{\pm} h_+^{\rm rec} =\Psi_\pm$ and ${\sf\hat T}_{\pm} h_-^{\rm rec} =\Psi_\pm$, leading to
\begin{eqnarray}
{\sf\hat T}_{\pm}{\rm Re\, }{\sf\hat S}^\dagger_{+} \Phi_{+} &=& \Psi_\pm \quad\text{(IRG)}, \label{inversionIRG}\\
{\sf\hat T}_{\pm}{\rm Re\, }{\sf\hat S}^\dagger_{-} \Phi_{-} &=& \Psi_\pm\quad\text{(ORG)}. \label{inversionORG}
\end{eqnarray}
These are the fourth-order ``inversion'' equations.
A Hertz potential $\Phi_{+}$ satisfying both the adjoint Teukolsky equation (\ref{Teukolski Hertz}) and either of the two  inversion relations in (\ref{inversionIRG}) will reproduce $h$ up to some perturbation $\Delta h_+$ that is in the Kernel of ${\sf\hat T}_{+}$; and a Hertz potential $\Phi_{-}$ satisfying both (\ref{Teukolski Hertz}) and either of the two  inversion relations in (\ref{inversionORG}) will reproduce $h$ up to some perturbation $\Delta h_-$ that is in the Kernel of ${\sf\hat T}_{-}$. That is,
\begin{equation}
h= h^{\rm rec}_{\pm} + \Delta h_{\pm},
\end{equation}
where 
\begin{equation}
{\sf\hat T}_{\pm} \Delta h_{\pm} = 0 . 
\end{equation}
Wald \cite{waldtheo} explored the Kernel of ${\sf\hat T}_{\pm}$, and hence the space of $\Delta h_{\pm}$, for vacuum perturbations in Kerr. He found that $\Delta h_{\pm}$ is spanned by pure gauge perturbations (which are also in the Kernel of $\sf\hat E$), in addition to exactly four types of stationary and axially symmetric (algebraically-special) vacuum perturbations: a mass perturbation, an angular-momentum perturbation, and perturbations away from Kerr into the Kerr--NUT or the C-metric solutions.
%In the vacuum case, regularity rejects the perturbations $\delta N$ and $\delta C$.

\subsection{Naive reconstruction with sources and its failure}

The above procedure returns a solution to the {\em vacuum} linearised Einstein's equation (\ref{EFEvacuum}). One may naively attempt to generalize this method to the nonvacuum case as follows. Suppose we wish to reconstruct a solution to 
\begin{equation}\label{EFEnonvacuum}
({\sf\hat E} h)_{\alpha\beta} =8\pi T_{\alpha\beta}, 
\end{equation}
where $T_{\alpha\beta}$ is some (first-order, perturbative) source of stress-energy. It is tempting to try a Hertz potential that satisfies a sourced equation of the form   
\begin{equation}\label{Hertznonvacuum}
{\sf\hat O}^\dagger_{\pm}\Phi_{\pm} = S_\pm ,
\end{equation}
in lieu of the vacuum equation (\ref{Teukolski Hertz}). Here the source $S_\pm$ is to be determined from the condition that a metric perturbation reconstructed via (\ref{h_rec}) is a solution to (\ref{EFEnonvacuum}). By operating on both sides of (\ref{Hertznonvacuum}) with ${\sf\hat T}_{\pm}^\dagger$ [the adjoint of the operator ${\sf\hat T}$ in (\ref{T})] and then using the adjoint of the operator identity (\ref{operator identity}), one obtains the condition 
\begin{equation}\label{Source_rec}
T^{\rm rec}_{\alpha\beta\pm}:={\rm Re\, }({\sf\hat T}_{\pm}^\dagger S_\pm)_{\alpha\beta} = 8\pi T_{\alpha\beta}.
\end{equation}
However, considering the form of the operators ${\sf\hat T}_{\pm}^\dagger$ in Eqs.\ (\ref{Tdagger+Expl}) and (\ref{Tdagger-Expl}), we immediately see that (\ref{Source_rec}) has no solutions for $S_\pm$, in general. For instance, observe that $T^{\rm rec}_{\alpha\beta\pm}$ is traceless, while $T_{\alpha\beta}$ need not be (and never is, in the case of a mass particle source, of interest to us here). Also observe that the tetrad components 
%\begin{equation}
%%{\rm tr\, } T^{\rm rec}_{\alpha\beta\pm} =0, 
%\end{equation}
%while ${\rm tr\, } T_{\alpha\beta} \ne 0$ in general (and always for a point mass).
%Also, we see that 
%\begin{equation}\label{inconsistency}
$T^{\rm rec}_{{\bf 1}\beta+}$,
%\quad\quad
$ T^{\rm rec}_{{\bf 2}\beta-}$ and
%\quad\quad
$T^{\rm rec}_{{\bf 34}\pm}$
%\end{equation}
all vanish identically, while the corresponding tetrad components of $T_{\alpha\beta}$ need not be zero. 
%\begin{eqnarray}\label{ldotT=0}
%\ell \cdot {\sf\hat T}_{+}^\dagger = 0 %= m \cdot {\sf\hat T}_{+}^\dagger
%,\\
%\label{ndotT=0}
%n \cdot {\sf\hat T}_{-}^\dagger = 0 . %= \bar m \cdot {\sf\hat T}_{-}^\dagger.
%\end{eqnarray}
%This means that (\ref{Seqn}) is consistent only if $\ell^\alpha (T_{\alpha\beta}-T_{\alpha\beta}^{\rm comp})=0$ (for $+$) or if $n^\alpha (T_{\alpha\beta}-T_{\alpha\beta}^{\rm comp})=0$ (for $-$). But this is generally not the case, and it is never the case for a $T_{\alpha\beta}$ corresponding to a massive point particle. 
Thus, in general, our naive reconstruction procedure fails in nonvacuum regions of spacetime.

In recent work \cite{Green_2020}, Green et al.\ prescribed a modification of the above naive approach, based on adding a certain ``corrector tensor'' in Eq.\ (\ref{Source_rec}) (together with a ``completion'' piece $\Delta h_{\pm}$), essentially to the effect of balancing out the component content of the two sides of the equation. They showed how, remarkably, the corrector can be obtained by integrating a certain hierarchical set of {\it ordinary} differential equations (ODEs) along null directions. There is ongoing work to demonstrate the applicability of this method in practice. 

\subsection{Point-particle source}

A more acute question is whether the standard vacuum reconstruction procedure works in {\em vacuum} regions of spacetime in the presence of sources elsewhere. It has long been known, from analysis of the point-particle source example \cite{gauge}, that this was not the case: a perturbation $h^{\rm rec}_{\pm}$ reconstructed as in Eq.\ (\ref{h_rec}) (with or without a completion piece $\Delta h_{\pm}$) develops singularities in the vacuum region {\em away} from the particle. This can be appreciated already from the simple example of a static particle in flat space---see Sec.\ V C. of \cite{gauge}, or the more detailed analysis in Sec.\ VI of \cite{PMB}. What one finds is that $h^{\rm rec}_{\pm}$ exhibits string-like singularities that emanate from the particle along radial null directions. By adjusting the residual gauge freedom (within the class of radiation gauges) one can arrange to confine the string to either outgoing or ingoing directions, but no choice of a radiation gauge can rid of the strings altogether. The leading-order singular form of the string is described in Table I of \cite{PMB}. The singularly is sufficiently strong that the perturbation field fails to be (absolutely) integrable over a two-dimensional surface intersecting the string, with the result that a multipole decomposition of the field is not even well defined. Thus a mode-by-mode reconstruction procedure cannot work in the entire vacuum part of spacetime containing the string. It should be presumed that an analysis based on the new corrector-tensor method of \cite{Green_2020} would reproduce this basic picture when applied to the point-particle case.

Let us describe the situation more precisely, using some new notation that will serve us through the rest of this work. We are interested in the case of a pointlike particle of mass $\mu$, moving outside a Kerr black hole (to be specialised to Schwarzschild further below) with mass $M\gg\mu$. We assume the particle's stress-energy is given by the distribution
\begin{equation}\label{Tmunu}
T_{\alpha\beta}= \mu \int_{-\infty}^{\infty} u_{\alpha}u_{\beta}\delta^4(x^\alpha-x_{\rm p}^\alpha(\tau))(-g)^{-1/2}d\tau ,
\end{equation}
where $x_{\rm p}^\alpha(\tau)$ describes the particle's timelike worldline ($\tau$ being proper time), $u^\alpha:=dx_{\rm p}^{\alpha}/d\tau$, and, as usual, indices are lowered using the background metric $g_{\alpha\beta}$ with determinant $g$. In Boyer-Lindquist coordinates (and a slight notational abuse) we write $x_{\rm p}^\alpha(t)=\left(t,R(t),\theta_{\rm p}(t),\varphi_{\rm p}(t)\right)$, so that $R(t)$ is the radial location of the particle at time $t$. 
We denote by $\cal S$ the 2+1-dimensional closed surface $r=R(t)$; this is a 2-sphere  through the particle at each given time. The surface $\cal S$ splits the exterior of the black hole into two disjoint regions, $r>R(t)$ and $r<R(t)$, which we call ${\cal S}^{>}$ and ${\cal S}^{<}$, respectively. 
 
As we have described, a reconstructed radiation-gauge metric $h^{\rm rec}_{\pm}$ generically exhibits a string singularity in both ${\cal S}^{>}$ and ${\cal S}^{<}$: it is a ``full-string'' solution, in the terminology of \cite{PMB}. It is not known how to calculate the physical self-force in such a pathological gauge, so the full-string reconstruction is not useful in the present context. As also described, there is a way to choose a radiation gauge such that the string is confined to ${\cal S}^{>}$ and the reconstructed perturbation, denoted here $h^{\rm <}_{\pm}$, is regular (smooth) anywhere in ${\cal S}^{<}$. Similarly, there is a choice of radiation gauge for which the string is confined to ${\cal S}^{<}$, and the perturbation, denoted $h^{\rm >}_{\pm}$, is regular (smooth) anywhere in ${\cal S}^{>}$. These are the two ``half-string'' solutions. Ref.\ \cite{PMB} showed how the physical self-force may be computed from either of the two half-string solutions using a procedure that involves taking a directional (radial) limit to the particle from its ``regular'' side. This procedure may be suitable for frequency-domain calculations, where one could (in principle) integrate the relevant radial ODE from boundary conditions either on the event horizon or at infinity, towards the particle, working in the ``regular'' side of spacetime. However, the half-string reconstructions are not suitable for time-domain calculations, where one evolves the field equations as PDEs on the full exterior of the black hole.

This brings us to the ``no-string'' reconstruction, first advocated in a series of papers by Friedman and collaborators \cite{Keidl:2006wk,GSFradgauge,Keidl:2010pm}, and later formulated in detail (and received its name) in \cite{PMB}. 
%Glue $h^{\rm <}_{\pm}$ to $h^{\rm >}_{\pm}$ along $\cal S$. 
The idea is simple: take the two ``regular'' sides of the two one-string solutions, and glue them together at $\cal S$. The resulting, ``no-string'' perturbation is given by
\begin{equation}
h_{\pm}^{\rm nos} = h^{<}_{\pm}\Theta(R(t)-r)+h^{>}_{\pm}\Theta(r-R(t)),
\end{equation}
where $\Theta(\cdot)$ is the Heaviside step function. The perturbation $h_{\pm}^{\rm nos}$ is regular (smooth) in both ${\cal S}^{<}$ and ${\cal S}^{>}$, where it solves the linearised vacuum Einstein's equations. On $\cal S$ itself $h_{\pm}^{\rm nos}$ is not a vacuum solution, even away from the particle, and even when allowing arbitrary completion pieces $\Delta h_{\alpha\beta}$ in and out of $\cal S$ [see Sec.\ VI.B.1 of \cite{PMB}, where it is shown that, at least in the flat-space example, the completed no-string solution differs from a vacuum solution by a singular perturbation with a distributional support (a delta function) on $\cal S$]. However, this failure of the no-string solution to be regular---or even a valid solution---on $\cal S$ turns out to be inconsequential in practice. Ref.\ \cite{PMB} obtained a formulation of the physical self-force, complete with a practical mode-sum formula, from a no-string metric perturbation. This formulation requires information about the perturbation field (and its derivatives) only in the one-sided radial limits $r\to R(t)^\pm$, which avoid $\cal S$. It is this formulation that forms the basis for Ref.\ \cite{vandeMeent:2017bcc}'s calculation of the gravitational self-force for generic orbits in Kerr spacetime, using a frequency-domain method.

Importantly for us here, the no-string reconstruction also, in principle, enables calculations in the time domain. The idea is to solve the relevant evolution equation in each of the two vacuum regions ${\cal S}^{<}$ and ${\cal S}^{>}$, with suitable jump conditions across $\cal S$. In our method, we solve directly for the Hertz potential in the two vacuum regions---$\Phi^{<}_{\pm}$ in ${\cal S}^{<}$ and $\Phi^{>}_{\pm}$ in ${\cal S}^{>}$---with suitable jump conditions that relate between $\Phi^{<}_{\pm}$ and $\Phi^{>}_{\pm}$ on $\cal S$. The key ingredient in this formulation are, indeed, the particular jumps necessary for $\Phi^\gtrless_{\pm}$ to reproduce the no-string perturbation via 
\begin{equation}\label{h_rec<>}
h^{\rm \gtrless}_{\pm}:={\rm Re\, }{\sf\hat S}^\dagger_{\pm} \Phi^\gtrless_{\pm}.
\end{equation}
The derivation of the required jumps, for generic geodesic orbits in a Schwarzschild geometry, will be described in Sec.\ \ref{Sec:Jumps}.

First, however, we present a formulation of the evolution problem for $\Phi^\gtrless_{\pm}$ via a 1+1D decomposition, henceforth specializing to the Schwarzschild case. 

%%%%%%%%%%%%%%%%%%%%%%%%%%%%%%%%%%%%%%%%%%%%%%%%%%%%%%%%%%%%%%%%%%%%%%%%
\section{1+1D evolution scheme for the no-string Hertz potential}
\label{Sec:HertzFormulation}
%%%%%%%%%%%%%%%%%%%%%%%%%%%%%%%%%%%%%%%%%%%%%%%%%%%%%%%%%%%%%%%%%%%%%%%%

\subsection{Multipole decomposition}

We recall the IRG fields $\Phi^\gtrless_{+}$ and ORG fields $\Phi^\gtrless_{-}$ have spin weights $s=-2$ and $s=+2$, respectively. We thus expand 
$\Phi^\gtrless_{\pm}$ in $s=\mp 2$ spin-weighted spherical harmonics:
\begin{equation} \label{expansionHertz}
\Phi_\pm=
\frac{\Delta^{\pm 2}}{r}
       \sum_{\ell=2}^{\infty}\sum_{m=-\ell}^{\ell}
       \phi_{\pm}^{\ell m}(t,r)\,  {}_{\mp2}\!Y_{\ell m}(\theta,\varphi),
\end{equation}
where for the time being we omit the labels $\gtrless$ for brevity. 
The normalization factor $\Delta^{\pm 2}/r$, where $\Delta:=r(r-2M)$, is introduced (following \cite{Barack:2017oir}) to regulate the behavior of the time-radial fields $\phi_{\pm}^{\ell m}$ at infinity and on the horizon: it is such that the physical solutions (satisfying physical boundary solutions) generally approach constant nonzero values at both ends. 
The spherical basis functions ${}_{\mp2}\!Y_{\ell m}$ can be derived from standard spherical harmonics $Y_{\ell m}(\theta,\varphi)$ via 
\begin{align}\label{Y2}
{}_{\pm 2}\!Y_{\ell m}=\sqrt{\frac{(\ell-2)!}{(\ell+2)!}}\left[\frac{\partial^2 Y_{\ell m}}{\partial\theta^2}
-\left(\frac{\cos\theta\pm 2m}{\sin\theta}\right)\frac{\partial Y_{\ell m}}{\partial\theta}\right.
\nonumber\\
+\left.\left(\frac{m^2\pm 2m\cos\theta}{\sin^2\theta}\right)Y_{\ell m}\right].
\end{align}
They satisfy the differential equation
\begin{align}\label{YslmEq}
\frac{1}{\sin\theta}\frac{\partial}{\partial\theta}\left(\sin\theta\frac{\partial {}_s\!Y_{\ell m}}{\partial\theta}\right) 
+\left(-\frac{m^2+2ms\cos\theta}{\sin^2\theta}\right.
\nonumber\\
 -s^2\cot^2\theta+s+(\ell-s)(\ell+s+1)\Big){}_s\!Y_{\ell m}=0 ,
\end{align}
and the symmetry relation
\begin{equation}\label{symmetry}
{}_{\pm 2}\!\bar Y_{\ell m}=(-1)^m {}_{\mp 2}\!Y_{\ell,-m}.
\end{equation}
We also note the symmetry under reflection by the equatorial plane,
\begin{equation}\label{reflection}
{}_{\pm 2}\!Y_{\ell m}(\theta,\varphi)=(-1)^\ell{}_{\pm 2}\!\bar Y_{\ell,-m}(\pi-\theta,\varphi),
\end{equation}
to become useful further below.  

For our derivation of the jump conditions across $\cal S$ (in Sec.\ \ref{Sec:Jumps} below) we will also need a decomposition of the Weyl scalars in the same basis. Recalling  $\Psi_{\pm}$ have spin weights $s=\pm 2$, we introduce  
\begin{equation} \label{expansionpsi}
\Psi_{\pm}=
\frac{\Delta^{\mp 2}}{r}
       \sum_{\ell=2}^{\infty}\sum_{m=-\ell}^{\ell}
       \psi^{\ell m}_{\pm}(t,r)\, {}_{\pm2}\!Y_{\ell m}(\theta,\varphi).
\end{equation}
%where for simplicity we have dropped the $\ell,m$ indices off $\phi$ and $\psi_{\pm 2}$. 

In what follows we frequently drop the labels $\ell,m$ off of $\phi^{\ell m}_{\pm}$ and $\psi^{\ell m}_{\pm}$ for notational economy; it should be remembered that $\Phi$ and $\Psi$ are the full 4D fields, while $\phi$ and $\psi$ are the corresponding 1+1D reductions.

\subsection{Bardeen-Press-Teukolsky equation in 1+1D}

With the substitution (\ref{expansionHertz}), the adjoint vacuum BPT equation (\ref{Teukolski Hertz}) separates into $\ell,m$ modes, with each modal function $\phi_{\pm}(t,r)$ satisfying the 1+1D wave equation
%%~~~~~~~~~~~~~~~~~~~~~~~~~~~~~~~~~~~~~~~~~~~~~~~~~~~~~~~~~~~~~~~~~~~~~~
\begin{align} \label{Teukolsky1+1phi}
\phi^{\pm}_{,uv} + U_s(r)\, \phi^{\pm}_{,u} + V_s(r)\, \phi^{\pm}_{,v}  + W_s(r)\phi^{\pm} =0 ,
\end{align}
%%~~~~~~~~~~~~~~~~~~~~~~~~~~~~~~~~~~~~~~~~~~~~~~~~~~~~~~~~~~~~~~~~~~~~~~
with $s=\mp 2$  for $\phi_{\pm}$. Here
%~~~~~~~~~~~~~~~~~~~~~~~~~~~~~~~~~~~~~~~~~~~~~~~~~~~~~~~~~~~~~~~~~~~~~~
\begin{equation} \label{UV}
U_s(r)=-\frac{sM}{r^2},\quad\quad
V_s(r)=\frac{sf}{r},
\end{equation}
%~~~~~~~~~~~~~~~~~~~~~~~~~~~~~~~~~~~~~~~~~~~~~~~~~~~~~~~~~~~~~~~~~~~~~~
%~~~~~~~~~~~~~~~~~~~~~~~~~~~~~~~~~~~~~~~~~~~~~~~~~~~~~~~~~~~~~~~~~~~~~~
\begin{align} \label{W_Sch}
W_s(r)=\frac{f}{4}\left(\frac{(\ell+s +1)(\ell- s)}{r^2}+\frac{2(1+s)M}{r^3}\right),
\end{align}
%~~~~~~~~~~~~~~~~~~~~~~~~~~~~~~~~~~~~~~~~~~~~~~~~~~~~~~~~~~~~~~~~~~~~~~
with 
\begin{equation}
f:=1-2M/r = \Delta/r^2.
\end{equation}
We have introduced here the Eddington-Finkelstein null coordinates $v=t+r_*$ and $u=t-r_*$, where $r_*= r + 2M\ln[r/(2M)-1]$. Our convention is that, when acting on a function of $u$ and $v$, $\partial_u$ and $\partial_v$ are always taken with fixed $v$ and fixed $u$, respectively.

Similarly, in vacuum, the modal functions $\psi_{\pm}(t,r)$ of the Weyl scalars satisfy the 1+1D BPT equations
%%~~~~~~~~~~~~~~~~~~~~~~~~~~~~~~~~~~~~~~~~~~~~~~~~~~~~~~~~~~~~~~~~~~~~~~
\begin{align} \label{Teukolsky1+1psi}
\psi^{\pm}_{,uv}+ U_{s}(r)\, \psi^{\pm}_{,u} + V_{s}(r)\, \psi^{\pm}_{,v}  + W_{s}(r)\psi_{\pm} =0 ,
\end{align}
%%~~~~~~~~~~~~~~~~~~~~~~~~~~~~~~~~~~~~~~~~~~~~~~~~~~~~~~~~~~~~~~~~~~~~~~
with $s=\pm 2$  for $\psi_{\pm}$.

%%~~~~~~~~~~~~~~~~~~~~~~~~~~~~~~~~~~~~~~~~~~~~~~~~~~~~~~~~~~~~~~~~~~~~~~
%\begin{align} \label{Teukolsky1-2}
%\phi_{,uv} + \frac{2M}{r^2}\phi_{,u} - \frac{2f}{r}\phi_{,v}  + \frac{f}{4}%\left[\frac{(\ell+2)(\ell-1)}{r^2}-\frac{2M}{r^3}\right]\phi =0 .
%\end{align}
%%~~~~~~~~~~~~~~~~~~~~~~~~~~~~~~~~~~~~~~~~~~~~~~~~~~~~~~~~~~~~~~~~~~~~~~
%The (adjoint) Bardeen-Press equations ${\sf\hat O}_{\pm}\Psi_{\pm}=0$ separate in a similar way, with 
%$\psi_{-2}(t,r)$ satisfying the same equation as $\phi$, and $\psi_{+2}(t,r)$ satisfying
%%~~~~~~~~~~~~~~~~~~~~~~~~~~~~~~~~~~~~~~~~~~~~~~~~~~~~~~~~~~~~~~~~~~~~~~
%\begin{align} \label{Teukolsky1+2}
%\psi_{+2,uv} - \frac{2M}{r^2}\psi_{+2,u} + \frac{2f}{r}\psi_{+2,v}  + \frac{f}{4}%\left[\frac{(\ell+3)(\ell-2)}{r^2}+\frac{6M}{r^3}\right]\psi_{+2} =0 .
%\end{align}
%%~~~~~~~~~~~~~~~~~~~~~~~~~~~~~~~~~~~~~~~~~~~~~~~~~~~~~~~~~~~~~~~~~~~~~~

\subsection{Inversion relations in 1+1D}
\label{Sec:InvRel}

In our method we solve for the (modal) Hertz potential $\phi$ directly, making use of neither the BPT equation (\ref{Teukolsky1+1psi}) for $\psi$, nor the inversion relations that link $\psi$ to $\phi$. However, we {\it will} make use of the inversion relations in deriving jump conditions for $\phi$ across $\cal S$ (this will be done in Sec.\ \ref{Sec:Jumps}), and for that purpose we need these relations in a 1+1D form. 

The inversion relations for $\Phi_+$ and $\Phi_-$ were given in Eqs.\ (\ref{inversionIRG}) and (\ref{inversionORG}), respectively. We recall there are two alternative relations for each of the two gauges, one linking (each of) $\Phi_\pm$ to $\Psi_+$, and another linking them to $\Psi_-$. In the Schwarzschild case, the relations read
\begin{subequations}\label{invertion_rad}
\begin{eqnarray}
{\boldsymbol D}_\ell^4\bar\Phi_+ &=& 2\Psi_{+}, 
\label{invertion_rad_IRG}
\\
\Delta^{2}\tilde{\boldsymbol D}_n^4\Delta^2\bar\Phi_- &=& 32\Psi_{-} 
\label{invertion_rad_ORG}
\end{eqnarray}
\end{subequations}
(``radial'' inversion), and
%where 
%$\tilde{\boldsymbol D}_n:=-(2\Sigma/\Delta){\boldsymbol D}_n$.
%Alternatively, instead of (\ref{invertion_rad_IRG}) or (\ref{invertion_rad_ORG}), one can require
\begin{subequations}\label{invertion_ang}
\begin{eqnarray}
\bar\eth_{-1}\bar\eth_0\bar\eth_{1}\bar\eth_{2}\bar\Phi_+-12M\partial_t\Phi_{+} &=& 8\Psi_{-}, 
\label{invertion_ang_IRG}
\\
\eth_{1}\eth_0\eth_{-1}\eth_{-2}\bar\Phi_-+12M\partial_t\Phi_- &=& 8\Psi_{+}
\label{invertion_ang_ORG}
\end{eqnarray}
\end{subequations}
(``angular'' inversion).
%where
%\begin{eqnarray}
%{\cal L}_s&:=& -\left(\partial_\theta-s\cot\theta+i\csc\theta\partial_\vf\right)-ia\sin\theta\partial_t,
%\nonumber\\ 
%\tilde{\cal L}_s&:=&-\left(\partial_\theta+s\cot\theta-%i\csc\theta\partial_\vf\right)+ia\sin\theta\partial_t.
%\end{eqnarray}
The differential operators ${\boldsymbol D}_\ell$ and $\tilde{\boldsymbol D}_n$, whose general definition is given just below Eq.\ (\ref{Ocompact}), are, in the Schwarzschild case,
\begin{equation}
{\boldsymbol D}_\ell = (2/f)\partial_v, \quad\quad
\tilde{\boldsymbol D}_n = -(2/f)\partial_u .
\end{equation}
The operators $\eth_{s}$ and $\bar\eth_{s}$ are the ``spin raising'' and ``spin lowering'' angular operators defined in Eq.\ (\ref{eth}), whose action on ${}_{s}\!Y_{\ell m}(\theta,\varphi)$ is described in Eq.\ (\ref{raising&lowering}).

To separate the radial inversion relations (\ref{invertion_rad}) into multipole modes, we first take the complex conjugate of (\ref{expansionHertz}) to obtain 
\begin{eqnarray}\label{expansionbarphi}
\bar\Phi_\pm =
%&=&
\frac{\Delta^{\pm 2}}{r}
       \sum_{\ell,m}
       \bar\phi_{\pm}^{\ell,-m}(-1)^m {}_{\pm 2}\!Y_{\ell m},
%       \nonumber \\
%    &=&  \frac{\Delta^{\pm 2}}{r}
%       \sum_{\ell,m}
%       \bar\phi_{\pm}^{\ell,-m}(-1)^m {}_{\pm 2}\!Y_{\ell m},
\end{eqnarray}
where use was made of the symmetry relation (\ref{symmetry}). The expansions (\ref{expansionbarphi}) and (\ref{expansionpsi}) then separate Eqs.\ (\ref{invertion_rad}) to give, for each $\ell,m$, the fourth-order ODEs
\begin{subequations}\label{radinversion1}
\begin{equation}\label{radinversion_IRG1}
r\Delta^2 {\boldsymbol D}_\ell^4 \left(\Delta^2 \phi_+^{\ell m}/r\right) = 2 (-1)^m \bar\psi_{+}^{\ell,-m},
\end{equation}
\begin{equation}\label{radinversion_ORG1}
r\tilde{\boldsymbol D}_n^4 \left(\phi_-^{\ell m}/r\right) = 32 (-1)^m \bar\psi_{-}^{\ell,-m}.
\end{equation}
\end{subequations}
These relations can be written in a tidier form when the perturbation possesses a symmetry of refection about the equatorial plane, as in the setup to be considered in this paper: a particle source moving in the equatorial plane of the Schwarzschild black hole. In this case we have the symmetry relation
\begin{equation}\label{psi_reflection}
\bar\psi_\pm^{\ell,-m}=(-1)^\ell \psi_\pm^{\ell m},
\end{equation}
which follows from the following argument. First, we note that under the reflection transformation $\theta\to \pi-\theta$ (with fixed $t,r,\varphi$) the tetrad legs $\ell^\alpha$ and $n^\alpha$ remain invariant, while $m^\alpha\to -\bar m^\alpha$ and $\bar m^\alpha\to -m^\alpha$ [see Eqs.\ (\ref{eq:kerrtetrad})]. Inspecting Eqs.\ (\ref{eq:psi}), we see this implies $\Psi_{\pm}\to \bar \Psi_{\pm}$, assuming the perturbed Weyl tensor $C_{\alpha\beta\gamma\delta}$ is invariant under such reflection. Thus, using (\ref{reflection}), we have
\begin{align}\label{expansionbarpsi}
\Psi_\pm(\theta)&=\bar\Psi_\pm(\pi-\theta) =
\frac{\Delta^{\mp 2}}{r}
       \sum_{\ell,m}
       \bar\psi_{\pm}^{\ell m}{}_{\pm 2}\!\bar Y_{\ell m}(\pi-\theta)
       \nonumber \\
    &=  \frac{\Delta^{\mp 2}}{r}
       \sum_{\ell,m}
       \bar\psi_{\pm}^{\ell,-m}(-1)^\ell {}_{\pm 2}\!Y_{\ell m}(\theta),
\end{align}
and a comparison with (\ref{expansionpsi}) then leads to (\ref{psi_reflection}). Using (\ref{psi_reflection}) we now write the 1+1D radial inversion relation (\ref{radinversion1}) in their final form,
\begin{subequations}\label{radinversion}
\begin{equation}\label{radinversion_IRG}
r\Delta^2 {\boldsymbol D}_\ell^4 \left(\Delta^2 \phi_+^{\ell m}/r\right) = 2 p \psi_{+}^{\ell m},
\end{equation}
\begin{equation}\label{radinversion_ORG}
r\tilde{\boldsymbol D}_n^4 \left( \phi_-^{\ell m}/r\right) = 32 p \psi_{-}^{\ell m},
\end{equation}
\end{subequations}
where 
\begin{equation}
p:=(-1)^{\ell+m}
\end{equation}
is the `parity' factor.
We note (\ref{radinversion}) implies the 1+1D Hertz potentials share the same refection symmetry as the 1+1D Weyl scalars:\footnote{More precisely, Eqs.\ (\ref{radinversion}) alone imply (\ref{phi_reflection}) only up to homogeneous solutions of (\ref{radinversion}). However, no homogeneous solution of (\ref{radinversion}) satisfies the BPT equations as required, so such solutions can be excluded.}
\begin{equation}\label{phi_reflection}
\bar\phi_\pm^{\ell,-m}=(-1)^\ell \phi_\pm^{\ell m}.
\end{equation}

Let us next separate the angular inversion formulas (\ref{invertion_ang}). Using (\ref{raising&lowering}) with (\ref{expansionbarphi}) and (\ref{phi_reflection}), we have
\begin{subequations}\label{spinlowering}
\begin{eqnarray}
\bar\eth_{-1}\bar\eth_0\bar\eth_{1}\bar\eth_{2}\bar\Phi_+
&=&\frac{\Delta^{2}}{r}
       \sum_{\ell,m}p\,\lambda_2\,
       \phi_+^{\ell m} {}_{-2}\!Y_{\ell m},\quad\quad
       \\
\eth_{1}\eth_0\eth_{-1}\eth_{-2}\bar\Phi_-
 &=& \frac{\Delta^{-2}}{r}
       \sum_{\ell,m}p\,\lambda_2\,
       \phi_-^{\ell m} {}_{+2}\!Y_{\ell m} ,\quad\quad
\end{eqnarray}
\end{subequations}
where 
\begin{equation}
\lambda_n := \frac{(\ell+n)!}{(\ell-n)!}. 
\end{equation}
With this substitution, Eqs.\ (\ref{invertion_ang}) separate to give, for each $\ell,m$, the first-order ODEs
\begin{equation}\label{anginversion}
\partial_t\phi_{\pm}^{\ell m}  \mp p\alpha \phi_{\pm}^{\ell m} = \mp \frac{2}{3M}\psi_{\mp}^{\ell m},
\end{equation}
where
\begin{equation}\label{alpha}
\alpha= \frac{\lambda_2}{12M}.
\end{equation}

We note that it is obviously possible to solve (\ref{anginversion}) in closed form in terms of a time integral involving $\psi_{\mp}$ (this was the main result of Ref.\ \cite{Lousto_2002}). That, however, would not serve our purpose here. Recall that the inversion relations (\ref{anginversion}) are only valid in vacuum, and cannot be used (despite temptation) to relate the distributional contents of $\psi_{\mp}$ on $\cal S$ to these of the no-string Hertz potentials $\phi_{\pm}$. The idea, instead, is to use the inversion relations {\em evaluated in the two vacuum domains} ${\cal S}^>$ and ${\cal S}^<$ in order to get information about the jumps in $\phi_{\pm}$ across $\cal S$, given the known jumps in $\psi_{\mp}$. As we show in the next section, with some further manipulation [which also involves the radial inversion relations (\ref{radinversion})] this procedure can completely determine the jumps in $\phi_{\pm}$ and all of their derivatives on $\cal S$.\footnote{Here we use $\cal S$ to represent the curve $r=R(t)$ in the $r,t$ plane, while in Sec.\  \ref{Sec:review} it was introduced as the 2+1D sphere $r=R(t)$ in spacetime. Throughout this work we will continue to use $\cal S$ in both ways; the relevant meaning in each instance should be clear from the context. A similar remark applies to ${\cal S}^<$ and ${\cal S}^>$.}

We also note the relation (\ref{anginversion}) means that (given $\psi_{\mp}$) all time derivatives of $\phi_{\pm}$ are determinable {\em algebraically} from $\phi_{\pm}$ itself. For example, taking $\partial_t$ of (\ref{anginversion}) and then substituting for $\partial_t \phi$ back from Eq.\  (\ref{anginversion}), we find
\begin{equation}\label{phitt}
\partial_{tt}\phi_{\pm} = \alpha^2 \phi_{\pm} \mp \frac{2}{3M}\left(\partial_t \psi_{\mp}\pm p\alpha \psi_{\mp}\right).
\end{equation}
Taking $\partial_r$ of (\ref{anginversion}) similarly determines $\partial_{tr}\phi_{\pm}$ algebraically from $\phi_{\pm}$ and $\partial_{r}\phi_{\pm}$.
With the help of the vacuum BPT equation (\ref{Teukolsky1+1phi}), we can then iteratively express $\partial_{rr}\phi_{\pm}$ and all higher derivatives of $\phi_{\pm}$ algebraically in terms of $\phi_{\pm}$ and $\partial_{r}\phi_{\pm}$ alone. The significance of this in the context of this work is as follows: It means we need only determine the jumps across $\cal S$ of $\phi_{\pm}$ and of its first $r$ derivative; the jumps in all $t$, $r$ and mixed derivatives to all orders are obtainable algebraically from these two alone. 
%As a consequence, we see that the jumps in all partial derivatives of $\phi_{\pm}$ across $\cal S$ ($t$, $r$ or mixed derivatives, at any order) are determinable algebraically from the jumps in $\partial_{r}\phi_{\pm}$ and $\phi_{\pm}$ alone.

%We can write this in the form 
%\begin{equation}\label{inv_2nd}
%(\partial_t^2-\alpha^2)\phi_{\pm} = \mp \frac{2}{3M}(\partial_t \pm \alpha )\psi_{\mp}.
%\end{equation}
%and observe that the operator $\partial_t^2-\alpha^2$ is the same one featuring in the (time-domain version of the) Teukolsky-Starobinsky identities,
%\begin{eqnarray}
%\label{TS1}
%\Delta^2 \tilde{D}_n^4 \Delta^2 D_\ell^4 \Psi_{-2} &=& -144 M^2(\partial_t^2-\alpha^2)\Psi_{-2},
%%\\
%%\label{TS2}
%%\tilde{D}_\ell^4 \Delta^2 \tilde D_n^4 \Delta^2\Psi_{+2} &=& -144 M^2(\partial_t^2-\alpha^2)%\Psi_{+2}
%\end{eqnarray}
%which, in terms of the 1+1D variables $\phi^{(\pm)}$, read
%\begin{equation}
%\Delta^2 \tilde{D}_n^4 \Delta^2 D_\ell^4 (\Delta^2\phi^{(\pm)}/r) = -144 M^2(\partial_t^2-\alpha^2)%(\Delta^2\phi^{(\pm)}/r),
%\end{equation}
%CHECK $\alpha$!!!
%Indeed, (\ref{inv_2nd}) can also be derived from the {\em radial} inversion relation (\ref{radinversion}) using the TS identities by applying  $\Delta^2 \tilde{D}_n^4 \Delta^2$ to (\ref{radinversion}). Thus, the TS identities immediately reduce the 4th-order radial inversion relation to a 2nd-order ODE (in $t$), Eq.\ (\ref{inv_2nd}); but that ODE is also a direct consequence of the angular inversion and does not give an additional constraint on $\phi$.  

\subsection{Initial/boundary-value formulation}

Our strategy is to solve the 1+1D vacuum hyperbolic equation (\ref{Teukolsky1+1phi}) directly as a time evolution problem from initial data outside the black hole. The fields $\phi^\gtrless_{+}$ are to be evolved on the respective vacuum domains ${\cal S}^\gtrless$, with suitable jump conditions imposed on the timelike interface $\cal S$ (cf.\ Fig.\ \ref{uvGridScatter} below for an illustration of this setup with characteristic initial data). In principle, it suffices to impose the jumps in the field and in its first normal derivative at $\cal S$. The solution is then uniquely determined once boundary conditions are imposed at null infinity (past or future, $\mathscr{I}^\pm$), and on the event horizon (past or future, $\mathscr{H}^\pm$). 

The specific form of boundary conditions for $\phi^\gtrless_{+}$ is inherited from that of the reconstructed no-string metric $h_{\pm}^{\rm rec}$. For the applications we have in mind (e.g., a self-force calculation) it is the {\em retarded} (hereafter ``physical'') perturbation that we are after, i.e., the one corresponding to the boundary conditions of having no radiation coming in from $\mathscr{I}^-$ or out of $\mathscr{H}^-$. These requirements can be translated into asymptotic conditions on the behavior of $\phi_{\pm}^>$ at infinity, and of $\phi_{\pm}^<$ on the horizon (we are assuming here an orbit that does not plunge into the black hole, so that $\cal S$ does not cross the horizon). This analysis was carried out in Ref.\ \cite{Barack:2017oir}, and we quote the results here. For a monochromatic physical perturbation that has the asymptotic form $\sim e^{-i\omega u}/r$ at $\mathscr{I}^+$ (in a suitable Lorentzian frame) and $\sim e^{-i\omega v}$ on $\mathscr{H}^+$ (in a suitable horizon-regular frame), for some frequency $\omega>0$, the corresponding Hertz potential modes admit
\begin{eqnarray}\label{asymptPhys}
\phi_\pm^>(r\to\infty) &\sim & e^{-i\omega u}\quad \text{(physical)}, \nonumber\\
\phi_\pm^< (r\to 2M)&\sim & e^{-i\omega v}\quad \text{(physical)},
\end{eqnarray} 
Note $\phi_\pm$ generically approach constant nonzero values at $\mathscr{I}^+$ ($r\to\infty$ with constant $u$) and on the $\mathscr{H}^+$ ($r\to 2M$ with constant $v$). To achieve this convenient behavior was the purpose of introducing the radial prefactors in Eq.\ (\ref{expansionbarphi}).

For the interpretation of numerical results in Sec.\ \ref{sec:vacuum}, it will be useful to also have at hand the asymptotic behavior of ``nonphysical'' monochromatic modes, which correspond to waves coming in from $\mathscr{I}^-$, $h_{\pm}^{\rm rec}\sim e^{-i\omega v}/r$, or to waves coming out of $\mathscr{H}^-$, $h_{\pm}^{\rm rec}\sim e^{-i\omega u}$. For such solutions, the asymptotic analysis in Ref.\ \cite{Barack:2017oir} finds
\begin{eqnarray}\label{asymptNonphys}
\phi_\pm^>(r\to\infty) &\sim & r^{\mp 4} e^{-i\omega v}\quad \text{(nonphysical)}, \nonumber\\
\phi_\pm^<(r\to 2M) &\sim & \Delta^{\mp 2} e^{-i\omega u} \quad \text{(nonphysical)}.
\end{eqnarray}

The main missing ingredient in the above formulation are the jump conditions across $\cal S$. We proceed, in the next section, with a derivation of these conditions for generic geodesic orbits.

%%%%%%%%%%%%%%%%%%%%%%%%%%%%%%%%%%%%%%%%%%%%%%%%%%%%%%%%%%%%%%%%%%%%%%%%
\section{Jump conditions for the no-string Hertz potential}
\label{Sec:Jumps}
%%%%%%%%%%%%%%%%%%%%%%%%%%%%%%%%%%%%%%%%%%%%%%%%%%%%%%%%%%%%%%%%%%%%%%%%

Ref.\ \cite{Barack:2017oir} sketched a method for obtaining the jumps across $\cal S$ for a generic geodesic orbit in Kerr spacetime, but the actual jumps were only calculated for circular orbits in the Schwarzschild case. In the general case (and even in the Schwarzschild limit) the method requires the solution of a complicated set of coupled fourth-order ODEs for the jumps in $\phi$ and in $\phi_{,r}$ along the orbit. There was no attempt to solve these equations (neither analytically nor numerically), except in the circular-orbit case, where they reduce to algebraic equations. 

Here we describe a different method for obtaining the jumps, and apply it to generic orbits in the Schwarzschild case. The method yields a single first-order ODE for the jump in $\phi$ along the orbit, which can be solved in closed form. The jumps in all partial derivatives of $\phi$, at any order, are then obtained algebraically from that solution. There were two key advances that made possible this much simpler and more effective approach: First, we have found a way of utilizing both radial and angular inversion formulas in tandem, in a particular way that simplifies the calculation. Second, we have observed certain algebraic simplifications that were overlooked (by one of us) in Ref.\ \cite{Barack:2017oir}.

We consider here only the IRG Hertz potential $\Phi_+$ (as also in \cite{Barack:2017oir}), but the jumps for the ORG potential $\Phi_-$ can be worked out in just the same way. We henceforth omit the label `$+$' for notational economy, taking $\Phi\equiv \Phi_+$ and $\phi^{\ell m}(t,r)\equiv \phi^{\ell m}_+(t,r)$. We let the interface $\cal S$ be described by the smooth function $r=R(t)$, and denote the jump in $\phi(t,r)$ across $\cal S$ by
\begin{equation}
[\phi]: = \lim_{\epsilon\to 0}\left[\phi^>(t,R(t)+\epsilon) -\phi^<(t,R(t)-\epsilon)\right].
\end{equation}
The jumps in other 1+1D fields are similarly defined: $[\phi_{,r}]$, $[\phi_{,t}]$, $[\psi_{\pm}]$, etc.
We think of $[\phi]$ as a function of coordinate time $t$ along the orbit, and note the relation 
\begin{equation}\label{chainrule}
\dot{[\phi]} =  [\phi_{,t}] +\dot{R} [\phi_{,r}],
\end{equation} 
where an overdot denotes $d/dt$.
 
In what follows we assume that the jumps across $\cal S$ of the modal Weyl scalars $\psi^{\ell m}_\pm(t,r)$ and of their first 3 derivatives are already known and are given. These jumps can be obtained in a straightforward way from the source of the Teukolsky equation. We carry out this calculation in Appendix \ref{App:WeylJumps} for generic (geodesic) orbits, and for both $\psi_-$ and $\psi_+$ (as both will be needed in our approach even if we restrict to the IRG potential $\phi_+$).

\subsection{Expressions for $[\phi_{,t}]$ and $[\phi_{,r}]$ in terms of $[\phi]$}

Our task is to express each of $[\phi_{,r}]$ and $[\phi_{,t}]$ in terms of [$\phi$] alone (and possibly the known jumps in the Weyl scalars). Substitution in (\ref{chainrule}) would then give a first-order ODE for $[\phi]$. The second half of this task can be accomplished immediately thanks to the angular inversion formula (\ref{anginversion}). We obtain 
\begin{equation}\label{Jphit}
[\phi_{,t}]   = p\alpha [\phi] -\frac{2}{3M}[\psi_{-}].
\end{equation}
The jump $[\psi_{-}]$ is given in Eq.\ (\ref{Jpsipm}) of Appendix \ref{App:WeylJumps}, and recall $\alpha=\lambda_2/(12M)$. To obtain $[\phi^{(\pm)}_{,r}]$ in terms of $[\phi]$ is harder, and utilizes the fourth-order radial inversion (\ref{radinversion_IRG}), using a procedure we now describe.

First, we write (\ref{radinversion_IRG}) more explicitly in terms of coordinate derivatives. Using ${\boldsymbol D}_\ell = 2(r^2/\Delta)\partial_v$ (taken with fixed $u$), a calculation yields 
\begin{align}\label{inversion_explicit}
\partial_v^4 \phi&
+\frac{2}{r^2}(3r-5M)\partial_v^3 \phi
+\frac{1}{r^4}(9r^2-26Mr+15M^2)\partial_v^2 \phi \nonumber \\
+& \frac{1}{2r^5}(6r^2-21 Mr+16 M^2)\partial_v \phi
= \frac{p}{8r^8}\psi_{+}.
\end{align} 
We now act with $\partial_u$ (fixed $v$) on both sides of (\ref{inversion_explicit}), and use the vacuum BPT equation (\ref{Teukolsky1+1phi}) to substitute for each mixed derivative $\phi_{,uv}$ in terms of $\phi_{,u}$, $\phi_{,v}$ and $\phi$. In the resulting expression we then substitute for $\partial_v^4 \phi$ from Eq.\ (\ref{inversion_explicit}). We arrive at a third-order ODE of the form 
\begin{equation}\label{inversion_3rd}
\sum_{n=0}^3 \hat A_n(r) \partial_v^n \phi
= \sum_{n=0}^1 \hat B_n(r)\partial_u^n \psi_{+},
\end{equation} 
where $\hat A_n(r)$ and $\hat B_n(r)$ are certain (rational) functions. Notably, no $u$ derivatives occur on the LHS. Repeating this procedure with a second application of $\partial_u$, this time replacing $\partial_v^3 \phi$ from Eq.\ (\ref{inversion_3rd}), yields a second-order ODE of the form 
\begin{equation}\label{inversion_2nd}
\sum_{n=0}^2 \tilde A_n(r) \partial_v^n \phi
= \sum_{n=0}^2 \tilde B_n(r)\partial_u^n \psi_{+},
\end{equation} 
with some other (rational) functions $\tilde A_n(r)$ and $\tilde B_n(r)$. Again, we find that no $u$ derivatives occur on the LHS. One last application of $\partial_u$ reduces the inversion relation to a first-order differential equation, which, however, is now a PDE, since (it turns out) is features both $\phi_{,u}$ and $\phi_{,v}$. We can, however, reduce this to an ODE by first converting to $r_*$ and $t$ derivatives using
$\phi_{,v}=\frac{1}{2}(\phi_{,t}+\phi_{,r_*})$ and $\phi_{,u}=\frac{1}{2}(\phi_{,t}-\phi_{,r_*})$, and then eliminating $\phi_{,t}$ using the angular inversion relation (\ref{anginversion}). This leads to a first-order ODE for $\phi$, which has the form
\begin{equation}\label{1stradial}
\phi_{,r_*} + A(r)\phi = p \sum_{n=0}^3  B_n(r)\partial_u^n \psi_{+} + B(r)\psi_{-}.
\end{equation}
An explicit calculation gives 
\begin{equation}
A(r)=-\alpha+\frac{2M\left[(2\lambda_1-3)r-6M\right]}{r^2(\lambda_1 r-6M)}
\end{equation}
for odd-parity modes ($p=-1$), and 
\begin{align}
&A(r)=\alpha+
\nonumber\\ 
& \ \ \frac{M\left[
4r^3\alpha^2(2\lambda_1+3)+2r^2\alpha \lambda_1(\lambda_1+4)+\lambda_1^2(3r-2M)\right]}
{r^2\left[2\alpha^2 r^2(\lambda_1 r+6M)+M\lambda_1(6\alpha r+\lambda_1)\right]} 
\end{align}
for even-parity modes ($p=+1$).
The other radial coefficients in Eq.\ (\ref{1stradial}) are found to be given by
\begin{eqnarray}
B_0 &=&\lambda_1 f^3 r(\lambda^2 r^2+3M\lambda r+6M^2)/C(r) \ ,
\nonumber\\
B_1 &=& \big[18\alpha\lambda M r^4 + 8M^2 r^3\alpha(9-7\lambda) \nonumber \\ 
&& \quad +M^2 r^2(4\lambda^3-9\lambda^2-31\lambda+24)  \nonumber \\ 
&& \quad +2M^3 r(\lambda+3)(7\lambda-13) \nonumber \\
&& \quad +12M^4(\lambda+5)\big]/C(r)\ ,
\nonumber\\
B_2 &=& 12M r^2 \big[\alpha\lambda r^3-\alpha(\lambda-5)Mr^2  \nonumber \\ 
&& \quad +2(1-2M\alpha)Mr -4M^2 \big]/C(r)\ ,
\nonumber\\
B_3 &=& 4Mr^3\left[\alpha r^2(\lambda r+6M)+3Mr f\right]/C(r) \ ,
\nonumber\\
B &=&  -4f^3 M^2 r^2\big[6\alpha r^4(\alpha^2 r^2+\lambda)-r^3\lambda(\lambda^2-4) \nonumber \\
&& \quad  +9M r^2(1-6M\alpha)-36M^2(r-M) \big]/C(r) , \nonumber\\
\end{eqnarray}
with
\begin{equation}
C=18 M^3 f^3 r^4\alpha \left[-2\alpha^2 r^4-\lambda r^2+2Mr(\lambda-1)+6M^2\right].
\end{equation}
Here we have introduced 
\begin{equation}
\lambda:=\lambda_2/\lambda_1 = (\ell+2)(\ell-1),
\end{equation}
and we remind $\lambda_1 = \ell(\ell+1)$ and $\alpha= \lambda_2/(12M)$.

Using Eq.\ (\ref{1stradial}) we can finally express the jump $[\phi_{,r_*}]=f(R)[\phi_{,r}]$ in terms of the jump $[\phi]$ (and the known jumps in the Weyl scalars $\psi_\pm$):
\begin{equation}\label{Jphir}
[\phi_{,r_*}]   = - A(R) [\phi] 
+p\sum_{n=0}^3  B_n(R)[\partial_u^n \psi_{+}] + B(R)[\psi_{-}].
\end{equation}

\subsection{First-order ODE for $[\phi]$ and its solution}

Substituting (\ref{Jphit}) and (\ref{Jphir}) in (\ref{chainrule}) now gives a simple first-order ODE for $[\phi]$ as a function along the orbit:
\begin{equation}\label{ODE}
\dot{[\phi]} + \Big(A(R)\dot{R}_*-p \alpha\Big)[\phi] = {\cal F}
\end{equation}
where $\dot{R}_*=\dot{R}/f(R)$. The source term here is 
\begin{equation}\label{calF}
{\cal F} =  p \dot{R}_*\sum_{n=0}^3  B_n(R)[\partial_u^n \psi_{+}] + \left(\dot{R}_* B(R)-\frac{2}{3M}\right)[\psi_{-}].
\end{equation}

Equation (\ref{ODE}) admits simple homogeneous solutions, given by (any constant multiple of)
\begin{equation}\label{[phi]AsympEven}
[\phi]_{\rm h}= \left(\frac{R(\lambda_1 R-6M)}{(R-2M)^2}\right)\times e^{-\alpha(t-R_*)}
\end{equation}
for odd-parity modes, or
\begin{align}\label{[phi]AsympOdd}
[\phi]_{\rm h}= \left(\frac{R^3\lambda^2\lambda_1+6MR^2\lambda^2+36M^2 R\lambda+72M^3}{R(R-2M)^2} \right)\nonumber\\ \times e^{\alpha(t-R_*)}
\end{align}
for even-parity modes.
The general inhomogeneous solution of (\ref{ODE}) reads
\begin{equation}
[\phi] = [\phi]_{\rm h}\int_{t_0}^t \frac{{\cal F}(t')}{[\phi]_{\rm h}}\, dt',
\end{equation} 
where $t_0$ is an a-priori arbitrary integration constant. We determine $t_0$ from the physical requirement that $[\phi]$ remains bounded for $t\to\pm\infty$. Observing that $[\phi]_{\rm h}$ blows up like $e^{\pm\alpha t}$ at  $t\to\pm \infty$ ($+$ for even parity modes, $-$ for odd-parity modes), while ${\cal F}(t)$ is at worst polynomial in $t$, it is easy to see that the requirement of boundedness necessitates $t_0=\pm\infty$ for $p=\pm 1$. 
%is to be determined from boundary conditions. Our choice of lower integration boundary ensures convergence of the integral, noting $[\phi^{\pm}]_{\rm h}$ blow up as $e^{\pm\alpha t}$ as $t\to \pm\infty$, while $F^{\pm}$ is  at worst polynomial in $t$. It is clear that the requirement that $[\phi^{(\pm)}]$ should not blow up exponentially neither at $t\to\infty$ nor at $t\to -\infty$ demands $C^\pm=0$.  
Hence, the unique physical solution of (\ref{ODE}) is  
\begin{equation}\label{HertzJump}
[\phi] = [\phi]_{\rm h}\int_{\pm\infty}^t \frac{{\cal F}(t')}{[\phi]_{\rm h}} dt' \quad \text{(for $p=\pm 1$)}.
\end{equation} 

Equation (\ref{HertzJump}) gives the jumps across $\cal S$ that the no-string Hertz potential modes must satisfy, for an arbitrary orbit in Schwarzschild spacetime. (It requires as input the jumps in the modes of the Weyl scalars, which in Appendix \ref{App:WeylJumps} we give explicitly specialised to {\em geodesic} orbits; but given the Weyl scalar jumps, there is no further assumption on whether the orbit is geodesic.) This is one of the main results of this paper.

We recall that the jumps in the field's derivatives, $[\phi_{,t}]$ and $[\phi_{,r}]$ (or $[\phi_{r_*}]$), can be obtained algebraically from $[\phi]$, using Eqs.\ (\ref{Jphit}) and (\ref{Jphir}), respectively. In principle, knowledge of the jumps in the field and its first derivatives should suffice in our formulation. However, in practice it is also useful to have at hand the jumps in higher derivative, which eases the formulation of finite-difference schemes that have high-order convergence properties. Once the jumps in the field and its first derivatives are known, it is straightforward to obtain the jumps in higher derivatives in an iterative manner using the procedure described in the last paragraph of Sec.\ \ref{Sec:InvRel} [the paragraph containing Eq.\ (\ref{phitt})]. The application of this procedure up to third derivatives is illustrated in Appendix \ref{app:WeylJumpHighDerivatives} (as applied to modes of the Weyl scalars). 

%\subsection{Jumps in higher derivatives of $\phi$}

%Given the solution to (\ref{ODE}) it is easy to obtain jump conditions for derivatives of $\phi$, such as ones needed in (\ref{RWODESource}). Using Eqns. (\ref{Jphit}) and (\ref{Jphir}) we can write
%\begin{align}
%[\phi^{(\pm)}_{,v}] =& \frac{1}{2} \left([\phi^{(\pm)}_{,t}] + f [\phi^{(\pm)}_{,r}] \right)\nonumber \\
%=& \frac{1}{2} \Big[\left(- p_\pm(R) \pm \alpha \right) [\phi^{(\pm)}] + \left(q(R)-\frac{2}{3M} \right) [\psi_{-2}^{(\pm)}] \nonumber \\
%& \pm \sum_{n=0}^3  q_n(R)[\partial_u^n \psi^{(\pm)}_{+2}] \Big].
%\end{align}
%We can then obtain the full jump condition using $[\phi_{,v}] = [\phi^{(+)}_{,v}] +[\phi^{(-)}_{,v}]$. Similarly, we can derive an expression for $[\phi^{(\pm)}_{,u}] = \frac{1}{2} ([\phi^{(\pm)}_{,t}] - f [\phi^{(\pm)}_{,r}] )$ and obtain $[\phi_{,u}]$. Any jumps in higher order derivatives we can get using the same method to get jumps in high order derivatives of the Weyl scalars as described in Appexdix \ref{app:WeylJumpHighDerivatives}.

\subsection{Large-$R$ asymptotics for scatter orbits}
\label{subsec:asymptotics}

We were not able to evaluate the integral in (\ref{HertzJump}) analytically for a generic orbit, but it is straightforward to compute $[\phi](t)$ numerically for any given geodesic orbit. In practice we find it easier to obtain $[\phi]$ by (numerically) solving the first-order ODE (\ref{ODE}). For the class of scatter orbits of interest to us in this paper, we need to integrate the equation over $-\infty < t< \infty$. We choose to do so forward in time for odd-parity modes but backward in time for even-parity modes, in each case going ``against" the direction of exponential growth of the homogeneous solutions (\ref{[phi]AsympEven}) and (\ref{[phi]AsympOdd}). This prevents the growth of nonphysical modes from numerical error. We now derive the leading-order asymptotic form of $[\phi]$ at $t\to \pm\infty$. One of these two asymptotic values will be used as an initial value for the ODE solver, and the other will be used to check the result of the numerical integration.

We consider a timelike scatter geodesic orbit in Schwarzschild spacetime, parametrized by specific energy $E>1$ (``gamma factor'') and angular momentum $L$ (a more detailed description of such orbits will be given in Sec.\ \ref{EoM} below). We let
\begin{equation}
\Rdotinf:= \pm\left|\dot{R}(t\to\pm\infty)\right| = \pm\frac{\sqrt{E^2-1}}{E}
\end{equation}
be the ``velocity at infinity'' (with respect to coordinate time $t$), so that $\Rdotinf$ is negative (positive) for the inbound (outbound) asymptotic states. We formally expand $[\phi]$ as a power series in $1/R$ at large $R(t)$, and seek to obtain the leading term of that expansion. 

To this end, we first obtain the large-$R$ asymptotic form of $\cal F$ in Eq.\ (\ref{calF}). Using as input the asymptotic expressions derived in Appendix \ref{WeylJumpsAsympt} for $[\psi_-]$ and $[\partial^n_u\psi_+]$ ($n=0,\ldots,3$), a direct calculation leads to
\begin{equation}\label{calFasy} 
{\cal F} = c_0 R^{-3} +O(R^{-4}),
\end{equation}
where
\begin{align}\label{c0_initial}
c_0 = &\frac{4\pi\mu(1+\Rdotinf)}{3\lambda_2}
\Big[
i(L/M)\lambda_2 \Big(\partial_\theta - m \Big){}_{-2}\! \bar{Y}_{\ell m} \nonumber \\ 
& + 6E\dot{R}_{\infty}
\Big(\partial_{\theta\theta} -2m\partial_{\theta} +(m^2-2)  \Big) {}_{-2}\! \bar{Y}_{\ell m}
\Big].
\end{align}
Here all angular functions are evaluated at $\theta=\pi/2$ and $\varphi=\varphi_{\rm in}$ (or $\varphi=\varphi_{\rm out})$, with $\varphi_{\rm in}$ ($\varphi_{\rm out}$) being the asymptotic value of the particle's azimuthal phase at $t\to -\infty$ ($t\to +\infty$). 
%where $\varphi_{\rm in}:=\varphi(\chi\to-\chi_\infty)$, and similarly for the first and second $\theta$ derivatives. 
%\begin{eqnarray}
%a_- &=& -\frac{4i\pi\mu L(1-\Rdotinf)}{3M}\Big(m {\cal Y}^{(-2)} - {\cal Y}_{\theta}^{(-2)}\Big),
%\nonumber\\
%a_+ &=& \frac{8\pi\mu E \vinf(1-\Rdotinf)}{\lambda_2}\Big( (2m^2-\lambda){\cal Y}^{(+2)}+2m {\cal Y}_\theta^{(+2)}\Big).
%\end{eqnarray}
Equation (\ref{c0_initial}) takes a neater form when written in terms of spin-$0$ spherical harmonics. With the aid of (\ref{Y2}), we find
\begin{align}
c_0 = \frac{4\pi\mu (1+\Rdotinf)}{3M\sqrt{\lambda_2}}
\left[\Big(6ME\Rdotinf -im\lambda L \Big)\bar Y -i\lambda L \bar Y_{\theta}\right], 
%\nonumber\\
%a_- &=& -\frac{4i\pi\mu \lambda L (1+\Rdot^{\infty})}{3M\sqrt{\lambda_2}}\, \bar Y_{\theta}.
\end{align}
where $\bar Y:= \bar Y_{\ell m}\!\left(\frac{\pi}{2},\varphi_{\rm in/out}\right)$ and $\bar Y_\theta:= \partial_\theta\bar Y_{\ell m}\!\left(\frac{\pi}{2},\varphi_{\rm in/out}\right)$.

%We also have $[\phi^-]_{\rm h}\simeq \lambda_1 e^{-\alpha u}$ and $[\phi^+]_{\rm h}\simeq \lambda_1 \lambda^2 e^{+\alpha u}$. 
The asymptotic form of $[\phi]$ can now be obtained either by evaluating (\ref{HertzJump}) with the asymptotic form (\ref{calFasy}), or directly from the ODE (\ref{ODE}) using a power-law ansatz. Either way, we arrive at 
\begin{align}
\label{JphiAsymp}
[\phi]_{R\to\infty} =&
-\frac{16\pi\mu}{\lambda_2^{3/2}}\left(\frac{1+\dot{R}_{\infty}}{1-\dot{R}_{\infty}}\right)
\Big[6M \dot{R}_{\infty}E \bar Y \nonumber \\ 
& \qquad + i \lambda L \Big(\bar Y_{\theta}- m \bar Y\Big)
 \Big] R^{-3} + O(R^{-4}).
\end{align}
We note $\bar Y_{\theta}=0$ for even-parity modes, and $\bar Y=0$ for odd-parity modes. 
Our result (\ref{JphiAsymp}) can be checked against the $m=0$, circular-orbit expression given in Eq.\ (87) of Ref.\ \cite{Barack:2017oir}, by setting $\dot{R}_{\infty}=0$, $r_0=R$, $\Omega=\sqrt{M/R^3}$ and ${\cal Y}_\theta = -\lambda\bar Y_{\theta}/\sqrt{\lambda_2} $. We find an agreement.

In Sec.\ \ref{sec:RWformulation}, for reasons that will become clear there, we will require also the asymptotic forms of the jumps $[\phi_{,v}]$ and $[\phi_{,vv}]$. The jump $[\phi_{,v}]=\frac{1}{2}([\phi_{,t}]+[\phi_{,r_*}])$ is obtained using Eqs.\ (\ref{Jphit}) and (\ref{Jphir}) with the known asymptotic expressions for $[\phi]$, $[\psi_-]$ and $[\partial^n_u\psi_+]$. The result is
\begin{equation}\label{JphivAsymp}
[\phi_{,v}]_{R\to\infty} = \frac{4\mu\pi E (1+\Rdotinf)}{\sqrt{\lambda_2}}\bar Y R^{-3} + O(R^{-4}).
\end{equation}
The asymptotic form of $[\phi_{,u}]$ is obtained in a similar way.
The jump $[\phi_{,vv}]$, in turn, can be written in terms of lower-derivative jumps as explained in the last paragraph of Sec.\ \ref{Sec:InvRel}, and, substituting the asymptotic expressions already obtained for these, one finds
\begin{equation}\label{JphivvAsymp}
[\phi_{,vv}]_{R\to\infty} = -\frac{4\mu\pi E (2+\Rdotinf)}{\sqrt{\lambda_2}}\bar Y R^{-4} + O(R^{-5}).
\end{equation}

%%%%%%%%%%%%%%%%%%%%%%%%%%%%%%%%%%%%%%%%%%%%%%%%%%%%%%%%%%%%%%%%%%%%%%%%
\section{Time-domain evolution of the Teukolsky equation: problem of growing modes} \label{sec:vacuum}
%%%%%%%%%%%%%%%%%%%%%%%%%%%%%%%%%%%%%%%%%%%%%%%%%%%%%%%%%%%%%%%%%%%%%%%%

\subsection{Numerical method}
\label{TeukNumMethod}

Our aim, in the remainder of this paper, is to demonstrate the applicability of our strategy with an end-to-end numerical implementation. First, in this section, we implement a simple finite-difference Teukolsky solver based on 1+1D characteristic evolution in $u,v$ coordinates. The basic architecture of the code is similar to that of the one developed in \cite{Barack:2017oir}, but our new code can handle any orbit (\cite{Barack:2017oir} had circular orbits hardwired into it), and can evolve both the IRG ($s=-2$) and the ORG ($s=+2$) Hertz potentials (\cite{Barack:2017oir} dealt only with $s=-2$).  We have produced two identical implementations, one in \texttt{Mathematica} and another in \texttt{C++}, to enable cross-checks.  

\begin{figure}[h!]
\centering
\includegraphics[width=\linewidth]{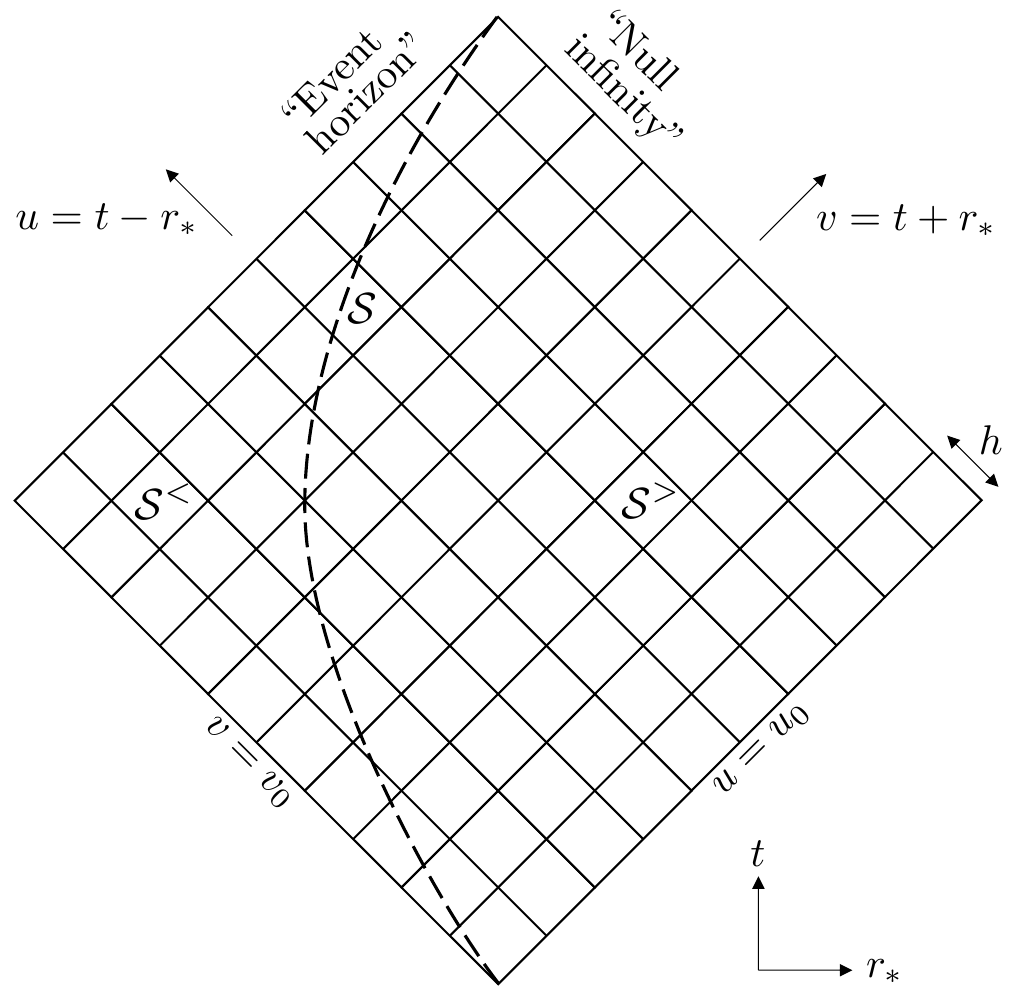}
\caption{Sketch of the 1+1D characteristic grid used in our numerical evolution of the no-string Hertz potentials $\phi^{\ell m}_{\pm}(t,r)$ outside a Schwarzschild black hole. The grid lines are uniformly spaced in Eddington-Finlkelstein coordinates $u,v$. Initial conditions are set on the rays $u=u_0$ and $v=v_0$. The dashed line represents the particle's worldline (or, equivalently, the 1+1D reduction of the surface $\cal S$ interfacing between the vacuum regions ${\cal S}^{\rm \gtrless}$) for a typical hyperbolic orbit. The evolution proceeds along successive $u={\rm const}$ rays, with appropriate jump conditions imposed across $\cal S$. }
\label{uvGridScatter}
\end{figure}

The numerical domain is depicted in Fig.\ \ref{uvGridScatter}. We use a fixed characteristic mesh, with uniform grid-cell dimensions $h\times h$, where $h$ is a small fraction of $M$ (typically $\sim M/10$ to $\sim M/100$ in our test runs). Characteristic initial data are set on two initial rays $v=v_0$ and $u=u_0$ (see the figure), chosen so that $\cal S$ intersects the initial vertex $(v_0,u_0)$. The data is evolved using a finite-difference version of Eq.\ (\ref{Teukolsky1+1phi}) that has a local discretisation error of $O(h^4)$, leading to a quadratic convergence globally (i.e., the accumulated error scales like $h^2$).  Our finite-difference scheme is precisely identical to the one used in \cite{Barack:2017oir} (as detailed in Appendix B therein) when applied to circular orbits. 
A detailed description of our scheme for generic orbits is provided in Appendix \ref{App:RWCode} (as applied to the modified version of the field equation that we end up solving in practice; see below). The appropriate jump conditions across $\cal S$ are implemented at the level of the finite-difference formula when it is applied to grid cells containing a segment of $\cal S$, as detailed in Appendix \ref{app:FDSNonVac}. Our code takes as input the spin $s=\pm 2$, modal numbers $\ell$, $m$, and orbital trajectory $R(t)$ (as well as a range of numerical parameters such as $h$ and the coordinate ranges), and returns the Hertz potential modes $\phi^+_{\ell m}(t,r)$ (IRG) or $\phi^-_{\ell m}(t,r)$ (ORG). 

The physical initial data for the evolution are not known in general, so we start with fictitious data on $v=v_0$ and $u=u_0$.  Specifically, we set $\phi^>(v,u_0)\equiv 0$ and $\phi^<(v_0,u)\equiv 0$, and let the field be sourced by the imposed jumps along $\cal S$. This produces an outburst of ``junk'' radiation at the initial vertex $(v_0,u_0)$. We expect such junk radiation to decay in time (with an inverse power law $\sim t^{-2\ell-3}$, theoretically \cite{Barack_1999}), leaving behind the desired physical solution at late time. The early, junk-contaminated part of the data is discarded. 

Since our characteristic numerical domain has no timelike boundaries, there is no need to impose boundary conditions, and no way to actively control violations away from the desired retarded solution. This is not a problem when all other (``nonphysical'') vacuum solutions of the field equation decay at late time, but can become a problem when there exist nonphysical solutions that fail to decay, or, worse, grow at late time. Past implementations of 1+1D characteristic schemes for the scalar field equation (e.g., \cite{Haas:2007kz}), electromagnetic vector potential \cite{Haas:2011np}, Regge-Wheeler-Zerilli equations (e.g., \cite{Martel:2003jj}) and the Lorenz-gauge metric perturbation equations (e.g., \cite{bsago2}) show no signs of such troublesome modes. In these cases, the numerical solutions always appear to converge to the true, physical solution at late time.\footnote{Sole exceptions known to us are certain monopole and dipole gauge modes of the Lorenz-gauge metric perturbations, which grow linearly in time \cite{Dolan}.} As we demonstrate below, the situation with the $|s|=2$ Teukolsky equation is less fortunate: In our simple $u,v$-coordinate-based evolution, nonphysical modes of the equation, seeded by numerical error, will grow unbounded at late time, spoiling the evolution.  We will discuss the origin of the problem and suggest ways around it.

%However, as we demonstrate in Sec.\ \ref{sec:vacuum}, the situation with the BPT equation (\ref{Teukolsky1+1phi}) is less fortunate. In a straightforward uni-grid characteristic evolution (without compactification), nonphysical modes of the BPT equation will grow unbounded at late time, spoiling the evolution.  We will discuss the origin of the problem and suggest a way around it.

Why has the problem not identified already in Ref.\ \cite{Barack:2017oir}, which used the same numerical method? 
Our new code, when run with a circular-orbit source and $s=-2$, does reproduce the numerical results of \cite{Barack:2017oir} in the early stage of the evolution, before the onset of growth.  We no longer have access to the code used in \cite{Barack:2017oir}, but it appears that the evolutions performed in that study were simply too short to reveal the problem: The calculation of the Hertz potential along circular orbits did not require very long runs, and evolutions were always terminated before the relatively slowly growing mode ($\sim t^4$ for $s=-2$; see below) had a chance to manifest itself in the data.  Calculations for scatter orbits require much longer evolutions, so here we must deal with the problem. The problem must be dealt with anyway if one is interested in an ORG reconstruction $(s=+2)$, where, as discussed below, the blow-up is exponential. 

In what follows we illustrate the problem of growing modes with numerical examples and describe the range of tests we performed to understand its origin. We then discuss possible remedies. Since the issue arises already in vacuum evolutions---indeed, it is more easily seen in the absence of a particle source---we restrict the discussion in the rest of this chapter to the vacuum case.

\subsection{The case $s=-2$ (vacuum)}

\begin{figure}[t]
\includegraphics[width=\linewidth]{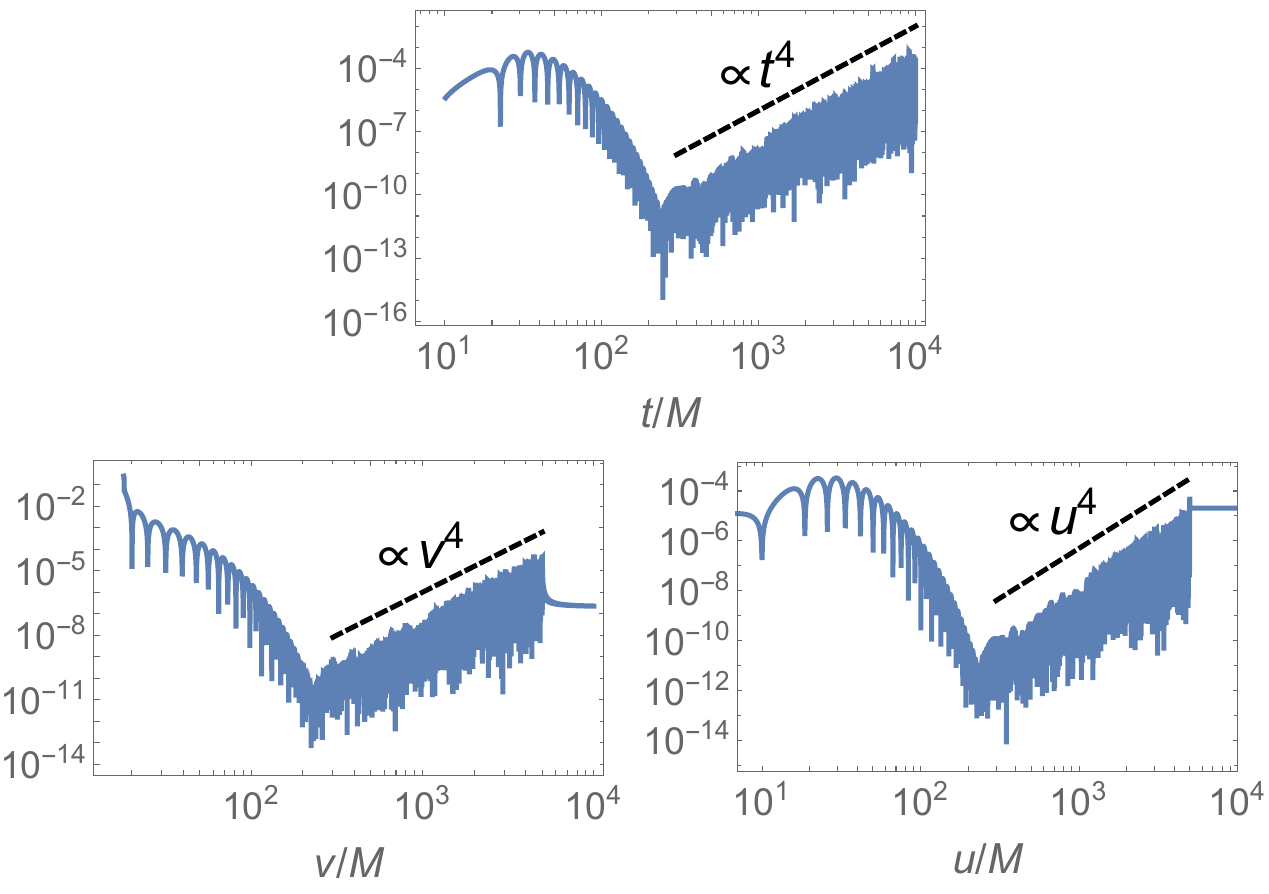}
\caption{Results from evolution of the $(\ell,m)=(2,0)$ mode of the vacuum 1+1D BPT equation with $s=-2$. The evolution is seeded with a narrow Gaussian near $(t,r_*)\sim (0,9M)$. We show, on a log-log scale, the field amplitude $|\phi^-_{20}|$ sampled along slices of constant $r_*=10M$ ({\it top}), $u=500M$ ({\it lower left}) and $v=500M$ ({\it lower right}). The dashed lines (${\rm const}\times t^4$, ${\rm const}\times v^4$ and ${\rm const}\times u^4$, respectively) are shown for reference.}
\label{TeukVacsm2}
\end{figure}

Figure \ref{TeukVacsm2} shows a typical output from an $s=-2$ numerical evolution in vacuum, i.e., setting all jumps across $\cal S$ to zero. We seed the evolution with a narrow Gaussian pulse near the initial vertex at $(u_0,v_0)=(-9M,9M)$ [corresponding to $(t,r)\sim (0,7.12M)$], and evolve out to $(u,v)=(10^4M,10^4M)$. After the initial spike of radiation (not shown in the figures), the field decays with characteristic quasinormal ringing. However, at around $t\sim 250M$ the solution becomes dominated by a noisy component, whose amplitude appears to grow approximately like $\sim t^4$. The growth seems to continue indefinitely towards future timelike infinity ($t\to\infty$ with fixed $r>2M$), but the solution settles to a finite value approaching null infinity ($v\to\infty$ with fixed $u$), and also approaching the event horizon ($u\to\infty$ with fixed $v$). A similar behavior is observed for all values of $\ell$ and $m$ and irrespectively of the choice of compact initial data. The evolution up to the onset of growth is numerically stable and the solution there converges quadratically in grid spacing $h$, as expected. The growing component, however, is not numerically stable: It displays noisy features on grid-size scale, and its amplitude appears to {\em increase} with decreasing $h$ (finer resolution). The $\sim t^4$ behavior, however, seems to be persistent and universal.      

We have performed a series of tests in attempt to understand these results. First, as mentioned, we have tried a variety of initial data, including a point seed at the initial vertex, smooth Gaussians of various configurations, and data corresponding to an exact static solution of the BPT equation.  Second, we have tried a range of alternative finite-difference formulas and stepping schemes. Third, we have used our code to solve for the Weyl scalar modes $\psi^-(t,r)$ with jump conditions on $\cal S$ corresponding to a circular geodesic orbits (the necessary jump conditions are derived in Appendix \ref{App:WeylJumps}); we have done so both with ``zero'' initial data, and with data corresponding to an exact static solution (for an $m=0$ mode). In all these tests, the $t^4$ growing mode developed just the same. Fourth, we note that the troublesome $t^4$ behavior is observed \cite{priv_comm_Conor} also in the application to the Teukolsky equation of the recently introduced approach by O'Toole {\it et al.}\ \cite{otoole2021characteristic}, in which the Green function (rather than the field itself) is evolved from exact characteristic initial data. Finally, we observe that we are, in fact, able to successfully suppress the $t^4$ growth (albeit at considerable computational cost) using our \texttt{Mathematica} implementation with very high working precision. All this supports the conclusions that (i) the $t^4$ behavior has a genuine dynamical origin, and (ii) the $t^4$ component is seeded by numerical roundoff error. 

We suggest that the $t^4$ mode represents nonphysical incoming radiation sourced by numerical roundoff error near $\mathscr{I}^+$. This can be seen from the following heuristics. First, recall from Eqs.\ (\ref{asymptPhys}) and (\ref{asymptNonphys}) the asymptotic form of monochromatic $s=-2$ solutions in the ``wave zone'' ($r\gg M$ with $v\gg u$): $\phi\sim e^{-i\omega u}$ for physical solutions (purely outgoing waves) and $\phi\sim r^{-4}e^{-i\omega v}$ for nonphysical solutions representing purely incoming waves. More generally, the time-domain solutions are superpositions of such monochromatic modes, and have the forms $\phi\sim F(u)$ (physical) and $\phi\sim r^{-4}G(v)$ (nonphysical) for some functions $F(u)$ and $G(v)$ that depend on the initial data. [These forms can be confirmed more directly by substituting the ans\"{a}tze $\phi=r^{\alpha}F(u)$ and $\phi=r^{\beta}G(v)$ into the BPT equation (\ref{Teukolsky1+1phi}) and solving at leading order in $M/r$ under the wave-zone assumptions $F'(u)\gg F(u)/r$ and $G'(v)\gg G(v)/r$, to obtain $\alpha=0$ and $\beta=-4$.] Consider an outgoing ray $u=\text{const}$ shortly after the start of the evolution. The field on this ray is composed mostly of outgoing radiation $\phi\sim F(u)$, which approaches a constant value at large $v$. However, roundoff error in the numerical data along this ray will inevitably source a small component of nonphysical high-frequency incoming radiation $\phi\sim r^{-4}G(v)$. Since the sourcing field is asymptotically constant at $v\to\infty$, the amplitude of the seeded incoming radiation is also expected to be asymptotically constant on the $u={\rm const}$ ray, i.e.\ $r^{-4}|G(v)|\sim {\rm const}$ for $v\to\infty$. This implies $|G(v)|\sim v^4 \sim(t+r)^4$ at large $v$, and it follows that the incoming-wave component has an amplitude $|\phi|\sim v^4/r^4\sim (t+r)^4/r^4$. At fixed $r$, this will exhibit a $\sim t^4$ growth, at least in the wave zone where our heuristic analysis applies. (To show that this wave-zone behavior might lead to a $t^4$ growth elsewhere at late time, as evident in the numerical data, would require a more detailed asymptotic matching analysis, which we have not attempted.)

This heuristic description explains the results of our various experiments. The $t^4$ behavior arises dynamically from roundoff error seeds, so it is persistent, universal and independent of initial data. The amplitude of the $t^4$ component can be suppressed by increasing the precision of the floating-point arithmetic, which reduces the roundoff error. For a fixed floating-point precision, increasing the grid resolution (decreasing $h$) enhances the amplitude of the incoming radiation component, by seeding more of its modes at higher frequencies.     

It is also possible to explain why the $t^4$ growth does not appear to plague other time-domain treatments of the $s=-2$ Teukolsky equation reported in the literature. In the 2+1D Cauchy evolution approach of Khanna {et al.}\ (legacy of \cite{LopezAleman:2003ik,Khanna:2003qv} and many works since), boundary conditions are actively imposed, which presumably suppress the growth of the unphysical component. In the compactified hyperboloidal slicing approaches of Refs.\ \cite{Racz:2011qu,Zenginoglu:2012us,Harms2013}, we suspect it is the {\it compactification} of $\mathscr{I}^+$ that averts the problem, since the wavezone for incoming waves is vastly underresolved on the compactified grid. In contrast, our simple $u,v$-coordinate-based approach resolves the wavezone equally well for both outgoing and incoming waves. Unfortunately, as we have seen, the resolution of incoming waves near null infinity is harmful in our case.

%We can also consider the asymptotic limit near null infinity. Substituting the ansatz $\phi = C_3 t^\alpha e^{-i\omega v}$, where $C_3$, $\alpha$ and $\omega$ are constants with $\omega \ll r$, into (\ref{Teukolsky1+1phi}) gives
%\begin{align}
%\frac{f (2M (1 \mp 2)+r (\lambda_1 -4 \pm 2 (1 +4 i r \omega )))}{4 r^3} &\\ \nonumber +\frac{ \mp 2 \alpha  (r-3 M)- i r^2 \alpha  \omega}{2r^2 t}+\frac{(\alpha -1) \alpha }{4 t^2} &= 0.
%\label{TeukEqnAnsatzI+}
%\end{align}
%Expanding this in powers of $1/v$, at fixed $u$, gives
% \begin{equation}
%\frac{i \omega (\alpha \mp 4)}{v} + {\cal O} \left( v^{-2} \right)= 0,
%\label{TeukAsympI+}
%\end{equation}
%hence the asymptotic solution at null infinity is
%\begin{equation}
%\phi_{\pm} \sim C_3 t^{\pm 4} e^{-i\omega v}.
%\label{TeukAsympI+}
%\end{equation}
%This is a non-physical solution as it corresponds to waves entering from null infinity.

\subsection{The case $s=+2$ (vacuum)}

\begin{figure}[t]
\includegraphics[width=\linewidth]{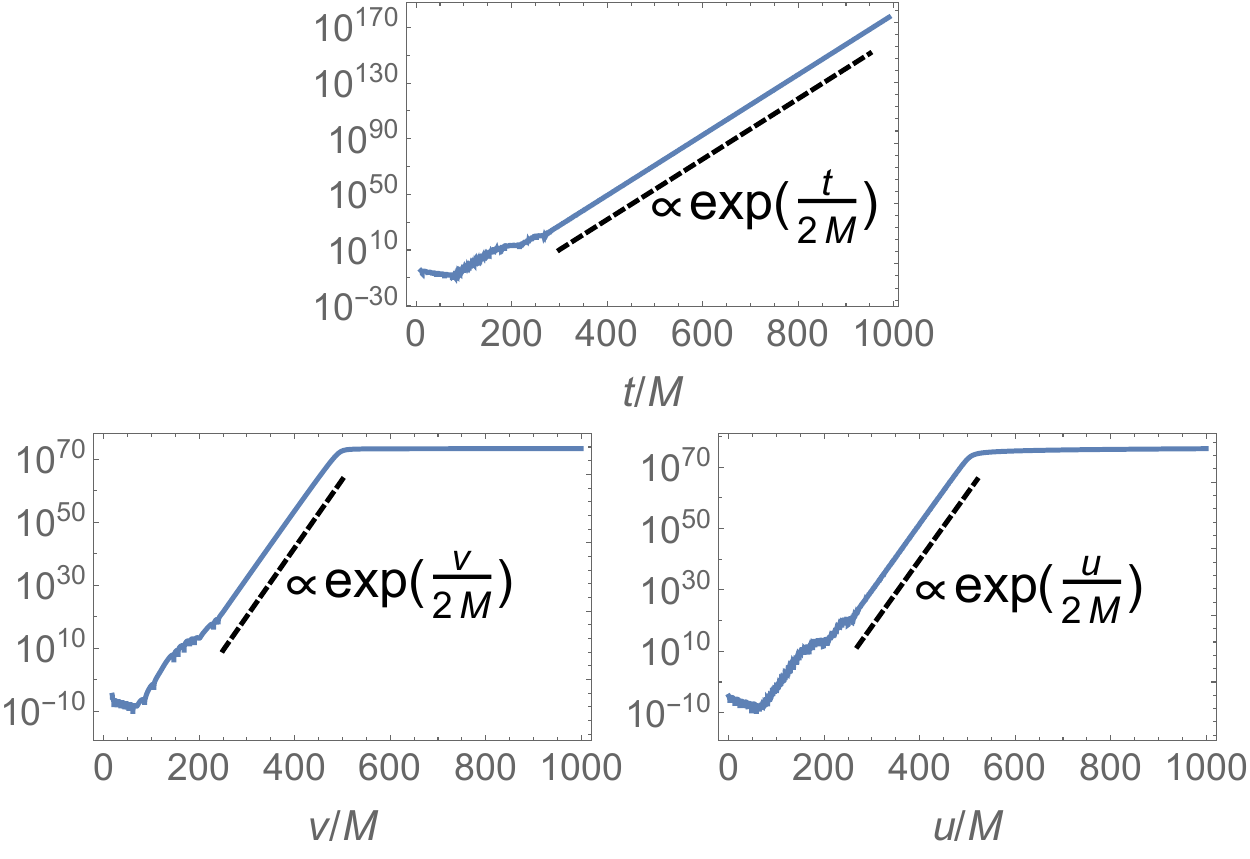}
\caption{Results for the evolution of the $(\ell,m)=(2,0)$ mode of the vacuum 1+1D BPT equation with $s=+2$. Other details are as in Fig.\ \ref{TeukVacsm2}, except here the scale is semilogarithmic. The dashed lines are for reference.}
\label{TeukVacs2}
\end{figure}

Figure \ref{TeukVacs2} shows a typical output from an $s=+2$ numerical evolution in vacuum. Again we start with a narrow Gaussian near the initial vertex at $(u_0,v_0)=(-9M,9M)$, and this time evolve out to $(u,v)=(10^3M,10^3M)$. In this case, after a short phase of quasinormal decay (harder to discern on the semilogarthmic scale of Fig.\ \ref{TeukVacs2}), there commences a rapid exponential growth, $\phi\sim\exp[t/(2M)]$. Again, the growth seems to continue indefinitely towards future timelike infinity ($t\to\infty$ with fixed $r>2M$), but the solution settles to finite values towards $\mathscr{I}^+$ and $\mathscr{H}^+$. A similar behavior is observed for all values of $\ell$ and $m$ and all choices of initial data we have tried, and the blow-up exponent $(1/2M)$ seems universal. The growing component is not numerically stable, increasing in amplitude with decreasing $h$ (finer resolution). We have performed similar tests to the ones described above for $s=-2$ and with similar results: The exponential growth is persistent, universal, and can be moderated (in amplitude) only with high-precision arithmetic. 

We again argue that the culprit is a nonphysical growing solution of the BPT equation seeded by roundoff error, this time an exponential mode of the $s=+2$ equation. We can see this most instructively from a simple asymptotic analysis near the horizon, as follows. Working at leading order in $\Delta$, and assuming $\phi$, $\phi_{,u}$ and $\phi_{,v}$ are all of the same order in $\Delta$ near the horizon [this is true of the $r\to 2M$ asymptotic expressions in Eq.\ (\ref{asymptPhys}) and (\ref{asymptNonphys})], the BPT equation (\ref{Teukolsky1+1phi}) reduces to
\begin{equation}
\phi_{,uv} -k\,\phi_{,u}  =0,
 \label{TeukEqnAsympH+}
\end{equation}
in which $k:=s/(4M)$, and
where we have retained the $s$ dependence to enable us to compare the situation between the two spin values. 
%\begin{equation}
%\frac{\partial^2\phi_{\pm}}{\partial v\partial u} \pm \frac{1}{2M} \frac{\partial% \label{TeukEqnAsympH+}
%\end{equation}
The general solution is
\begin{equation}
\phi = C_1 (u)e^{kv} + C_2 (v),
\label{TeukAsympH+}
\end{equation}
where $C_1(u)$ and $C_2(v)$ are arbitrary functions. Solutions of the form $\phi= C_2(v)$ represent physical perturbations that are purely ingoing at the horizon [compare with Eq.\ (\ref{asymptPhys})], while solutions of the form $\phi = C_1 (u)e^{kv}$ represent nonphysical perturbations coming out of the past horizon [compare with Eq.\ (\ref{asymptNonphys}), noting $\Delta^{s}\sim (2M)^{2s}e^{2kr_*}\sim (2M)^{2s}e^{kv}$ near the horizon]. For $s=+2$ the nonphysical solution blows up exponentially in $v$ along the horizon, while for $s=-2$ it is exponentially suppressed. 

The situation now mirrors what we had near null infinity for the $s=-2$ growth: As the main physical perturbation, of the form $\phi= C_2(v)$, reaches the horizon, roundoff error along the incoming ray seeds a nonphysical component $\sim C_1 (u)e^{kv}$, which, for $s=+2$, blows up exponentially along the horizon. The predicted rate of exponential growth is consistent with that observed in the numerical data: $\sim e^{sv/(4M)}=e^{v/(2M)}$.  To understand the propagation of this exponential growth into other areas of the black hole's exterior would require a detailed asymptotic matching analysis, but it would not be surprising to find a similar exponential growth in time anywhere outside the black hole, as seen in the numerics. We note the fortunate situation in the $s=-2$ case, where all nonphysical modes are exponentially suppressed in a dynamical manner, with no need to actively impose boundary conditions. 

There are in the literature several successful time-domain numerical methods for the $s=+2$ Teukolsky equation (e.g., \cite{Harms2013,Burko2016,Burko2018,Burko2019}), all incorporating horizon-penetrating coordinates in some form. The use of such coordinates (effectively a compactification of our $u,v$ coordinates) under-resolves any outgoing component of the perturbation field near the horizon, thereby avoiding the problem encountered here.  

\subsection{Mitigation}

Although initially surprising to us, it is clear that a standard uni-grid characteristic evolution based on $u,v$ coordinates does not work well for either $s=+2$ or $s=-2$ Teukolsky equations. A remedy based on the use of very high precision arithmetic is clearly impracticable. The preceding discussion and evidence from the literature suggest that compactification of the two asymptotic domains ($\mathscr{I}^+$ and $\mathscr{H}^+$) can offer a solution that is both computationally efficient and practicable. This has already been achieved, e.g., by Harms {\it et al.}\ \cite{Harms2013}, using asymptotically null compactified spacelike slices. It is perceivable that the same could also be achieved within the convenient framework of a fully double-null architecture. This approach is worth exploring. 

Here we choose to apply a different strategy. Instead of tackling the BPT equation directly, we will introduce a transformation of the Hertz potential to a new field variable, which satisfies a field equation free of the above difficulties. From the preceding discussion it is clear that the culprit term in the BPT equation (in both $s=-2$ and $s=+2$ cases) is the one involving $\phi_{,t}$, so we seek a transformation that eliminates that term. The simplest such transformation is a time-domain version of the familiar Chandrasekhar transformation \cite{Chandra1975}, which takes solutions of the BPT equation to solutions of the Regge-Wheeler (RW) equation. As we have mentioned, the RW equation evolves without a problem on a simple uniform mesh based on $u,v$ coordinates (see, e.g., \cite{Martel:2003jj}), so this approach would require no radical architectural changes to our numerical method.  

In the next section we reformulate the 1+1D no-string evolution problem in terms of a RW-like variable. Then, in Sec.\ \ref{sec:hyperevol}, we demonstrate a full numerical calculation of the Hertz potential for a scatter orbit.

%%%%%%%%%%%%%%%%%%%%%%%%%%%%%%%%%%%%%%%%%%%%%%%%%%%%%%%%%%%%%%%%%%%%%%%%
\section{Reformulation in terms of a RW-like variable} 
%%%%%%%%%%%%%%%%%%%%%%%%%%%%%%%%%%%%%%%%%%%%%%%%%%%%%%%%%%%%%%%%%%%%%%%%
\label{sec:RWformulation}

Let the field $X(t,r)$ be a solution of the vacuum Regge-Wheeler equation 
\begin{equation}\label{RWeq}
X_{,uv}+\frac{f}{4}\left(\frac{\lambda_1}{r^2}-\frac{6M}{r^3} \right) X =0 ,
\end{equation}
where, recall, $f=1-2M/r$ and $\lambda_1=\ell(\ell+1)$.
%\begin{equation}
%{\cal W}(r)=\frac{1}{4}f\left[\frac{\lambda_1}{r^2}-\frac{6M}{r^3} \right].
%\end{equation}
%This equation has the form of the Regge--Wheeler equation but with a slightly modified potential. We can see that the RW--like equation lacks the single derivative terms which drove the numerical instabilities when evolving the BPT equation.
Then, as can be easily checked,
\begin{equation}\label{transformationIRG}
\phi_+ = \frac{1}{4r} \tilde{\boldsymbol D}_n^2(r X) 
%\frac{1}{r}f^{-1}\partial_u f^{-1}\partial_u(rX) 
= f^{-2}\left(X_{,uu}-\frac{r-3M}{r^2}\, X_{,u} \right)
\end{equation}
and 
\begin{equation}\label{transformationORG}
\phi_- = \frac{f^2 r^3}{4}{\boldsymbol D}_\ell^2(r X) 
%\frac{1}{r}f^{-1}\partial_u f^{-1}\partial_u(rX) 
= r^4\left(X_{,vv}+\frac{r-3M}{r^2}\, X_{,v} \right)
\end{equation}
are solutions of the vacuum BPT equation (\ref{Teukolsky1+1phi}) with $s=-2$ and $s=+2$, respectively. This means that we can use the RW variable $X$ as a generating function for both IRG Hertz potential $\phi_+$ and ORG Hertz potential $\phi_-$ in each of the vacuum regions ${\cal S}^{\gtrless}$. The advantage, of course, is that the RW equation (\ref{RWeq}), unlike the BPT equation (\ref{Teukolsky1+1phi}), does lend itself to a straightforward characteristic evolution in $u,v$ coordinates. The idea now would be to formulate a suitable characteristic initial-value problem for the RW variable $X$, from which the no-string 
%(That comes with some computational cost: Generating $\phi_\pm$ by taking derivatives of the numerical field $X$ inevitably leads to loss of numerical accuracy.)   
Hertz potential $\phi_+^{\gtrless}$ (or $\phi_-^{\gtrless}$) can be obtained by applying the transformation (\ref{transformationIRG}) [or (\ref{transformationORG})] to vacuum RW solutions $X^{\gtrless}$ in the corresponding vacuum domains ${\cal S}^{\gtrless}$. To achieve this, $X$ must satisfy appropriate jump conditions along $\cal S$, and suitable boundary conditions at ${\mathscr{I}}^+$ and ${\mathscr{H}}^+$. 

\subsection{Boundary conditions}

Let us consider boundary conditions first. In both asymptotic regions $r\to\infty$ and $r\to 2M$, monochromatic solutions of the RW equation (\ref{RWeq}) are superposition of modes $X\sim e^{-i\omega u}$ and $X\sim e^{-i\omega v}$ (with some generally nonzero constant coefficients at leading order). The ``retarded'' monochromatic solution has the behavior $X\sim e^{-i\omega u}$ near ${\mathscr{I}}^+$ and $e^{-i\omega v}$ near ${\mathscr{H}}^+$. It it easily seen that this retarded solution generates the physical IRG Hertz potential $\phi_+\sim e^{-i\omega u}$ near ${\mathscr{I}}^+$ and the physical ORG Hertz potential $\phi_-\sim e^{-i\omega v}$ near ${\mathscr{H}}^+$ [here we recall Eq.\ (\ref{asymptPhys})]. It is harder to show that the retarded RW solution necessarily generates the physical field $\phi_+$ near ${\mathscr{H}}^+$ or the physical field $\phi_-$ near ${\mathscr{I}}^+$ (this would require a higher-order asymptotic analysis), but we can circumvent this with the following observation: From Eq.\ (\ref{asymptNonphys}) we see that nonphysical modes of the the IRG potential $\phi_+$ diverge (like $\Delta^{-2}$) near ${\mathscr{H}}^+$, and that the nonphysical modes of the ORG potential $\phi_-$ diverge (like $r^4$) near ${\mathscr{I}}^+$. Thus, in either case, a nonphysical Hertz potential announces its presence loudly in the form of a strong asymptotic divergence. This is a point made already in Ref.\ \cite{Barack:2017oir}: A solution $\phi_+$ that is regular (bounded) at ${\mathscr{H}}^+$ is automatically the physical one, and so is a solution $\phi_-$ that is regular (bounded) at ${\mathscr{I}}^+$. We can establish a posteriori that our numerical solutions $\phi_+^<$ or $\phi_-^>$ satisfy physical boundary conditions simply by checking they are bounded. 

In summary, we propose that the required vacuum RW solutions on ${\cal S}^>$ and ${\cal S}^<$ are the ones satisfying standard, retarded boundary conditions at ${\mathscr{I}}^+$ and ${\mathscr{H}}^+$, respectively. For the IRG potential $\phi_+$, this is guaranteed on ${\cal S}^>$, and can be easily checked a posteriori on ${\cal S}^<$. For the IRG potential $\phi_-$, this is guaranteed on ${\cal S}^<$, and can be easily checked a posteriori on ${\cal S}^>$. 

\subsection{Jumps across ${\cal S}$}

It remains to translate the jumps in $\phi$ and its derivatives across $\cal S$, obtained in Sec.\ \ref{Sec:Jumps}, to jumps in $X$ and its derivatives there. For brevity we only discuss here the IRG case, but jumps for the ORG case can be obtained in a similar manner. 

We could not find an explicit inverse of the transformation (\ref{transformationIRG}), but (given $\phi_+$) it is easy to obtain two independent algebraic relations between $X$, $X_{,u}$ and $X_{,v}$, which will suffice for our purpose. First, taking $\partial_v$ of (\ref{transformationIRG}), and using $(\ref{RWeq})$ and (\ref{transformationIRG}) to substitute for $X_{,uv}$ and $X_{,uu}$, respectively, leads to 
\begin{align}\label{chiu}
X_{,u}=\frac{f}{r(\lambda r+6M)}\left(3M X-8Mr^2 \phi_+
-4r^4 \phi_{+,v}\right),
\end{align}
where, recall, $\lambda=(\ell+2)(\ell-1)$.
Second, taking $\partial_{vv}$ of (\ref{transformationIRG}), then using the $u$ and $v$ derivatives of the RW equation (\ref{RWeq}) to replace for $X_{,uvu}$ and $X_{,uvv}$, and finally using $(\ref{RWeq})$ and (\ref{transformationIRG}) again to replace for $X_{,uv}$ and $X_{,uu}$, we obtain  
\begin{eqnarray}\label{chiv}
X_{,v}&=&-\left(\alpha+\frac{3Mf}{r(\lambda r+6M)}\right)X + \frac{8M r (\lambda+3)}{3(\lambda r+6M)}\,\phi_+
\nonumber \\
&& +\frac{4r^3[\lambda r+(\lambda+9)M]}{3M(\lambda r+6M)}\, \phi_{+,v}
+\frac{4r^4}{3M}\,\phi_{+,vv} .
\end{eqnarray}
By applying Eqs.\ (\ref{chiu}) and (\ref{chiv}) in both vacuum sides of $\cal S$ in the limit to $\cal S$, we obtain relations between the jumps $[X_{,u}]$ and $[X_{,v}]$ on the one hand, and the jumps $[X]$, $[\phi_+]$, $[\phi_{+,v}]$ and $[\phi_{+,vv}]$ (the latter 3 assumed known) on the other hand; these relations are obtained by simply replacing $X\mapsto [X]$ etc.\ in Eqs.\ (\ref{chiu}) and (\ref{chiv}), and setting $r\mapsto R(t)$ there. 

Now, along the orbit we also have the relation 
\begin{equation}\label{chainruleX}
\dot{[X]} =  [X_{,t}] +\dot{R}_* [X_{,r_{*}}],
\end{equation} 
where, recall, an overdot denotes $d/dt$. Using $[X_{,t}]=[X_{,v}]+[X_{,u}]$ and $[X_{,r_*}]=[X_{,v}]-[X_{,u}]$ and substituting $[X_{,u}]$ and $[X_{,v}]$ from Eqs.\ (\ref{chiu}) and (\ref{chiv}), we thus obtain a simple first-order ODE for the jump $[X]$ along the orbit, of the form [compare with (\ref{ODE})]
\begin{equation}\label{RWODE}
\dot{[X]} + \Big(A_X(R)\dot{R}_*+\alpha\Big)[X] = {\cal F}_X .
%\left([\phi_+],[\phi_{+,v}], [\phi_{+,vv}]\right).
\end{equation}
The coefficient $A_X$ here is given by 
\begin{equation}
A_X= \alpha+\frac{6M(R-2M)}{R^2 (\lambda  R+6M)},
%\frac{\lambda_2 \left(f_R+ R_t \right)}{12 M f_R}+\frac{6 M R_t}{(\lambda  R+6M)R},
\end{equation}
and the sourcing function ${\cal F}_X$ involves the known jumps in the Hertz potential and its derivatives:
\begin{align}
{\cal F}_X =& \frac{8 M \left[6 M \left(f_R - \dot R \right)+ R \left(\lambda  (f_R+\dot{R})+6 \dot{R}\right)\right]}{3 f_R (\lambda  R+6M)} \J{\phi_+} \nonumber \\
&+ \frac{4 R^2}{3 M f_R (\lambda  R+6M)} \Big[6 M^2 (f_R-\dot{R})\nonumber \\
& +\lambda  R (M+R) (f_R+\dot{R})+6 M R (f_R+2\dot{R})\Big] \J{\phi_{+,v}} \nonumber \\
&+\frac{4 R^4 (f_R+\dot{R})}{3 M f_R} \J{\phi_{+,vv}},
\label{RWODESource}
\end{align}
where $f_R:=f(R)=1-2M/R$.

Equation (\ref{RWODE}) admits simple homogeneous solutions, given by (any constant multiple of) 
\begin{equation}\label{X_hom}
[X]_{\rm h}= \left(\lambda+\frac{6M}{R}\right)\, e^{-\alpha(t+R_*)}.
\end{equation}
Note that all these homogeneous solutions (except the trivial zero one) blow up exponentially at $t\to -\infty$. There is a unique particular solution of the full inhomogeneous equation (\ref{RWODE}) that remains bounded; it is given by
\begin{equation}\label{XJump}
[X] = [X]_{\rm h}\int_{-\infty}^t \frac{{\cal F}_X(t')}{[X]_{\rm h}} dt' .
\end{equation} 
This describes the jump in the RW variable needed for it to reproduce the no-string Hertz potential. 

\begin{figure}[htb]
	\begin{center}
        \includegraphics[width=\linewidth]{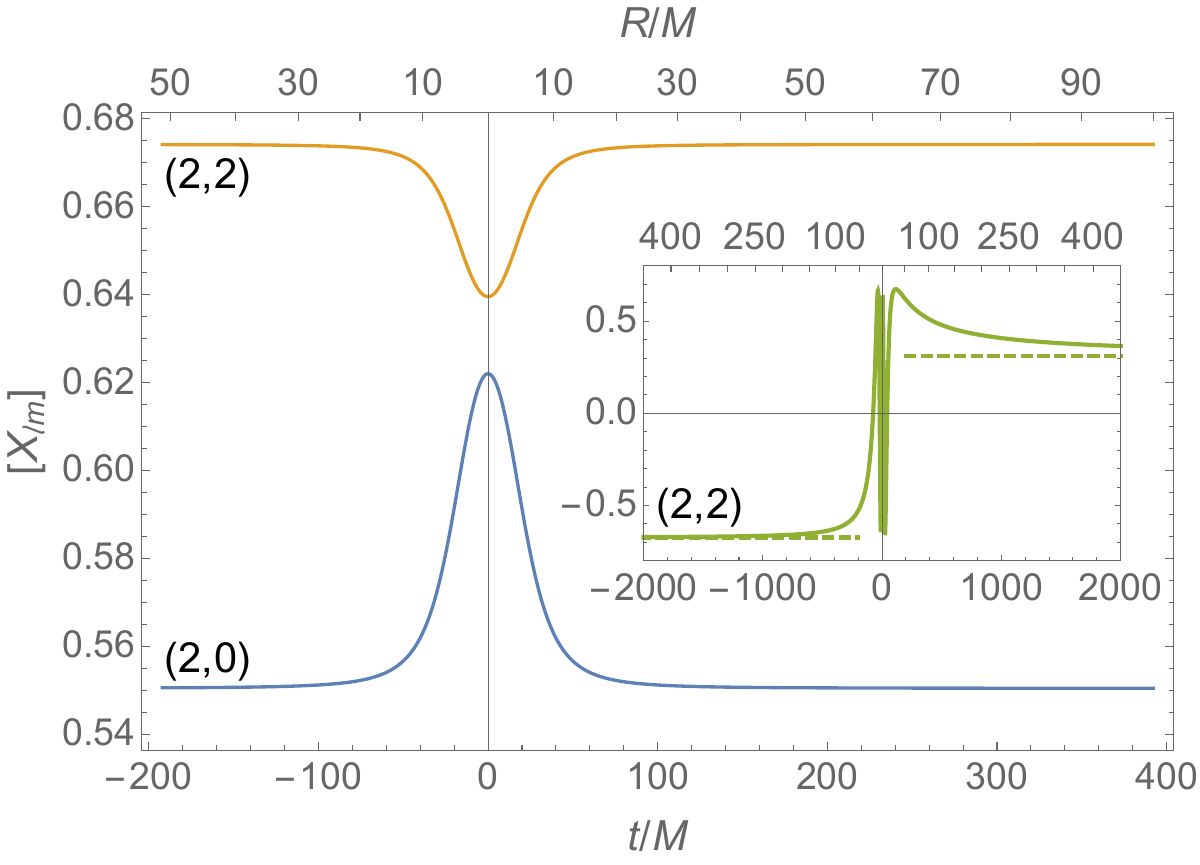} 
\caption{The modulus of the jump $[X_{\ell m}]$ in the ``no-string'' field $X_{\ell m}$ along the geodesic scatter orbit depicted in Fig.\ \ref{orbit}, for the $(\ell,m)=(2,0)$ and $(2,2)$ modes. This is obtained by numerically integrating the first-order ODE (\ref{RWODE}) forwards in time with the initial condition (\ref{IC}) at large negative $t$ and $\varphi_{\rm in}=0$. As a check, the solution approaches the asymptotic value given in (\ref{IC}) with $\varphi_{\rm out}\simeq481^\circ$ (obtained by integrating the geodesic equation). The inset plot demonstrates this for the real part of $[X_{22}]$, with dashed lines indicating the analytical asymptotic values. This jump function inputs into our 1+1D characteristic evolution scheme, whose application is illustrated in Sec.\ \ref{sec:hyperevol}.
}  
        \label{Fig:X_jump}
	\end{center}
\end{figure}

In practice we find it easier to calculate $[X]$ not from Eq.\ (\ref{XJump}) but via a numerical integration of the first-order ODE (\ref{RWODE}). It is best to integrate forward in time from $t\to -\infty$, going ``against'' the direction of exponential growth of the homogeneous solutions (\ref{X_hom}), in order to prevent the growth of such nonphysical modes from numerical error. For this integration we need an initial condition at $t\to -\infty$, which in the case of a scatter orbit corresponds to $R\to\infty$ (with $\dot{R}_\infty<0$). The condition is obtained from a simple asymptotic analysis of the ODE (\ref{RWODE}): Assuming $[X]$ has a power-law behavior at infinity, we have $\dot{[X]}\ll [X]$ at large $R$, so the derivative term in (\ref{RWODE}) may be dropped at leading order. Then, using the $R\to\infty$ limits $A_X \to \alpha$ and 
\begin{equation}
{\cal F}_X \to -\frac{16\pi\mu E}{3M\sqrt{\lambda_2}}(1+\dot{R}_\infty) \bar Y 
\end{equation}
[obtained with the help of Eqs.\ (\ref{JphiAsymp})--(\ref{JphivvAsymp})], we arrive at 
\begin{equation}\label{IC}
[X]_{R\to\infty} = -\frac{64\pi\mu E}{\lambda_2^{3/2}}\, \bar Y_{\ell m}\!\left(\frac{\pi}{2},\varphi_{\rm in/out}\right) ,
\end{equation}
which applies with $\varphi_{\rm in}$ for $t\to -\infty$, and with $\varphi_{\rm out}$ for $t\to +\infty$.
In our implementation we use (\ref{IC}) to set the initial value of $[X]$ at $t\to -\infty$,\footnote{In practice, we impose our boundary condition at a sufficiently large negative value of $t$ with $\varphi_{\rm in}$ calculated at the start point by integrating the geodesic equation (\ref{phidot}) from $t=-\infty$.} integrate the ODE (\ref{RWODE}) forward in time, and then use (\ref{IC}) again to check the result of integration at $t\to +\infty$. Figure \ref{Fig:X_jump} illustrates the result of applying this procedure along a particular strong-field scatter orbit (the one depicted in Fig.\ \ref{orbit} further below).

Once we have $[X]$, it is straightforward to get jumps in derivatives of the field, also needed for our 1+1D evolution scheme. This can be done algebraically. From Eqs.\ (\ref{chiu}) and (\ref{chiv}) one immediately gets $[X_{,u}]$ and $[X_{,v}]$, and using the RW equation (\ref{RWeq}) one gets $[X_{,uv}]$. The jump $[X_{,uu}]$ is subsequently obtained from the transformation equation (\ref{transformationIRG}), and $[X_{,vv}]$ can be found from the $v$ derivative of Eq.\ (\ref{chiv}). The jumps in third and higher derivative can be found recursively in a similar manner.  
%using
%\begin{equation}\label{Xvv}
%[X_{,vv}] = \frac{\dot{[X_{,v}] - u_t [X_{,uv}]}{v_t}.
%\end{equation}

%%%%%%%%%%%%%%%%%%%%%%%%%%%%%%%%%%%%%%%%%%%%%%%%%%%%%%%%%%%%%%%%%%%%%%%%
\section{Numerical implementation for scatter orbits} 
%%%%%%%%%%%%%%%%%%%%%%%%%%%%%%%%%%%%%%%%%%%%%%%%%%%%%%%%%%%%%%%%%%%%%%%%
\label{sec:hyperevol}

We now present a full implementation of our method for a strong-field geodesic scatter orbit. Our code takes as input the parameters of the geodesic orbit, along with multipolar numbers $\ell,m$, and returns the generating-function fields $X^\gtrless_{\ell m}(t,r)$ and the IRG no-string modal Hertz-potential fields $\psi^\gtrless_{\ell m}(t,r)$. In what follows we first review hyperbolic geodesic orbits in Schwarzschild spacetime, then describe the details of our numerical algorithm, and finally present a sample of numerical results.  

\subsection{Hyperbolic-type geodesic orbits}\label{EoM}

In Schwarzschild spacetime, we consider a timelike geodesic orbit that starts and ends at infinity.  We set our Schwarzschild coordinates so that the orbit lies in the equatorial plane, $\theta=\pi/2$. The orbit is then described by the two functions $r=R(t)$ and $\varphi=\varphi_{\rm p}(t)$, which satisfy 
\begin{eqnarray}
\dot{R}&=& \pm\frac{f_R}{E}\sqrt{E^2-\frac{f_R(R^2+L^2)}{R^2}},  \label{rdot}\\
\dot{\varphi}_{\rm p} &=& \frac{f_R L}{R^2 E}, \label{phidot}
\end{eqnarray}
where (recall) an overdot denotes $d/dt$.
Here we have parametrized the geodesic orbit with the two constants of motion $E$ and $L$, respectively the specific energy and angular momentum. For a scatter orbit $E>1$, and without loss of generality we have taken $L>0$, and, correspondingly, $\dot{\varphi}_{\rm p}>0$. 
The sign in (\ref{rdot}) switches from $-$ to $+$ at the periastron, where $\dot{R}=0$.
The particle scatters back to infinity (and does not fall into the black hole) if and only if $L>L_{\rm crit}(E)$, where 
\begin{equation}
L_{\rm crit} = M\left(\frac{27E^4+9\zeta E^3-36 E^2-8\zeta E+8}{2(E^2-1)}\right)^{1/2},
\end{equation}
with $\zeta:=\sqrt{9E^2-8}$. The function $L_{\rm crit}(E)$ is monotonically increasing, so the minimum possible value of $L$ is $L_{\rm crit}(E\to 1)=4M$.

An alternative, more physically intuitive parametrization is provided by the initial incoming speed at infinity,
\begin{equation}
\vinf:=\left|\dot{R}(t\to -\infty)\right|=\frac{\sqrt{E^2-1}}{E},
\end{equation}
and impact parameter, 
\begin{equation}
b:=\lim_{t\to -\infty} R(t)\sin\left|\varphi_{\rm p}(t)-\varphi_{\rm in}\right| = \frac{L}{\sqrt{E^2-1}},
\end{equation}
where $\varphi_{\rm in}:=\varphi_{\rm p}(t\to -\infty)$.
Note $E = (1-v_\infty^2)^{-1/2}$, so $E$ is the usual ``gamma factor'' at infinity. 
For a scatter orbit we need $b>b_{\rm crit}=L_{\rm crit}(E)/\sqrt{E^2-1}$.
The function $b_{\rm crit}(E)$ is monotonically decreasing, so the minimal possible value of $b$ is 
$b_{\rm crit}(E\to\infty) = 3\sqrt{3}M \simeq 5.196 M $.

For a scatter orbit, the equation $\dot{R}=0$ admits a single real root outside the horizon, corresponding to the periastron radius  $R=R_{\rm min}$. Given $E$ and $L$, this radius can be expressed in the convenient form \cite{priv_comm_Maarten}
\begin{equation}\label{Rmin}
R_{\rm min} = \frac{6M}{1-2\sqrt{1-12M^2/L^2}\, \sin\left(\frac{\pi}{6}-\frac{1}{3}\arccos \beta\right)},
\end{equation}
where
\begin{equation}
\beta = \frac{1+(36-54E^2)M^2/L^2}{(1-12M^2/L^2)^{3/2}}.
\end{equation}
It can be checked that, for any fixed $E>1$, $\beta$ is a monotonically increasing function of $L$, varying between $\beta=-1$ for $L=L_{\rm crit}(E)$ and $\beta=+1$ for $L\to\infty$; hence the expression in (\ref{Rmin}) is manifestly real.  $R_{\rm min}$ increases monotonically with $L$ at fixed $E$, and decreases monotonically with $E$ at fixed $L$. The smallest periastron distance achievable is $R_{\rm min}\to 3M$, for $L\to L_{\rm crit}(E)$ with $E\to\infty$.  Without loss of generality, we set $t=0$ at the periastron passage, i.e.\ take $R(0)=R_{\rm min}$.

Another convenient parametrization of scatter orbits employs the eccentricity, $e>1$, and (a-dimensionalized) semilatus rectum, $p>6+2e$, defined from
\begin{equation}\label{rminpe}
R_{\rm min}= \frac{M p}{1+e}, \quad\quad R_{-}= \frac{M p}{1-e},
\end{equation}
where $R_-$ is the negative root of $\dot{R}=0$. [For a scatter orbit, i.e.\ one with $E>1$ and $L>L_c(E)$, $\dot{R}=0$ admits three real roots: $R_{\rm min}$, $R_-$ and (say) $R_3$ (in addition to the trivial root at $R=2M$), satisfying $R_-<0<R_3<2M<R_{\rm min}$.]
The relations with $E$ and $L$ are 
\begin{equation}
E^2 =
\frac{(p-2)^2-4e^2}{p(p-3-e^2)}, \quad\quad
L^2 =
\frac{p^2M^2}{p-3-e^2}.
\label{ELep}
\end{equation} 
In terms of $e$ and $p$, the radial motion takes the convenient Keplerian-like form
\begin{equation}\label{rofchi}
R(t)=\frac{Mp}{1+e\cos\chi(t)}.
\end{equation}
The ``anomaly'' parameter $\chi$ increases monotonically in $t$ along the orbit, running over $\chi\in(-\chiinf,\chiinf)$ for $t\in(-\infty,\infty)$, with
%\begin{equation}\label{chiinf}
$\chi_\infty=\arccos(-1/e)$
%\end{equation}
and $\chi=0$ at the periastron passage.
The relations $t(\chi)$ and $\varphi_{\rm p}(\chi)$ can be determined by integrating
\begin{align}
\frac{dt}{d\chi} =& \frac{Mp^2}{(p-2-2e\cos\chi)(1+e\cos\chi)^2} \sqrt{\frac{(p-2)^2-4e^2}{p-6-2e\cos\chi}}, \label{eq:dt_dchi} \\
\frac{d\varphi_{\rm p}}{d\chi} =& \sqrt{\frac{p}{p-6-2e\cos\chi}}, \label{eq:dphi_dchi}
\end{align}
with the initial conditions $t(-\chiinf)=-\infty$ [or $t(0)=0$] and $\varphi_{\rm p}(-\chiinf)=\varphi_{\rm in}$. Finally, defining the scatter (or deflection) angle by $\delta\varphi := \varphi_{\rm out}-\varphi_{\rm in}-\pi$, one obtains, using Eq.\ (\ref{eq:dphi_dchi}),
\begin{eqnarray}
\delta\varphi = \int_{-\chiinf}^{\chiinf} \frac{d\varphi_{\rm p}}{d\chi}d\chi  - \pi
=2k\sqrt{p/e}\, \El_1\Big(\frac{\chi_\infty}{2};-k^2\Big) -\pi ,
\nonumber\\
%&=& 
%2k\sqrt{p/e}\, \El_1\Big(\frac{\chi_\infty}{2};-k^2\Big) -\pi,
\label{Deltaphi}
\end{eqnarray}
where $k:=2\sqrt{e/(p-6-2e)}$
%\begin{equation}\label{k}
%k=2\sqrt{\frac{e}{p-6-2e}},
%\end{equation}
and $\El_1$ is the incomplete elliptic integrals of the first kind,
\begin{equation}
%\hat K(k)=\El_1(\pi/2;k), \quad\quad
\El_1(\varphi;k)=\int_0^{\varphi} (1-k\sin^2 x)^{-1/2}dx.
\end{equation}

\begin{figure}[t]
\centering
\includegraphics[width=\linewidth]{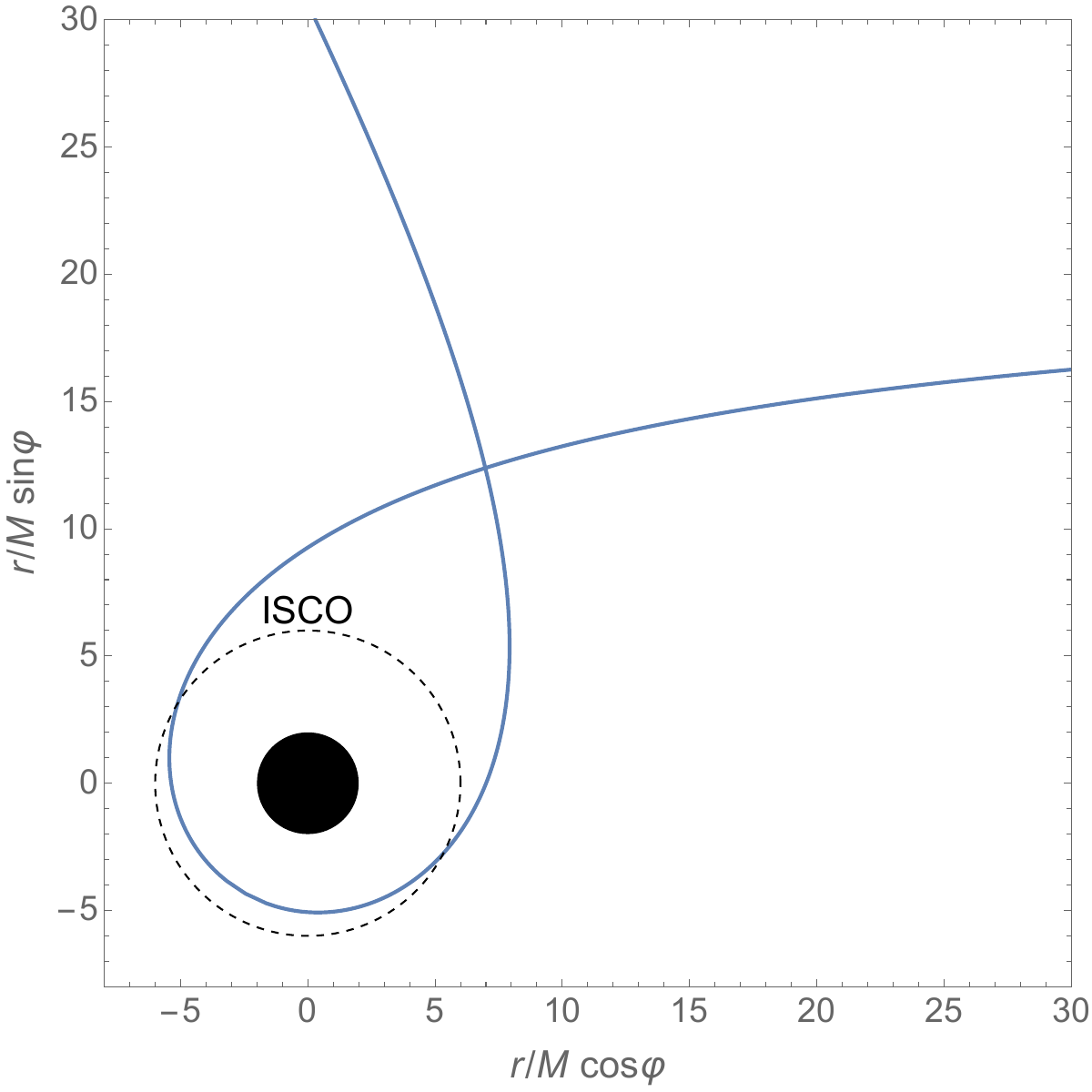}
\caption{The sample scatter geodesic orbit used for our numerical illustration, with parameters given in Eqs.\ (\ref{sampleorbit_vb}) and (\ref{sampleorbit}). The orbit is plotted in the equatorial plane using Cartesian-like coordinates $(x,y)=(r\cos\varphi,r\sin\varphi)$.  The location of the innermost stable circular orbit (ISCO) is shown for reference. The deflection angle of this strong-field orbit is $\delta\varphi \simeq 301^\circ$.} 
%The orbit is parameterized by $\vinf=0.2$ and $b=21M$ giving $r_{\rm min}=4.9228M$, $E=1.02062$, $L=4.28661M$, $p=10.9351M$, $e=1.1948$ and $\Delta\varphi = 8.39896$.}
\label{orbit}
\end{figure}

For the numerical demonstration to be presented below we have picked a sample strong-field scatter geodesic with 
\begin{equation}\label{sampleorbit_vb}
\vinf = 0.2\quad \text{and}\quad b=21M,
\end{equation} 
corresponding to 
\begin{eqnarray}\label{sampleorbit}
R_{\rm min}&\simeq& 4.98228M,\quad  E\simeq 1.02062,\quad L\simeq 4.28661M, \nonumber\\
e&\simeq& 1.1948, \quad  p\simeq 10.9351 \quad\text{and}\quad \delta\varphi \simeq 301^\circ . \nonumber\\
\end{eqnarray}
The orbit is depicted in Fig.\ \ref{orbit}.

\subsection{Numerical algorithm}

Our method is based on a characteristic numerical evolution in $u,v$ coordinates, as described in Sec.\ \ref{TeukNumMethod}---only here we are evolving the RW equation (\ref{RWeq}) instead of the BPT equation, and we impose suitable jump conditions along $\cal S$ (see Fig.\ \ref{uvGridScatter}) compatible with the no-string IRG solution for our sample scatter orbit. A detailed description of our finite-difference scheme is given in Appendix \ref{App:RWCode}, where we also explain how the jump conditions are incorporated into the scheme so as to achieve a (global) quadratic rate of numerical convergence. Here we lay out the main steps of the numerical algorithm.

{\it Input.} The code takes as input the two orbital parameters $\vinf$ and $b$, the orbital radius $r=R_{\rm init}$ at the start (and end) of the numerical evolution, the field's multiple numbers $\ell,m$, and the finite-difference interval $h:=\Delta u=\Delta v$. 

{\it Step 1: Calculate geodesic orbit.} Given $\vinf$ and $b$, the code calculates $E$ and $L$ and from these $e$ and $p$, as well as $R_{\rm min}$. The functions $R(t)$ and $\varphi_{\rm p}(t)$ are then derived in the range $R_{\rm min}\leq R\leq R_{\rm init}$ by numerically integrating $\dot{R}$ and $\dot{\varphi}_{\rm p}$ as obtained from Eqs.\ (\ref{rofchi})--(\ref{eq:dphi_dchi}), with initial conditions $R(0)=R_{\rm min}$ and $\varphi_{\rm p}(-\infty)=0$. The code also calculates $t_{\rm tot}$, the time it takes the particle to get from $R_{\rm init}$ back to $R_{\rm init}$ after being scattered.  

{\it Step 2: Set characteristic grid.} The code then prepares a $2\times 2$ array of $u,v$ coordinate values representing the nodes of the characteristic mesh shown in Fig.\ \ref{uvGridScatter}. For the initial rays we take $v_0=-t_{\rm tot}/2+R^*_{\rm init}$ and $u_0=-t_{\rm tot}/2-R^*_{\rm init}$ with $R^*_{\rm init}:=r_*(R_{\rm init})$. This is so that the initial vertex $(u,v)=(u_0,v_0)$ is crossed by the particle at $(t,r)=(-t_{\rm tot}/2,R_{\rm init})$. The stepping interval is set at $h$, and the grid's dimensions are taken such that the final characteristic rays are at $u=t_{\rm tot}/2-R^*_{\rm init}$ and $v=t_{\rm tot}/2+R^*_{\rm init}$---so that the particle crosses the upper vertex at $(t,r)=(t_{\rm tot}/2,R_{\rm init})$ on its way out. Finally, the coordinate values of all intersections of the orbit with grid lines are calculated and stored. 

{\it Step 3: Obtain $[\phi]$ along the orbit.} Using the analytical expressions in Appendix \ref{App:WeylJumps}, we calculate the jumps in the Weyl scalars $\psi_{\pm}$ and their derivatives along the orbit, for the $\ell,m$ mode in question. Specifically, we compute $[\psi_-]$ and $[\partial_u^n\psi_+]$ for $n=0,\ldots,3$, and from these, using (\ref{calF}), we analytically construct the source function ${\cal F}(t)$ in Eq.\ (\ref{ODE}). We next numerically integrate the first-order ODE (\ref{ODE}) with the initial condition (\ref{JphiAsymp}), to obtain the jump $[\phi_{+}]$ in the Hertz potential along the orbit. From $[\phi_{+}]$, we algebraically obtain also $[\phi_{+,v}]$ and $[\phi_{+,vv}]$ using the procedure described in the last paragraph of Sec.\ IV.B.

{\it Step 4: Obtain $[X]$ along the orbit.} We now construct the source function ${\cal F}_X$ using Eq.\ (\ref{RWODESource}), and then numerically solve the first-order ODE (\ref{RWODE}) for $[X]$ with the initial condition (\ref{IC}). From $[X]$ we algebraically obtain also $[X_{,v}]$, $[X_{,u}]$, $[X_{,vv}]$, $[X_{,uv}]$ and $[X_{,uu}]$, using the procedure described in the last paragraph of Sec.\ VI.B. The jump values are computed at all intersections of the particle's worldline with grid lines, and stored as vector datasets.

{\it Step 5: Obtain the generating function $X^{\gtrless}_{\ell m}$.} We evolve the RW equation (\ref{RWeq}) using the second-order-convergent finite-difference scheme described in Appendix \ref{App:RWCode}. The scheme requires as input the field jumps calculated in the previous step at intersections of the worldline with grid lines. The evolution starts with zero initial data along $v=v_0$ and $u=u_0$ and proceeds along successive lines of $u=$const. The outcome is a finite-difference approximation to the generating field $X$ in each of the vacuum regions ${\cal S}^>$ and ${\cal S}^<$.

{\it Step 6: Derive the Hertz potential $\phi^{\gtrless}_{\ell m}$.} Given $X$, the Hertz potential mode $\phi$ is calculated in each of the two vacuum regions using Eq.\ (\ref{transformationIRG}), where derivatives are taken numerically. 

{\it Output.} In principle, the code can make available the Hertz potential $\phi$ anywhere in the computational domain. For our initial tests and for the purpose of illustration in this paper, we output 
both $X$ and $\phi$ as functions of $t$ along the orbit (on either of its sides); and as functions of $u$ along the final $v=$const ray (approximating ${\cal J}^+$).  

\subsection{Sample results}

All of the results displayed here are for the orbit shown in Fig.\ \ref{orbit}, with parameters given in Eqs.\ (\ref{sampleorbit_vb}) and (\ref{sampleorbit}). In all of the figures shown below we have set $\mu=1$ and $M=1$ for convenience; as a result, in particular, $t$, $R$ and $h$ are expressed in units of $M$. 
 
\begin{figure}[h!]
\centering
\includegraphics[width=\linewidth]{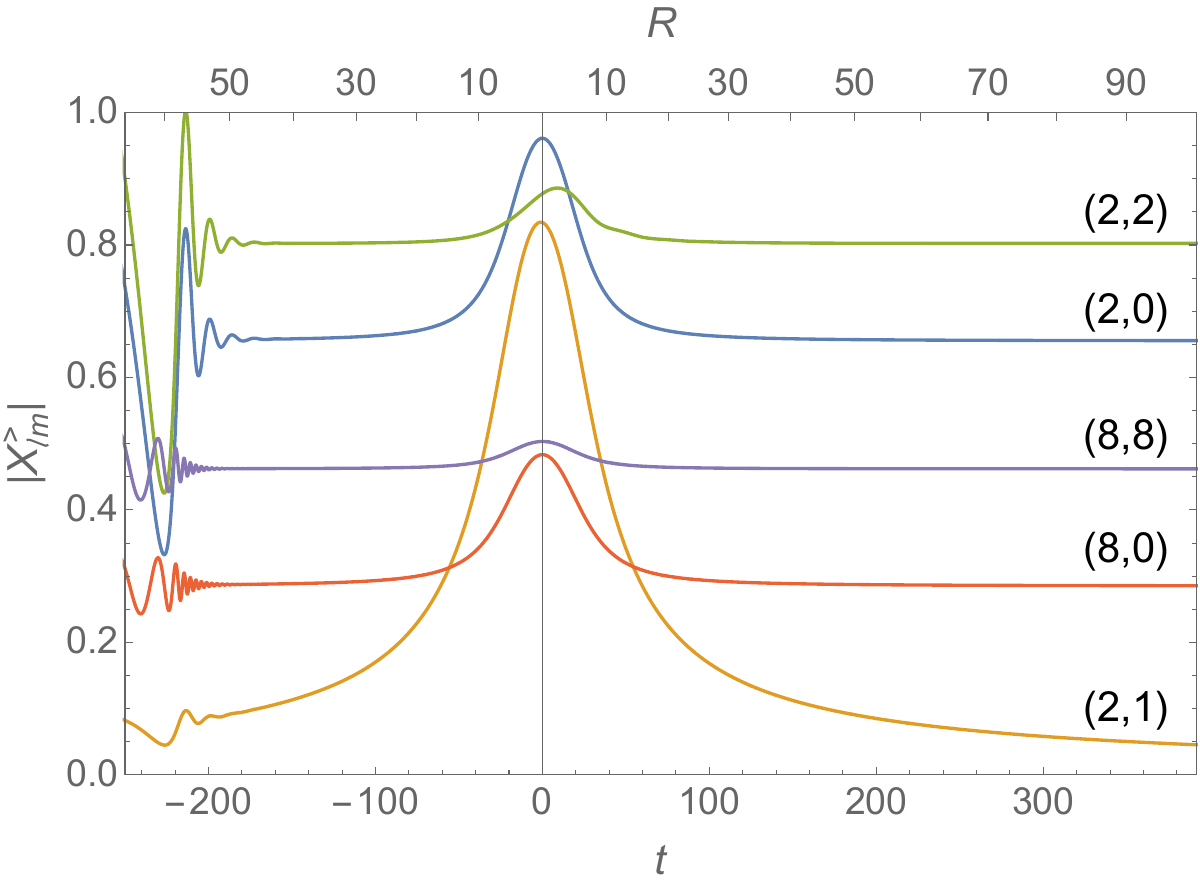}
\caption{The RW field $X_{\ell m}$ along the particle's worldline for the orbit shown in Fig.\ \ref{orbit} and  a sample of $(\ell,m)$ values. Here we show $|X^>_{\ell m}(t,R(t))|$ as a function of time $t$ (lower scale) and orbital radius $R$ (upper scale). Curves are labelled with their $(\ell,m)$ values, with $\ell=8$ data shown amplified by a factor $\times 600$. The periastron location at $t=0$ is indicated with a vertical line. The early part of the data is contaminated by initial junk radiation, and it is to be discarded.
}
\label{RWField}
\end{figure}

Figure \ref{RWField} demonstrates the behaviour of the field $X^>_{\ell m}$ along the worldline of the particle, for a sample of $\ell,m$ values (the field $X^<_{\ell m}$ has a similar behavior). The evolution begins when the incoming particle is at $R_{\rm init}=100M$, and ends when the outflying particle is back at $100M$. 

\begin{figure}[h!]
\centering
\includegraphics[width=\linewidth]{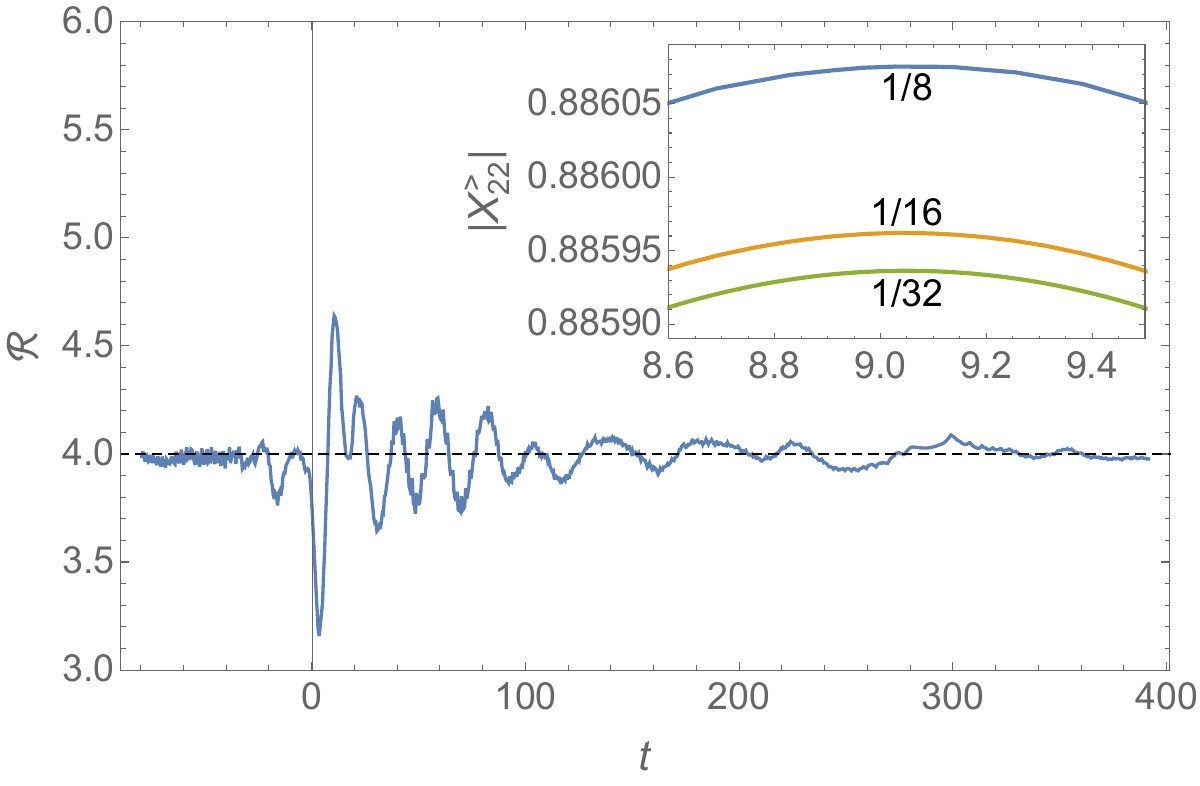}
\caption{Convergence test for the $(\ell,m)=(2,2)$ numerical solution. The inset shows a detail from the $|X^>_{22}|$ (green) curve in Fig.\ \ref{RWField}, for a sequence of runs with decreasing grid spacing, 
$h=\left\{\frac{1}{8},\frac{1}{16},\frac{1}{32}\right\}M$. The main plot quantifies the convergence rate: It shows the ratio ${\cal R}:=\left|X^>_8-X^>_{16}\right|/\left|X^>_{16}-X^>_{32}\right|$ as a function of $t$ along the orbit, where a subscript `8' (e.g.)\ denotes a calculation with grid spacing $h=M/8$. A ratio of ${\cal R}=4$ is indicative of quadratic convergence.
}
\label{RWConv}
\end{figure}

We have performed convergence tests to confirm that our code exhibits a quadratic global convergence rate in $h$, as it is designed to do. An example is shown in Fig.\ \ref{RWConv}. The global rate of convergence is very sensitive to the implementation details of the jump conditions in the finite-difference scheme (see Appendix \ref{App:RWCode}), so the observed quadratic convergence provides important reassurance that these jumps are implemented correctly. 

\begin{figure}[t]
\centering
\includegraphics[width=\linewidth]{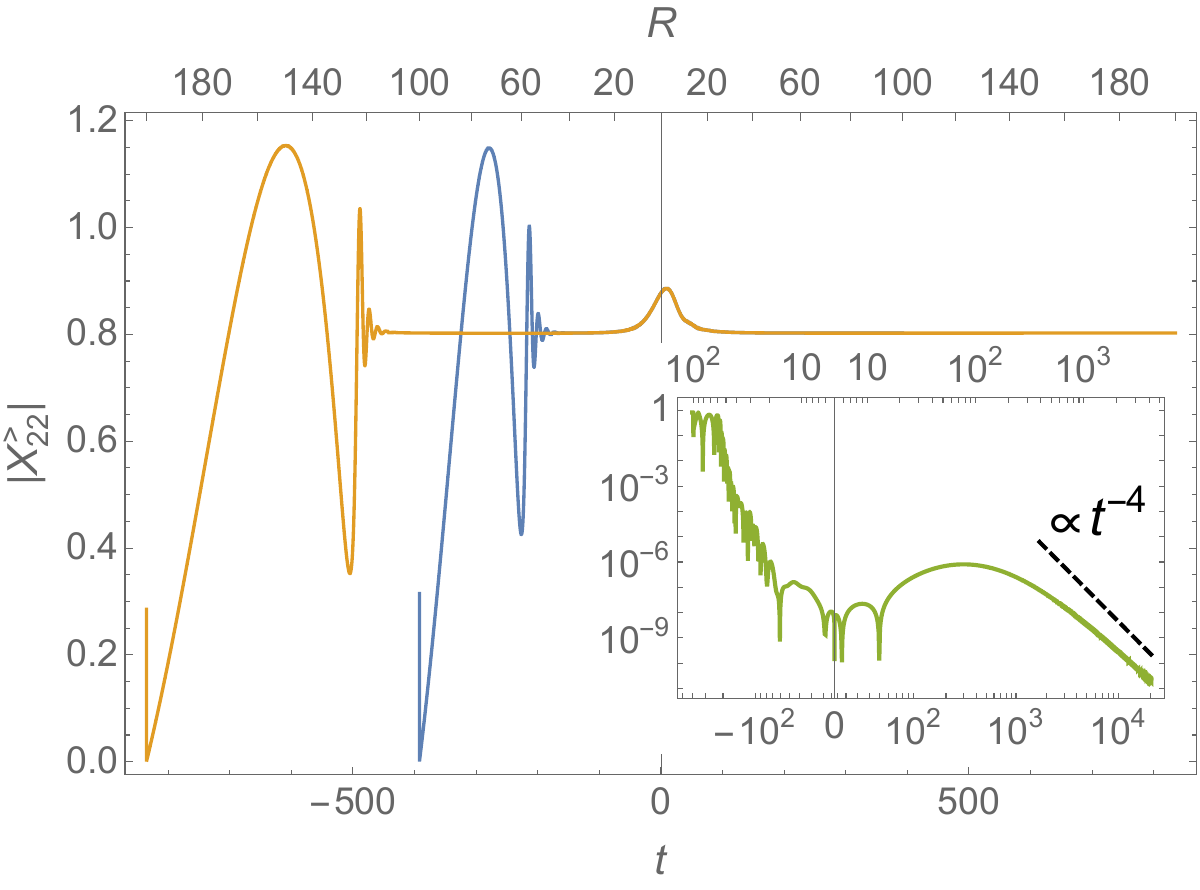}
\caption{Numerical results for $|X^>_{22}|$ on the particle's worldline, as calculated with $R_{\rm init}=100M$ (blue) and with $R_{\rm init}=200M$ (orange).  The comparison illustrates how, reassuringly, the ``clean'' portion of the data is insensitive to $R_{\rm init}$, up to a small error that dies off in time. The inset displays the relative difference between the two curves, showing a $t^{-4}$ fall-off at late time, consistent with the theoretically predicted $t^{-\ell-2}$ decay rate for compact vacuum perturbation along a curve $r=R\propto t$ [see, for instance, Eq.\ (89) of Ref.~\cite{Barack:1998bw}].
}
\label{RWRMax}
\end{figure}

As can be seen in Fig.~\ref{RWField}, initially the data is contaminated by spurious waves, which, however, decay over time to reveal the true, physical solution.  The decay appears faster for higher values of $\ell$, as expected from theory. Figure \ref{RWRMax} illustrates how, reassuringly, the ``clean'' part of the data appears to be insensitive to the value of $R_{\rm init}$, up to a small decaying difference. As the figure demonstrates, using $R_{\rm init}$ as a control parameter enables us  in practice to evaluate the level of residual contamination from initial junk.  

\begin{figure}[h!]
\centering
\includegraphics[width=\linewidth]{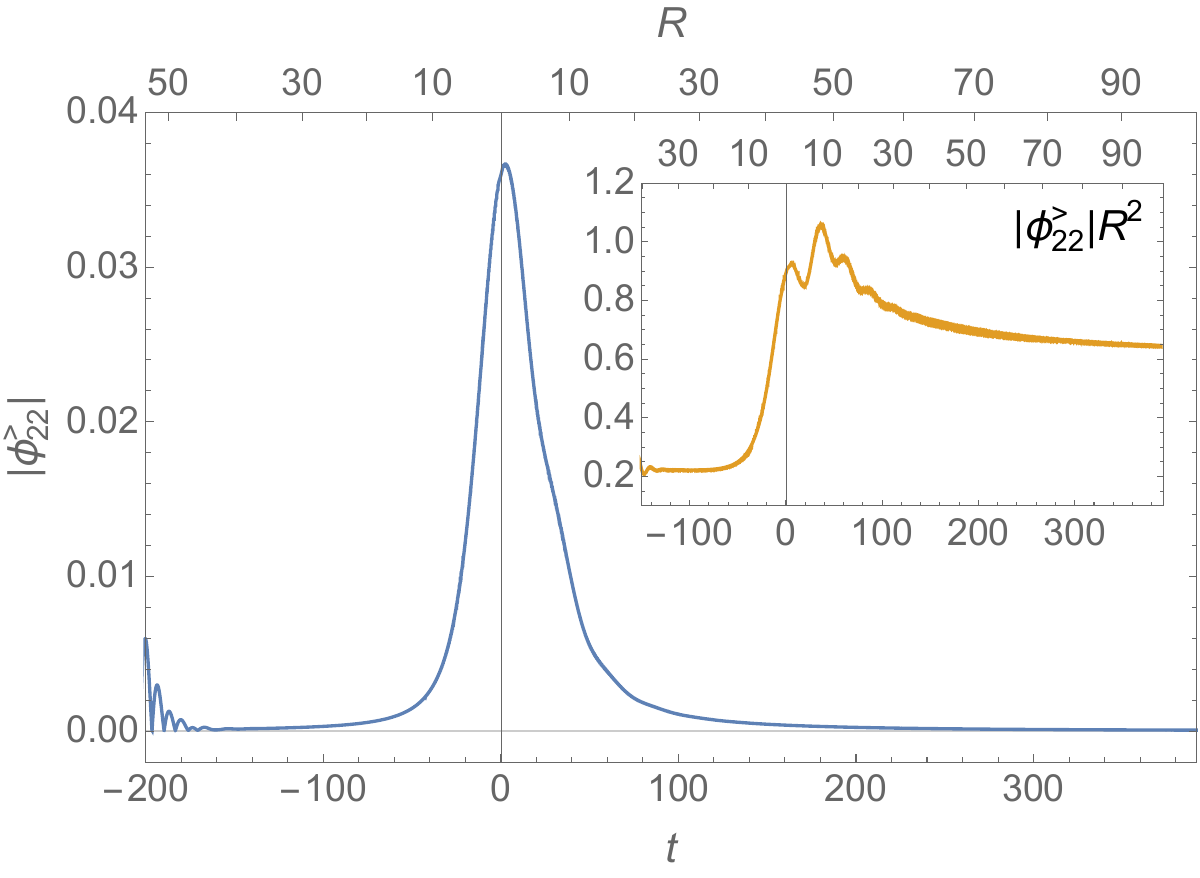}
\caption{The modulus of the ``no-string" IRG Hertz potential $\phi^>_{22}$ along the particle's worldline. The field falls off as $X\sim t^{-2}$ at large $R$. The inset shows the same data rescaled by a factor $(R/M)^2$. The field exhibits the lagging peak and post-periastron undulation features discussed in the text. (The multiplication by $R^2$ makes more distinct the undulation feature, only barely visible in the main plot.)
}
\label{HertzField}
\end{figure}

Figure \ref{HertzField} shows the no-string IRG Hertz potential $\phi^>_{22}$ derived from $X^>_{22}$, as a function along the orbit. Notable physical features include (i) the small lag between the peak of the field and the periastron passage, and (ii) the small undulation in the field amplitude not long after periastron passage. (Both features are visible already at the level of the generating function $X$, and are numerically stable.) The periastron lag has been observed before in calculations along eccentric orbits (see, e.g., \cite{bsago2}); it is attributed to the effect of ``tail'' contributions to the self-field, which peak in amplitude soon after periastron. The undulation, we suggest, is a weak manifestation of the quasinormal-mode excitation phenomenon observed in self-field calculations for highly eccentric orbits \cite{Nasipak:2019hxh,Thornburg:2019ukt}. Both features are associated with ``tail'' contributions to the self-field, and are less visible at larger $\ell$, where the ``direct'' part of the field is more dominant.  We have not conducted here a more detailed study of the above physical features. 

\begin{figure}[h!]
\centering
\includegraphics[width=\linewidth]{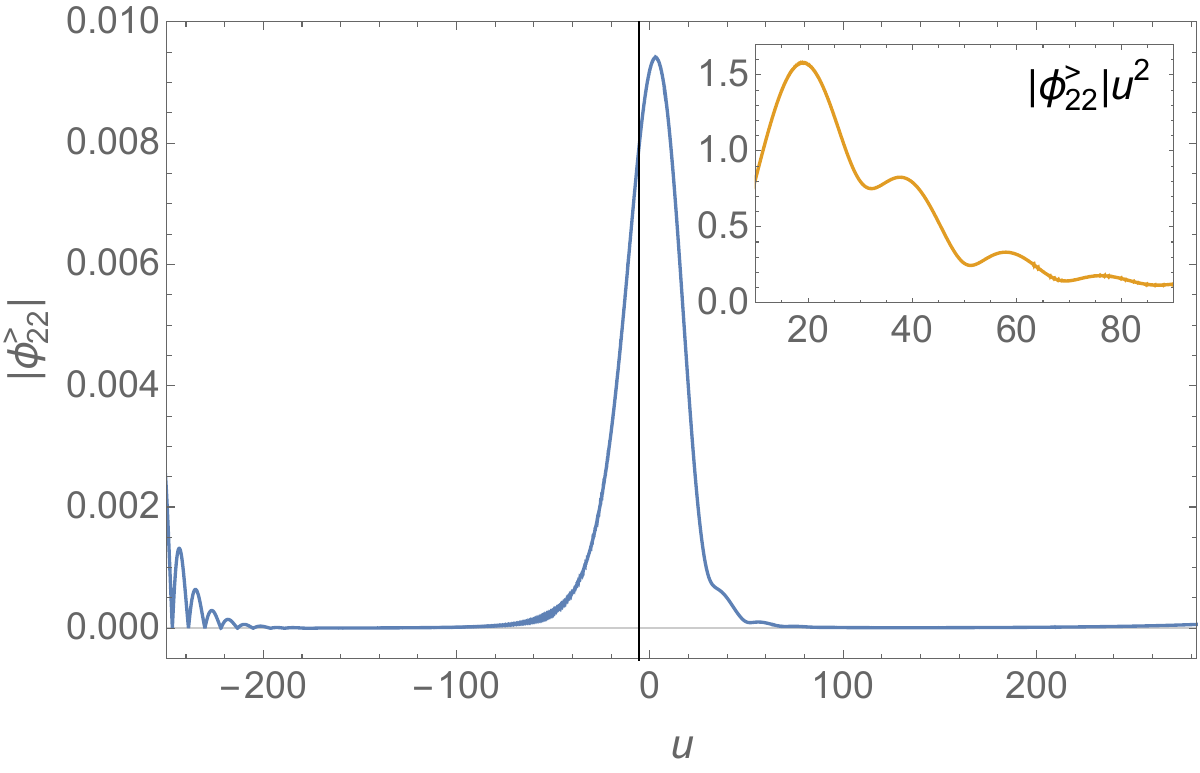}
\caption{The modulus of the Hertz potential $\phi_{22}^>$ as a function of $u$ at $v={\rm const}=499M$ (approximating $\mathscr{I}^+$). The vertical line represents $u$ at periastron. The inset shows the same data rescaled by a factor $(u/M)^2$ to again highlight the post-periastron undulation feature.
}
\label{Hertzscri}
\end{figure}

Finally, Fig.\ \ref{Hertzscri} shows the behavior of $\phi^>_{22}$ near $\mathscr{I}^+$, as a function of retarded time $u$. The periastron lag and post-periastron undulation are also visible in the radiation field in this domain.

%%%%%%%%%%%%%%%%%%%%%%%%%%%%%%%%%%%%%%%%%%%%%%%%%%%%%%%%%%%%%%%%%%%%%%%%
\section{Conclusion and outlook} 
%%%%%%%%%%%%%%%%%%%%%%%%%%%%%%%%%%%%%%%%%%%%%%%%%%%%%%%%%%%%%%%%%%%%%%%%
\label{Sec:conclusions}

The main results of this work are threefold. First, we have provided the details of a practical method for a time-domain calculation of the Hertz potential for point-particle metric perturbations in Schwarzschild spacetime. The main ingredients were jump conditions that the Hertz potential must satisfy along the particle's worldline (in a 1+1D multipolar reduction of the problem), which we derived in explicit form for generic geodesic orbits. Second, considering the numerical implementation strategy, we have demonstrated that a straightforward approach based on evolution of the Teukolsky equation in $u,v$ coordinated does not work (even for vacuum problems), and explained the reason for that failure. Third, we have proposed a way around the problem and demonstrated its applicability with an end-to-end numerical calculation of the Hertz potential for a scatter orbit. 

The specific  application that motivates this work is the calculation of self-force effects on scatter orbits, and in particular the self-force correction to the scatter angle. This calculation we intend to report in forthcoming work. Let us review here the additional steps necessary towards such a calculation, starting from the baseline of the computational method and code developed here.
\begin{itemize}
\item Given the Hertz potential, the no-string radiation-gauge metric perturbation is reconstructed (mode by mode) via Eq.\ (\ref{h_rec}). This involves taking two derivatives of the numerical variables $\phi_{\ell m}$ (and hence four derivatives of $X_{\ell m}$) along the orbit, on either side of it. For the eventual self-force calculation one requires the gradient of the metric perturbation, which therefore requires {\em three} derivatives of $\phi_{\ell m}$ (and hence {\em five} derivatives of $X_{\ell m}$). The computational implications are discussed further below. 
\item One has to separately compute the ``completion'' piece of the metric perturbation, which is not accounted for by the Hertz potential \cite{Merlin:2016boc,vandeMeent:2017fqk}. In the Schwarzschild problem this corresponds precisely to the determination of the $\ell=0,1$ perturbation modes. Of these, the axially-symmetric modes $(l,m)=(0,0)$ and $(1,0)$, which describe mass and angular-momentum perturbations, are easily determinable using the results of \cite{Merlin:2016boc}. The modes $(l,m)=(1,\pm 1)$, which regulate the centre-of-mass location, require a more careful analysis, similar to the one performed in \cite{Barack_2019} for marginally-bound orbits. 
\item Once all the modes of the metric perturbation and its gradient are available, the self-force along the orbit is straightforwardly obtained via the no-string radiation-gauge version of the mode-sum formula, prescribed in \cite{PMB}. It is also easy to separately extract the dissipative and conservative components of the self-force, utilizing the symmetries of the geodesic scatter orbit about the periastron point (see, e.g., Sec.\ 8.1 of \cite{barack}).
\item One can then calculate the self-force correction to the scatter angle (say, at fixed initial velocity $\vinf$ and impact parameter $b$) as certain integrals of the self-force along the orbit. The relevant formulas are straightforwardly derived from the geodesic equations with a self-force term. Additional physical quantities, such as the time delay induced by the self-force, or the integrated particle's spin precession and tidal-field invariants, may also be calculated, though the latter two would require evaluating higher derivatives of the metric perturbation. The self-force information allows calculation of all these effects with or without dissipation. 
\end{itemize}

We have noted above that a calculation of the metric perturbation and self-force involves taking high-order derivatives (fourth and fifth, respectively) of the numerical evolution field $X_{\ell m}$. This is an obvious computational disadvantage of our approach. It can be mitigated if a method is employed that allows a direct evolution of the Teukolsky equation for the Hertz potential $\phi_{\ell m}$, which would reduce the required number of derivatives to only two for the metric perturbation, and three for the self-force. As mentioned in Sec.\ \ref{sec:vacuum}, there already exist such methods, based on compactification of $\mathscr{I}^+$ and the use of horizon-penetrating coordinates---which appear to automatically eliminate the problematic non-physical growing solutions of the Teukolsky equation. Existing codes employ asymptotically null (hyperboloidal) Cauchy slicing of the numerical domain. We propose that, in our context, it might be advantageous to retain the convenience and simplicity of a fully double-null treatment, taking advantage of the domain split across $\cal S$. What we have in mind is a scheme where on ${\cal S}^+$ we use the original Eddington-Finklestein coordinate $u$ with a compactified $v$ coordinate, while on ${\cal S}^-$ we use the original $v$ coordinate with a compactified $u$. The coordinate discrepancy along $\cal S$ is then incorporated into the jump conditions. We intend to explore this route in future work. 

Since this work concentrates on basic method development, we have not explored in detail the performance of our code near the extremes of the parameter space for scatter orbits. Relevant asymptotic domains of interest are that of large $R_{\rm min}$ (weak-field regime) and that of large $\vinf$ (ultrarelativistic regime), where useful comparisons can be made with analytical approximations. Preliminary experiments suggest that, as expected, the performance of our code gradually deteriorates with larger $R_{\rm min}$ and/or larger $\vinf$. In large-$R_{\rm min}$ runs we are penalized by the longer evolution time required, and in the large-$\vinf$ case the slower decay of initial junk along the orbit requires a larger value of $R_{\rm init}$ (and again a longer run). We estimate, nonetheless, that our current (admittedly suboptimal) method and code can comfortably handle $R_{\rm min}\lesssim 50M$ and $\vinf\lesssim 0.6$. Note that we virtually have no limit on how large the impact parameter $b$ can be raken to be (indeed, in the marginally-bound case studied in \cite{Barack_2019} via a similar time-domain method one has $b\to\infty$).

It is natural to ask about the prospect of an extension to orbits in Kerr geometry. This has been discussed in some detail in \cite{Barack:2017oir}. A 1+1D treatment of the Teukolsky equation in Kerr is still possible, albeit with the additional complication of coupling between $\ell$ modes. The field equation, together with jump conditions on $\cal S$, can be recast in a narrow band-diagonal matrix form, and solved for all $\ell$ modes simultaneously (with a cutoff at a sufficiently high $\ell_{\rm max}$). The application of this mode-coupling approach has been demonstrated in vacuum problems \cite{Dolan_2013,Barack:2017oir}, but it is yet to be applied with a particle source, and the appropriate no-string jump conditions are yet to be derived. In the Kerr case there is no known way of transforming to a RW-like variable in the time domain (the Sasaki-Nakamura formulation achieves that in the frequency domain only), which further motivates an approach based on a direct evolution of the Hertz potential with a suitable form of domain compactification. 

\acknowledgments
We thank Gaurav Khanna for useful correspondence in relation to the issues discussed in Sec.\ \ref{sec:vacuum}. OL acknowledges support from EPSRC through Grant No.~EP/R513325/1. LB acknowledges support from STFC through Grant No.~ST/R00045X/1. This work makes use of the Black Hole Perturbation Toolkit.
%%%%%%%%%%%%%%%%%%%%%%%%%%%%%%%%%%%%%%%%%%%%%%%%%%%%%%%%%%%%%%%%%%%%%

%%%%%%%%%%%%%%%%%%%%%%%%%%%%%%%%%%%%%%%%%%%%%%%%%%%%%%%%%%%%%%%%%%%%%%%%%%%%%%
\appendix
%%%%%%%%%%%%%%%%%%%%%%%%%%%%%%%%%%%%%%%%%%%%%%%%%%%%%%%%%%%%%%%%%%%%%%%%%%%%%%

\section{Bardeen--Press--Teukolsky equation and metric reconstruction}\label{App:convention}

We give here a more detailed technical account of the background material presented 
%We supply here some essential technical details left out of the review discussion 
in Sec.\ \ref{Sec:review}, and in particular we give explicit expressions for the operators ${\sf\hat T}_{\pm}$, ${\sf\hat O}_{\pm}$ and ${\sf\hat S}_{\pm}$, and their adjoints. Our sign conventions for the Newman--Penrose formalism are adopted from Ref.\ \cite{Merlin:2016boc};  Appendix A therein gives a useful summary.   

In this paper we use Kinnersley's null tetrad basis on a Schwarzschild background with metric $g_{\alpha\beta}$ and mass parameter $M$. In Schwarzschild coordinates $(t,r,\theta,\varphi)$, the tetrad legs are given by
%\begin{subequations} 

\begin{align}%\label{eq:kerrtetradm}
\label{eq:kerrtetrad}
e_{\bm 1}^{\alpha}= \ell^\alpha &=\SP{\frac{r^2}{\Delta},1,0,0},\nonumber \\
e_{\bm 2}^\alpha= n^\alpha &=\frac{1}{2}\SP{1,-\frac{\Delta}{r^2},0,0 },\nonumber \\
e_{\bm 3}^\alpha = m^\alpha &=\frac{1}{\sqrt{2}\, r}\bP{0,0,1,\frac{i}{\sin\theta}}, \nonumber \\
e_{\bm 4}^\alpha = \bar m^\alpha &=\frac{1}{\sqrt{2}\,r}\bP{0,0,1,-\frac{i}{\sin\theta}},
\end{align}
%\end{subequations}
where $\Delta:=r(r-2M)$, and overbars denote complex conjugation. We have $g_{\alpha\beta}e_{\bm a}^\alpha e_{\bm b}^{\beta}=0$ for all ${\bm a}$ and ${\bm b}$, except $\ell^\alpha n_\alpha=-1$ and  $m^\alpha\bar{m}_\alpha=1$. 
%In what follows, Greek indices refer to spacetime components while bold Latin indices denote tetrad components. The directional derivatives along the tetrad legs are denoted ${\boldsymbol D}_\ell=\ell^\mu\nabla_\mu$, ${\boldsymbol D}_n=n^\mu\nabla_\mu$ and ${\boldsymbol  D}_m=m^\mu\nabla_\mu$. Their relations to the standard, less transparent symbols are
%\begin{equation}
%{\boldsymbol D}_\ell = {\boldsymbol D},
%\quad\quad
%{\boldsymbol D}_n = {\boldsymbol \Delta},
%\quad\quad
%{\boldsymbol D}_m = {\boldsymbol \delta}.
%\end{equation}
%Since much of the NP literature uses the $(+---)$ metric signature, it would be useful to first review here our sign conventions for the NP formalism with a $(-+++)$ signature. 
The corresponding spin coefficients are 
$
\gamma_{\bm{abc}} := g_{\mu\lambda}\tet{a}{\mu}\tet{c}{\nu}\nabla_{\nu}\tet{b}{\lambda}.
$
Up to trivial index permutations, the only nonzero coefficients in the Schwarzschild case are 
%\begin{subequations}
\begin{align}
\varrho &:= -\gamma_{314}= -\frac{1}{r},\nonumber\\
\mu &:= -\gamma_{243} = -\frac{\Delta}{2r^3},\nonumber\\
\gamma &:=-\frac{1}{2}(\gamma_{212}+\gamma_{342}) =  \frac{M}{2r^2},\nonumber\\
\beta &:= -\frac{1}{2}(\gamma_{213}+\gamma_{343})= \frac{\cot\theta}{2\sqrt{2}\, r},\nonumber\\
\alpha &:= -\frac{1}{2}(\gamma_{214}+\gamma_{344})=-\frac{\cot\theta}{2\sqrt{2}\, r}.
\end{align}
%\end{subequations}
The Weyl curvature scalars $\Psi_0$ and $\Psi_4$ are defined in terms of the Weyl tensor $C_{\alpha\beta\gamma\delta}$  as
%\begin{subequations}
\begin{align}
\Psi_0=&C_{\alpha\beta\gamma\delta}\,\ell^\alpha m^\beta \ell^\gamma m^\delta , \nonumber\\
\Psi_4=&C_{\alpha\beta\gamma\delta}\, n^\alpha \bar m^\beta n^\gamma \bar m^\delta .\label{eq:psi}
\end{align}
%\end{subequations}
Both $\Psi_0$ and $\Psi_4$ vanish in the Schwarzschild background, and so, for the sake of economy but in a slight abuse of notation, we use these symbols to represent the linear perturbations in these quantities. 
We define $\Psi_{+}:=\Psi_0$ and $\Psi_{-}:=\varrho^{-4}\Psi_4$ for notational ease. In terms of the metric perturbation $h_{\alpha\beta}$, we have
${\sf\hat T}_{\pm} h_{\alpha\beta} =\Psi_\pm$ [Eq.\ (\ref{T})], where the second-order differential operators ${\sf\hat T}_{\pm}$ are given by
\begin{eqnarray}\label{hatT}
(\sf\hat T_{+})^{\alpha\beta} &=& \frac{1}{2}\left(
 \ell^{(\alpha} m^{\beta)} m^\gamma\ell^\delta
-m^\alpha m^\beta \ell^\gamma\ell^\delta
-\ell^\alpha \ell^\beta m^\gamma m^\delta
\right. ,
\nonumber\\
&& \left.
+m^{(\alpha} \ell^{\beta)} \ell^\gamma m^\delta
\right)\nabla_\delta\nabla_\gamma , \nonumber \\
(\sf\hat T_{-})^{\alpha\beta} &=& \frac{1}{2}\left(
 n^{(\alpha} \bar m^{\beta)} \bar m^\gamma n^\delta
-\bar m^\alpha \bar m^\beta n^\gamma n^\delta
-n^\alpha n^\beta \bar m^\gamma \bar m^\delta
\right.
\nonumber\\
&& \left.
+\bar m^{(\alpha} n^{\beta)} n^\gamma \bar m^\delta
\right)\nabla_\delta\nabla_\gamma .
\end{eqnarray}
Here $\nabla_\alpha$ is the covariant derivative compatible with the Schwarzschild background metric $g_{\alpha\beta}$, and parenthetical indices are symmetrised, as in $A_{(\alpha\beta)}=\frac{1}{2}(A_{\alpha\beta}+A_{\beta\alpha})$.

The perturbation fields $\Psi_{\pm}$ satisfy the Teukolsky equation with spin parameter $s=\pm 2$, whose Schwarzschild reduction is sometimes referred to as the Bardeen--Press equation. Here we refer to it as the Bardeen--Press--Teukolsky (BPT) equation. It has the form
\begin{equation}\label{eq:kerrteuk} 
{\sf\hat O}_{\pm}\Psi_{\pm}= {\cal T}_\pm,
\end{equation}
%\begin{equation}
%\begin{aligned}
%\label{eq:kerrteuk} % Overall Sign reversed wrt to Teukolsky73
%{\sf\hat O}_{\pm}\Psi_{s}=
%&-\frac{r^4}{\Delta} \frac{\partial^2\Psi_s}{\partial t^2}
%+\Delta^{-s}\frac{\partial}{\partial r}\BP{\Delta^{s+1}\frac{\partial\Psi_s}{\partial r}}
%+2s\BP{\frac{Mr^2}{\Delta}-r}\frac{\partial\Psi_s}{\partial t} 
%+\frac{1}{\sin\theta}\frac{\partial}{\partial\theta}\BP{\sin\theta \frac{\partial\Psi_s}%{\partial\theta}}
%+\frac{1}{\sin^2\theta}\, \frac{\partial^2\Psi_s}{\partial\vf^2}
%\\
%&
%+\frac{2is\cos\theta}{\sin^2\theta}\frac{\partial\Psi_s}{\partial\vf}
%-(s^2\cot^2\theta -s)\Psi_s = {\cal T}_s,
%\end{aligned}
%\end{equation}
where the differential operators on the left are
\begin{align}\label{Ocompact}
{\sf\hat O}_{+}=&\Delta\left(\boldsymbol{D}_\ell+2\frac{r-M}{\Delta}\right)
\left(\tilde{\boldsymbol{D}}_n+4\frac{r-M}{\Delta}\right)
\nonumber\\
& +\eth_1 \bar\eth_2 -6r\partial_t ,
\nonumber\\
{\sf\hat O}_{-}=&\Delta\left(\tilde{\boldsymbol{D}}_n-2\frac{r-M}{\Delta}\right)
\boldsymbol{D}_\ell
+\bar\eth_{-1} \eth_{-2} +6r\partial_t.
\end{align}
Here 
$\boldsymbol{D}_\ell:=\ell^\alpha\nabla_\alpha$, 
$\boldsymbol{D}_n:=n^\alpha\nabla_\alpha$,
$\tilde{\boldsymbol D}_n:=-(2r^2/\Delta){\boldsymbol D}_n$, 
and we have introduced the ``spin raising and lowering'' operators, respectively
%\begin{subequations}
\begin{align} \label{eth}
\eth_s&:=-\partial_\theta-{i\csc\theta}\partial_\vf+s \cot\theta, 
\nonumber\\
\bar\eth_s&:=-\partial_\theta+{i\csc\theta}\partial_\vf-s \cot\theta,
\end{align}
%\end{subequations}
whose action on spin-weighted spherical harmonics ${}_{s}\!Y_{\ell m}(\theta,\varphi)$ is described by  
%\begin{subequations}
\begin{align}\label{raising&lowering}
\eth_s\,{_s\!Y_{\ell m}} &= +\sqrt{(\ell-s)(\ell+s+1)} \, {_{s+1}\!Y_{\ell m}},
\nonumber \\
\bar\eth_s\,{_s\!Y_{\ell m}} &= -\sqrt{(\ell+s)(\ell-s+1)} \, {_{s-1}\!Y_{\ell m}}.
\end{align}
%\end{subequations}
%It is generally not separable in the $a\ne 0$ case, because of the $\pm ia\sin\theta\partial_t$ term in the angular operators, and because of the $a\cos\theta$ dependence of $\varrho$ in the $\pm \frac{6}{\bar\varrho}\partial_t$ terms. Note, however, that stationary perturbations ($\partial_t\Phi=0$) {\em are} separable. 
The source terms ${\cal T}_\pm$ in (\ref{eq:kerrteuk}) are obtained from the energy-momentum tensor $T_{\alpha\beta}$ using 
\begin{align} \label{eq:kerrsource}
{\cal T}_{+}&= {\sf\hat S}_+ T_{\alpha\beta}=8\pi r^2 \times
\nonumber\\  &
\Big[(\boldsymbol{D}_m  -2\beta )\boldsymbol{D}_m   T_{\boldsymbol{11}}
 - (\boldsymbol{D}_\ell -5\varrho)(\boldsymbol{D}_m -2\beta ) T_{\boldsymbol{13}}
 \nonumber\\  &
-(\boldsymbol{D}_m  -2\beta )(\boldsymbol{D}_\ell -2\varrho ) T_{\boldsymbol{13}}
 +(\boldsymbol{D}_\ell -5\varrho)(\boldsymbol{D}_\ell -\varrho ) T_{\boldsymbol{33}} 
 \Big], \nonumber\\
 \end{align}
\begin{align}
{\cal T}_{-}&= {\sf\hat S}_- T_{\alpha\beta}=8\pi r^6  \times
\nonumber\\ & 
\Big[
(\boldsymbol{D}_{\bar m}-2\beta )\boldsymbol{D}_{\bar m}  T_{\boldsymbol{22}} 
-(\boldsymbol{D}_n +2\gamma +5\mu )(\boldsymbol{D}_{\bar m} -2\beta )T_{\boldsymbol{24}}
\nonumber\\  &
 -(\boldsymbol{D}_{\bar m} -2\beta )(\boldsymbol{D}_n +2\gamma +2\mu ) T_{\boldsymbol{24}}
 \nonumber\\  &
+(\boldsymbol{D}_n +2\gamma +5\mu)(\boldsymbol{D}_n +\mu )T_{\boldsymbol{44}} \Big],
\label{eq:kerrsource-2}
\end{align}
where $T_{\boldsymbol{11}}=T_{\alpha\beta}e_{1}^{\alpha}e_1^{\beta}$, etc., and we have also introduced 
$\boldsymbol{D}_m:=m^\alpha\nabla_\alpha$ and $\boldsymbol{D}_{\bar m}:=\bar m^\alpha\nabla_\alpha$,

%The form (\ref{Ocompact}) makes it clear that the BPT equation is separable into spin-weighted spherical harmonics. 

As described in Sec.\ \ref{Sec:review}, the metric reconstruction procedure involves the operators adjoint to ${\sf\hat O}_{+}$, ${\sf\hat T}_{+}$ and ${\sf\hat S}_{+}$. These adjoint operators can be obtained by integrating each operator against a suitable test function and manipulating using integrations by parts. In this fashion it is straightforward to show that 
\begin{equation}
{\sf\hat O}^\dagger_{\pm}={\sf\hat O}_{\mp},
\end{equation}
i.e., solutions $\Phi_{\pm}$ to the adjoint BPT equation with spin $s=\pm 2$ are also solutions to the standard BPT equation with spin $s=\mp 2$.  
For the metric reconstruction operators [see Eq.\ (\ref{h_rec})] a calculation gives 
\begin{eqnarray} \label{eq:kerrh+}
{\sf\hat S}_+^\dagger &=&  -2\ell_\alpha \ell_\beta
\SP{\boldsymbol{D}_m +2\beta}\SP{\boldsymbol{D}_m+4\beta } 
\nonumber\\
 & +&2\ell_{(\alpha} m_{\beta)} \BB{
 \SP{\boldsymbol{D}_m +4\beta}\SP{\boldsymbol{D}_\ell+3\varrho}
+\boldsymbol{D}_\ell \SP{\boldsymbol{D}_m +4\beta}
}
\nonumber\\
&-& 2m_\alpha m_\beta \SP{\boldsymbol{D}_\ell-\varrho}\SP{\boldsymbol{D}_\ell +3\varrho},
\end{eqnarray}
\begin{eqnarray}\label{eq:kerrh-}
{\sf\hat S}_-^\dagger &= &  -2r^4\Big[
n_\alpha n_\beta
\SP{\boldsymbol{D}_{\bar m}+2\beta}\SP{\boldsymbol{D}_{\bar m}+4\beta} 
\nonumber\\
 && -n_{(\alpha} \bar m_{\beta)} \Big(
 \SP{\boldsymbol{D}_{\bar m} +4\beta}\SP{\boldsymbol{D}_n+\mu-4\gamma}
\nonumber\\ 
&& \hspace{12mm} +\SP{\boldsymbol{D}_n +4\mu-4\gamma}\SP{\boldsymbol{D}_{\bar m} -2\beta}
\Big)
\nonumber\\ 
&& +\bar m_\alpha \bar m_\beta \SP{\boldsymbol{D}_n+5\mu-2\gamma}\SP{\boldsymbol{D}_n +\mu-4\gamma}
\Big].
\nonumber\\
\end{eqnarray}
Finally, for the ``source reconstruction'' operators [see Eq.\ (\ref{Source_rec})]
one finds
\begin{eqnarray}\label{Tdagger+Expl}
({\sf\hat T}^\dagger_{+})^{\alpha\beta} &=&
-\frac{1}{2}\ell^\alpha \ell^\beta (\boldsymbol{D}_m+2\beta)(\boldsymbol{D}_m+4\beta)
\nonumber\\
&+&\frac{1}{2}\ell^{(\alpha} m^{\beta)}\Big[\boldsymbol{D}_\ell\left(\boldsymbol{D}_m+4\beta\right) 
+\left(\boldsymbol{D}_m +4\beta\right)\left(\boldsymbol{D}_\ell-\varrho\right)\Big]
\nonumber\\
&-& \frac{1}{2}m^\alpha m^\beta \left(\boldsymbol{D}_\ell-\varrho\right)^2 ,
\end{eqnarray}
\begin{eqnarray}\label{Tdagger-Expl}
({\sf\hat T}^\dagger_{-})^{\alpha\beta} &=&
-\frac{1}{2}n^\alpha n^\beta(\boldsymbol{D}_{\bar m}+2\beta)(\boldsymbol{D}_{\bar m}+4\beta)
\nonumber\\
&+&\frac{1}{2}n^{(\alpha} \bar m^{\beta)}\Big[\left(\boldsymbol{D}_{n}-4\gamma\right)\left(\boldsymbol{D}_{\bar m}+4\beta \right)
\nonumber\\
&& \hspace{12mm}+\left(\boldsymbol{D}_{\bar m}+4\beta\right)\left(\boldsymbol{D}_{n}+\mu-4\gamma\right)\Big]
\nonumber\\
&-&\frac{1}{2}\bar m^\alpha \bar m^\beta(\boldsymbol{D}_{n}+\mu -2\gamma)(\boldsymbol{D}_{n}+\mu-4\gamma).
\nonumber\\
\end{eqnarray}

%%%%%%%%%%%%%%%%%%%%%%%%%%%%%%%%%%%%%%%%%%%%%%%%%%%%%%%%%%%%%
\section{Jumps in the Weyl-scalar modes $\psi^{\pm}_{\ell m}$}
%%%%%%%%%%%%%%%%%%%%%%%%%%%%%%%%%%%%%%%%%%%%%%%%%%%%%%%%%%%%%
\label{App:WeylJumps}

In this appendix we derive the jumps across $\cal S$ in the Weyl-scalar modal fields $\psi^{\pm}_{\ell m}(t,r)$ and their first 3 derivatives, for a generic geodesic orbit. We do so analytically, and for both spins $s=\pm 2$. These jumps are necessary input for the calculation of the no-string Hertz-potential jumps $[\phi_{\pm}]$ in Sec.\ \ref{Sec:Jumps}. In our method we require both $[\psi_+]$ and $[\psi_-]$ for either $[\phi_+]$ (IRG potential) or $[\phi_-]$ (ORG potential); cf.\ Eq.\ (\ref{ODE}) with (\ref{calF}). At the end of this appendix we derive asymptotic expressions for $[\psi_{\pm}]$ at large radii in the case of scatter orbits; these are used in the asymptotic analysis of Sec.\ \ref{subsec:asymptotics}.

\subsection{BPT equation with a point-particle source}

Let $\Psi_{+}\equiv \Psi_{0}$ and $\Psi_{-}\equiv r^4\Psi_{4}$ be the Weyl scalars associated with the physical metric perturbation sourced by a geodesic point particle with stress-energy as in Eq.\ (\ref{Tmunu}).
%\begin{equation}
%T^{\alpha\beta}(x^\mu)=\mu \int_{-\infty}^{\infty}u^\alpha u^\beta \delta^4(x^\mu-x_{\rm p}^\mu(\tau))(r^2\sin%\theta)^{-1}d\tau .
%\end{equation}
We recall our notation: $\mu$ is the particle's mass, and $x^\mu=x_{\rm p}^\mu(\tau)$ describes its geodesic worldline, with proper time $\tau$ and four-velocity $u^\alpha:= dx_{\rm p}/d\tau$. For convenience, we set the Schwarzschild coordinates so that the orbit lies in the equatorial plane ($\theta_{\rm p}\equiv \pi/2$), and write $R(\tau)\equiv r_{\rm p}(\tau)$. The conserved (specific) energy and angular momentum of the orbit are $E=f_Ru^t$ and $L=r^2 u^\varphi$, respectively, where $f_R:=1-2M/R$. The Schwarzschild components of the four-velocity are
\begin{equation}
u^\mu_{\rm p} = \left(E/f_R,(E/f_R)\dot{R},0, L/r^2\right),
\end{equation}   
where an overdot denotes $d/dt$.

The Weyl scalars $\Psi_{\pm}$ satisfy the $s=\pm 2$ BPT equations (\ref{eq:kerrteuk}), where the source $\cal T_\pm$ is derivable from $T^{\alpha\beta}$ by means of Eqs.\ (\ref{eq:kerrsource}) and (\ref{eq:kerrsource-2}).
Expanding both $\Psi_{\pm}$ and $\cal T_\pm$ in $s=\pm 2$ spherical harmonics, as in Eq.\ (\ref{expansionpsi}), separates  the BPT equation into modal equations for each of the time-radial fields $\psi^{\pm}_{\ell m}(t,r)$. The modal equations are (dropping the indices $\ell m$ for brevity)
%\begin{equation} \label{expansionWeylinhom}
%\Psi_{s=2}=
%(r\Delta^{2})^{-1}
%       \sum_{\ell=2}^{\infty}\sum_{m=-\ell}^{\ell}
%       \psi_{2\ell m}(t,r) {}_2\!Y_{\ell m}(\theta,\varphi).
%\end{equation}
%The time-radial fields $\psi_{\pm 2}^{\ell m}$ satisfies the $1+1$D $s=\pm 2$ inhomogeneous Teukolsky equation with a suitable distributional source term $T$ corresponding to the energy-momentum of our geodesic pointlike particle:   
%%~~~~~~~~~~~~~~~~~~~~~~~~~~~~~~~~~~~~~~~~~~~~~~~~~~~~~~~~~~~~~~~~~~~~~~
\begin{align} \label{Teukolsky1+1psi_take2}
\psi^\pm_{,uv} + U_s(r)\psi^\pm_{,u} + V_s(r)\psi^\pm_{,v}  + W_s(r)\psi^\pm =T_\pm(t,r) ,
\end{align}
%%~~~~~~~~~~~~~~~~~~~~~~~~~~~~~~~~~~~~~~~~~~~~~~~~~~~~~~~~~~~~~~~~~~~~~~
where the radial functions on the left are those given in Eqs.\ (\ref{UV}) and (\ref{W_Sch}), with 
$s=\pm 2$ for $\psi_{\pm}$. The modal source term $T_\pm$ can be written in the form 
\begin{align}\label{Tsource}
T_\pm(t,r)=&s^\pm_0(t)\delta[r-R(t)]+s^\pm_1(t)\delta'[r-R(t)] \nonumber\\
&+s^\pm_2(t)\delta''[r-R(t)],
\end{align}
where a prime denotes a derivative with respect to the argument, and the source functions $s^\pm_n(t)$ are certain functions along the orbit. The explicit expressions for $s^\pm_n(t)$ are rather unwieldy, unfortunately, but they are essential within our method, so we give them here. They are 

\begin{align} \label{s0}
s^+_0=&
\frac{\pi\mu}{E}\bigg\{
-L^2 m f_R^2 R \left(i \ddot\varphi_{\rm p}+m \dot\varphi_{\rm p}^2\right)
\nonumber\\
&
+2Lf_R\Big[iL m\left((2+y)f_R-(1+2y)\Rdot\right) \nonumber\\
& + m^2ER(f_R-\Rdot) \Big]\dot\varphi_{\rm p} \nonumber\\
& +Lf_R\left[L(1+2y)-2imER\right]\ddot R
\nonumber\\
&
+\left[-2iLEm+8L^2y^2/R-(m^2-2)E^2 R\right]\Rdot^2
\nonumber\\
&
+2f_R \Big[-(L^2/R)y(6y+7)+2iLmE(1+3y) \nonumber\\
& 
+(m^2-2)E^2 R\Big]\Rdot + 12 f_R^2(L^2/R)y
\nonumber\\
&
-f_R^2\left[2iLmE(1+4y)+(m^2-2)E^2 R\right]
\bigg\} {\cal Y}^+(t) 
\nonumber\\
&
+\pi\mu\bigg\{
2  R(f_R-\Rdot)m\left[Lf_R \dot\varphi_{\rm p}-E(f_R-\Rdot)\right] 
\nonumber\\
&
+2i L \Big[-f_R R \ddot R-\Rdot^2 +2f_R(1+3y)\Rdot\nonumber\\
&-f_R^2(1+4y)\Big]
\bigg\} {\cal Y}^+_\theta(t)
-\pi\mu E R (f_R-\Rdot)^2{\cal Y}^+_{\theta\theta}(t),
%\nonumber\\
%&&
\end{align}
\begin{eqnarray}\label{s1}
s^+_1&=&
2i\pi\mu L f_R R(f_R-\Rdot)\Big[(f_R-\Rdot)\left[{\cal Y}^+_\theta(t)+m{\cal Y}^+(t)\right] 
\nonumber\\
&& \qquad \qquad \qquad \qquad 
-mf_R(L/E)\dot\varphi_{\rm p}{\cal Y}^+(t)\Big]
\nonumber\\
&&
+\pi\mu (L^2/E) f_R \Big[-f_R R \ddot R-2(1+2y)\Rdot^2 
\nonumber\\
&& \qquad \quad 
+2f_R(3+5y)\Rdot-4f_R^2(1+y)\Big] {\cal Y}^+(t),
\end{eqnarray}
\begin{equation} \label{s2}
s^+_2= \pi\mu(L^2/E) f_R^2 R (f_R-\Rdot)^2 {\cal Y}^+(t),
\end{equation}
and
%Eq.\ (\ref{calF}) shows that the only jump condition we need from the $s=-2$ field is $[\psi_{-2}]$. This means from Eqns.\ (\ref{psidelta}) and (\ref{Jpsipm}) that we only need the source components
\begin{align}
s^-_0 =& -\frac{\pi \mu }{4 E f_R^2 R^9} \bigg\{ L^2 m f_R^2 R^6 \left(m \dot{\varphi }_{\rm p}^2+i \ddot{\varphi }_{\rm p}\right) \nonumber\\
& -2 L m R^5 f_R \Big[E m R \dot{R}+f_R \Big(E m R \nonumber\\ 
&+i L \left(3 \dot{R}+2-5y\right)\Big)\Big] \dot{\varphi }_{\rm p} \nonumber\\
& + L f_R R^5 \left(3 L f_R-2 i E m R\right) \ddot{R} \nonumber\\
& + R^4 \Big[-12 L^2 f_R^2+E R \Big(E \left(m^2-2\right) R \nonumber\\
& +2 i L m (3-4 y)\Big)\Big]\dot{R}^2 + 2 R^4 \Big[-L^2 (8-25 y) f_R^2 \nonumber\\ 
&+E R \Big(E \left(m^2-2\right) R f_R+2 i L m (3-7 y) f_R\Big)\Big]\dot{R} \nonumber \\
& + R^4 \Big[-4 L^2 (1-5y) f_R^3+E R \Big(E \left(m^2-2\right) R f_R^2 \nonumber\\
&+2 i L m (3-8 y) f_R^2\Big)\Big] \bigg\} {\cal Y}^-(t) + \frac{\pi  \mu }{2 f_R^2 R^7} \bigg\{ \nonumber\\ 
& m R^4 (f_R+\dot{R}) \left(E (f_R+\dot{R})-L f_R \dot{\varphi }_{\rm p}\right) \nonumber\\
& -i L R^3 \bigg(f_R R \ddot{R} - (3-4y)\dot{R}^2 -2f_R (3-7 y)\dot{R} \nonumber\\
&- (3-8y) f_R^2 \bigg)\bigg\} {\cal Y}^-_\theta(t) -\frac{\pi  E \mu  (f_R+\dot{R})^2}{4 f_R^2 R^3} {\cal Y}^-_{\theta\theta}(t),
\end{align}
\begin{align}\label{s1psi4leor}
s_1^- =&
\frac{i\pi\mu L (f_R+\Rdot)}{2f_RR^3}\Big[(f_R+\Rdot)\left[{\cal Y}^-_\theta(t)-m{\cal Y}^-(t)\right]\nonumber\\
& +mf_R(L/E)\dot\varphi_{\rm p}{\cal Y}^-(t)\Big]
+\frac{\pi\mu L^2}{4E R^4} \Big[-R \ddot R+6\Rdot^2 \nonumber\\
& +2(5-13y)\Rdot+4f_R(1-3y)\Big] {\cal Y}^-(t),
\end{align}
\begin{equation} \label{s2psi4}
s_2^- = \frac{\pi  \mu  L^2 \left(f_R+\Rdot\right){}^2}{4 E R^3} \mathcal{Y}^-(t) .
\end{equation}
Here we have introduced 
\begin{equation}\label{calY}
y:=\frac{M}{R},\quad\quad
{\cal Y}^\pm(t):={}_{\pm2}\!\bar Y_{\ell m}\left(\frac{\pi}{2},\varphi_{\rm p}(t)\right),
\end{equation}
with ${\cal Y}^\pm_\theta$ and ${\cal Y}^\pm_{\theta\theta}$ being the first and second derivatives of ${}_{\pm2}\!\bar Y_{\ell m}\left(\theta,\varphi_{\rm p}(t)\right)$ with respect to $\theta$, evaluated at $\theta=\pi/2$.

Note that in Eqs.\ (\ref{s0})--(\ref{s2psi4}) we have not yet specialised to a timelike geodesic. With such specification, the time derivatives featuring in these expressions can be expressed in terms of $R(t)$ alone (as well as $E$ and $L$), as follows: 
\begin{align}\label{dotR}
\Rdot &= \pm (f_R/E)\left[E^2-f_R(1+L^2/R^2)\right]^{1/2}, \\
\ddot R&=\frac{f_R^2(1-5y)L^2+yR^2f_R(2E^2-3f_R)}{R^3 E^2},
\end{align}
\begin{equation}\label{dotvarphi}
\dot\varphi_{\rm p}= \frac{f_RL}{R^2 E},\quad\quad
\ddot\varphi_{\rm p}=-\frac{2L(1-3y)\Rdot}{ER^3}.
\end{equation}
The sign in (\ref{dotR}) is $(-)$ for the incoming leg of the orbit and $(+)$ for the outgoing leg.

\subsection{The jumps in $\psi_\pm$ and their first derivatives}

The jumps in the 1+1D Weyl-scalar fields $\psi_\pm$ are determined by requiring that (\ref{Teukolsky1+1psi_take2}) is satisfied as a distributional equation, with the ansatz
\begin{align}
\psi_\pm=&\psi_\pm^>(t,r)\Theta[r-R(t)]+\psi_\pm^<(t,r)\Theta[R(t)-r]\nonumber \\
&+\psi_\delta^\pm(t)\delta[r-R(t)].
\end{align}
Here $\Theta[\cdot]$ is the Heaviside step function, and $\psi^\pm_\delta(t)$ is to be determined.
Balancing the coefficients of $\Theta[r-R(t)]$ and $\Theta[R(t)-r]$ imply that $\psi_\pm^>(t,r)$ and $\psi_\pm^<(t,r)$ are homogeneous solutions of (\ref{Teukolsky1+1psi_take2}). The remaining terms are supported on $r=R(t)$ only, and are each proportional to either $\delta$, $\delta'$ or $\delta''$. We use the distributional identities 
\begin{eqnarray}
F(r)\delta(r-R) &=& F(R)\delta(r-R),\nonumber\\
F(r)\delta'(r-R) &=&F(R)\delta'(r-R)-F'(R)\delta(r-R),\nonumber\\
F(r)\delta''(r-R)&=&F(R)\delta''(r-R)-2F'(R)\delta'(r-R)\nonumber\\ 
 && +F''(R)\delta(r-R)
\end{eqnarray} 
[valid for any smooth function $F(r)$] to eliminate the $r$ dependence of the coefficients of each of these terms, and then compare the coefficient values across the two sides of Eq.\ (\ref{Teukolsky1+1psi_take2}), recalling the form of $T_\pm$ in Eq.\ (\ref{Tsource}).

From the coefficient of $\delta''$ we immediately obtain
\begin{equation}\label{psidelta}
\psi^\pm_\delta(t)=-\frac{4s^\pm_2(t)}{f^2_R-\Rdot^2} . 
\end{equation}
%From the geodesic equation of motion we have the relation 
%\begin{equation}
%f_R^2-\Rdot^2 =\frac{f_R^3(R^2+L^2)}{E^2 R^2}.
%\end{equation}
The coefficient of $\delta'$ then determines the jump:
%\begin{equation}
%[\psi](t):=\psi^>(t,R(t))-\psi^<(t,R(t)).
%\end{equation}
%We find
\begin{align}\label{Jpsipm}
\J{\psi_\pm}= &
-\frac{1}{f_R^2-\Rdot^2} \Big\{4s_1(t)+2 \dot{\psi}^\pm_{\delta}\Rdot \nonumber \\ 
&+ \left[-2f_R\big(s(1-y)+3y\big)+2s(1-3y)\Rdot+R\ddot R\right]\frac{\psi^\pm_\delta}{R} \Big\},
\end{align}
with $s=\pm 2$ for $\psi_\pm$.
%\begin{equation}\label{Jpsi}
%{
%\J{\psi}=
%\frac{-4s_1(t)-2(d\psi_{\delta}/dt)\Rdot+\psi_\delta
%\left[2(2R+M)f_R/R^2-4\Rdot(R-3M)/R^2-\ddot R\right]}
%{f^2(R)-\Rdot^2}},
%\end{equation}
Finally, comparing the coefficients of $\delta$ gives a relation between the jumps $\J{\psi^\pm_{,t}}$ and $\J{\psi^\pm_{,r}}$:
%\begin{eqnarray}\label{Jeq1}
%s_0(t)&=& 
%-\frac{1}{2}\Rdot\J{\psi_{,t}}
%-\frac{1}{4}(f^2+\Rdot^2)\J{\psi_{,r}}
%+\frac{1}{4}\ddot{\psi_\delta}
%+\left(\frac{R-3M}{R^2}\right)\dot{\psi_\delta}
%+\left(\frac{\lambda R f-6M-4(R-6M)\Rdot}{4R^3}\right)\psi_\delta
%\nonumber\\
%&&+
%\left(\frac{2(R^2-5MR+2M^2)-4R(R-3M)\Rdot-R^3\ddot R}{4R^3}\right)\J{\psi},
%\end{eqnarray}
\begin{equation}\label{Jeq1}
s^\pm_0(t)= 
-\frac{1}{2}\Rdot\J{\psi^\pm_{,t}}
-\frac{1}{4}(f_R^2+\Rdot^2)\J{\psi^\pm_{,r}} +P_\pm(t),
\end{equation}
where
\begin{align}
P_\pm=
& \frac{1}{4} \ddot{\psi}^\pm_\delta
+\frac{s}{2R}(1-3y)\dot{\psi}^\pm_\delta\nonumber \\ 
&
+\frac{\psi^\pm_\delta}{4R^2}\Big[\lambda f_R +2+s-s^2+2y(1+s^2-4s)\nonumber \\ 
& \qquad \qquad +8y^2(s-2) -2s(1-6y)\Rdot\Big]\nonumber \\ 
& +\frac{\J{\psi_\pm}}{4R}\left[2f_R\big(s(1-y)+y\big)-2s(1-3y)\Rdot-R \ddot R\right],
\end{align}
with $s=\pm 2$ for $P_\pm$. Recall $\lambda=(l+2)(l-1)$.

A second relation between $\J{\psi^\pm_{,t}}$ and $\J{\psi^\pm_{,r}}$ is obtained by writing 
\begin{equation}\label{Jeq2}
\dot{[\psi_\pm]}=\J{\psi^\pm_{,t}}+\Rdot\J{\psi^\pm_{,r}}.
\end{equation}
Solving (\ref{Jeq1}) and (\ref{Jeq2}) as a simultaneous set then gives
\begin{equation}
{
\J{\psi^\pm_{,r}}=\frac{4(P_\pm-s^\pm_0)-2\Rdot \dot{[\psi_\pm]}}{f_R^2-\Rdot^2},
}
\end{equation}
and
\begin{equation}
{
\J{\psi^\pm_{,t}}=\frac{-4(P_\pm-s^\pm_0)\Rdot+ (f_R^2+\Rdot^2)\dot{[\psi_\pm]} }{f_R^2-\Rdot^2}.
}
\end{equation}
The corresponding jumps in the $u$ and $v$ derivatives are
%Using $\J{\psi_{,v}}=\frac{1}{2}\left(\J{\psi_{,t}}+f_R\J{\psi_{,r}}\right)$ and $\J{\psi_{,u}}=\frac{1}{2}\left(\J{\psi_{,t}}-f_R\J{\psi_{,r}}\right)$, we also get 
\begin{equation}\label{Jpsiv}
{
\J{\psi^\pm_{,v}}=\frac{4(P_\pm-s^\pm_0)+ (f_R-\Rdot)\dot{[\psi_\pm]} }{2(f_R+\Rdot)},
}
\end{equation}
\begin{equation}\label{Jpsiu}
{
\J{\psi^\pm_{,u}}=\frac{-4(P_\pm-s^\pm_0)+ (f_R+\Rdot)\dot{[\psi_\pm]} }{2(f_R-\Rdot)}.
}
\end{equation}

Equations (\ref{Jpsipm}), (\ref{Jpsiv}) and (\ref{Jpsiu}) give the jumps in $\psi_\pm$ and its first derivatives for a generic geodesic orbit.

\subsection{Jumps in the second and third derivatives of $\psi_\pm$}
\label{app:WeylJumpHighDerivatives}

We can get $\J{\psi^\pm_{,uv}}$ directly from the vacuum BPT equation:
\begin{align} 
\J{\psi^\pm_{,uv}}= - U_{\pm 2}(R)\J{\psi^\pm_{,u}} - V_{\pm 2}(R)\J{\psi^\pm_{,v}}  - W_{\pm 2}(R)\J{\psi_\pm} ,
\end{align}
where the jumps $\J{\psi^\pm_{,v}}$, $\J{\psi^\pm_{,u}}$ and $\J{\psi^\pm}$ can be substituted for from Eqs.\ (\ref{Jpsiv}), (\ref{Jpsiu}) and (\ref{Jpsipm}).
Then $\J{\psi^\pm_{,uu}}$ and $\J{\psi^\pm_{,vv}}$ can be obtained from the chain rules
\begin{eqnarray}
\dot{\J{\psi^\pm_{,u}}} &=& \dot v\J{\psi^\pm_{,uv}} + \dot u\J{\psi^\pm_{,uu}}, \nonumber\\
\dot{\J{\psi^\pm_{,v}}} &=& \dot v\J{\psi^\pm_{,vv}} + \dot u\J{\psi^\pm_{,uv}},
\end{eqnarray} 
where 
\begin{equation}
\dot v= 1 +\Rdot/f_R,\quad\quad
\dot u= 1 -\Rdot/f_R.
\end{equation}
%and for the rest of this subsection we omit the $\pm$ label for notational simplicity. 
%We thus obtain
%\begin{equation} \label{Jpsi,uu}
%\J{\psi_{,uu}} = \frac{\dot{\J{\psi_{,u}}}-\dot v\J{\psi_{,uv}}}{\dot u}.
%\end{equation}
%Similarly,
%\begin{equation}
%{
%\J{\psi_{,vv}} = \left(\J{\psi_{,v}}_t-\dot u\J{\psi_{,uv}}\right)/\dot v.
%}
%\end{equation}

The jumps in the third derivatives are obtained in a similar fashion: First, we obtain $\J{\psi^\pm_{,uvu}}$ and $\J{\psi^\pm_{,uvv}}$ from the $u$ and $v$ derivatives of the vacuum BPT equations. Then, $\J{\psi^\pm_{,uuu}}$ and $\J{\psi^\pm_{,vvv}}$ are determined from the appropriate chain rule; e.g.,
\begin{equation}
\J{\psi^\pm_{,uuu}} = \frac{\dot{\J{\psi^\pm_{uu}}}-\dot v\J{\psi^\pm_{,uuv}}}{\dot u},
\end{equation} 
where the jumps on the right-hand side are known from previous steps. We may proceed in this recursive manner to determine the jumps in all higher derivatives.  In the calculation performed in this paper we require jump information only up to the third derivatives. 

\subsection{Large-$R$ asymptotics for scatter orbits}\label{WeylJumpsAsympt}
 
For our asymptotic analysis in Sec.\ \ref{subsec:asymptotics} (where we derive initial conditions for the Hertz potential's jump equations), it is useful to have at hand the large-$R$ asymptotic form of the Weyl-scalar jumps calculated above, in the case of a scatter orbit coming from infinity (i.e., the class of geodesic scatter orbits described in Sec.\ \ref{EoM}). Specifically, we need the asymptotic forms of $\J{\psi_{\pm}}$ as well as $\J{\psi^+_{,u}}$, $\J{\psi^+_{,uu}}$ and $\J{\psi^+_{,uuu}}$.
 
As input for this calculation, we need the asymptotic form of the source coefficients $s_n^\pm$ in Eqs.\ (\ref{s0})--(\ref{s2psi4}). Specializing to scatter geodesics and working at leading order in $y=M/R$ (at fixed $E,L$), we find
\begin{eqnarray}
s_n^+ &=& \sigma_n^+ R + O(R^0), \nonumber\\
s_n^- &=& \sigma_n^- R^{-3} + O(R^{-4}),
\end{eqnarray}
for $n=0,1,2$. The coefficients needed for our purpose are given explicitly by 
\begin{eqnarray}\label{s^+_nAsymp}
\sigma_0^+ &=& -\mu E\pi(1-\Rdotinf)^2\left[(m^2-2){\cal Y}^+
+2m{\cal Y}^+_{\theta} +{\cal Y}^+_{\theta\theta} \right]_{\infty},
\nonumber\\
\sigma_1^+ &=&\mp 2i \mu L\pi(1-\Rdotinf)^2\left(m{\cal Y}^++{\cal Y}^+_{\theta} \right)_{\infty},
%\nonumber \\
%\sigma_2^+ &=&\frac{\mu \pi L^2(1-\Rdotinf)^2}{E}\, {\cal Y}^+_{\infty},
\end{eqnarray}
and 
\begin{eqnarray}\label{s^-_nAsymp}
\sigma_1^- &=& \pm \frac{1}{2} i \mu L (1+\Rdotinf)^2\left(m \mathcal{Y}^- -\mathcal{Y}^-_{\theta}\right)_\infty ,
%\nonumber \\
%\sigma_2^- &=& \frac{\mu \pi L^2 }{4E}\, \mathcal{Y}^-_{\infty} 
\end{eqnarray}
%(the coefficient $\sigma_0^-$ is not needed for the IRG calculation performed in this paper).
where subscripts `$\infty$' imply an evaluation at $t\to \pm\infty$, 
%or, equivalently, at $\varphi_{\rm p}(t)\to \varphi_{\rm in/out}$, 
depending on whether it is the `in' or 'out' states being considered. In the expressions for $\sigma_1^+$ and $\sigma_1^-$, the upper sign is for the `out' state ($\Rdotinf>0$) and the lower sign is for the `in' state ($\Rdotinf<0$).

A straightforward leading-order calculation now gives 
\begin{eqnarray}
\J{\psi_{+}} &=& -4 \sigma_{1}^{+}E^2 R+ O(R^0), \\
\J{\psi_{-}} &=& -4 \sigma_{1}^{-}E^2 R^{-3}+ O(R^{-4}),
\end{eqnarray}
as well as 
\begin{equation}
\J{\psi^{+}_{,t}} = -\Rdotinf \J{\psi^{+}_{,r}} = 4 \sigma_{0}^{+}E^2 \Rdotinf R+ O(R^0),
\end{equation}
\begin{equation}\label{Ju}
\J{\psi^{+}_{,u}} = \frac{2 \sigma_{0}^{+}}{1-\Rdotinf}R+ {\cal O}(R^0),
\end{equation}
\begin{equation}\label{Juu}
\J{\psi^{+}_{,uu}} = -\frac{2 \sigma_{0}^{+}(2-3\Rdotinf)}{(1-\Rdotinf)^2}+ {\cal O}(R^{-1}),
\end{equation}
\begin{equation}\label{Juuu}
\J{\psi^{+}_{,uuu}} = \frac{\sigma_{0}^{+} \left[8(1-2\Rdotinf) + \lambda(1+\Rdotinf)\right]}{2 (1-\Rdotinf)^2}\, R^{-1} + O(R^{-2}).
\end{equation}

%%%%%%%%%%%%%%%%%%%%%%%%%%%%%%%%%%%%%%%%%%%%%%%%%%%%%%%%%%%%%
\section{Finite-difference scheme}
%%%%%%%%%%%%%%%%%%%%%%%%%%%%%%%%%%%%%%%%%%%%%%%%%%%%%%%%%%%%%
\label{App:RWCode}

In this appendix we detail the finite-difference (FD) scheme employed in Sec.\ \ref{sec:hyperevol} for solving the 1+1D vacuum RW field equation (\ref{RWeq}), with jump conditions on $\cal S$ corresponding to a no-string Hertz potential for a scatter orbit. The basic architecture of our characteristic evolution scheme was described in Sec.\ \ref{TeukNumMethod}. Here we focus on the FD scheme itself, at the grid-cell level. In deriving the scheme we follow the method of Ref.\ \cite{bsago2} (which itself builds on a long history of time-domain work in the self-force literature, e.g.\ \cite{Lousto05, Haas:2007kz}).

Recall Fig.\ \ref{uvGridScatter}, which shows the 1+1D numerical grid, based on $u,v$ coordinates with uniform cell dimensions $h\times h$. Consider an arbitrary grid point $c$ with coordinates $(u,v)=(u_c,v_c)$, and in reference to that point  denote by $X_{nk}$ the value of the numerical field variable $X$ at the grid point with coordinates $(u,v)=(u_c- nh,v_c-kh)$. Our goal is to prescribe a FD expression for $X_{00}$ (the field at $c$), given the values $X_{nk}$ for all $n,k>0$, assumed to have been obtained in previous steps of the characteristic evolution. We wish to achieve a global quadratic convergence, i.e.\ an accumulated error in $X$ that scales as $h^2$. Since the total number of grid points over which the error accumulates is $\propto h^{-2}$, this demands a local (single-point) FD error not larger than $O(h^4)$ in general. 

Our grid is traversed by the curve $\cal S$ representing the timelike geodesic trajectory of the particle. The curve is fixed and known in advance, and the coordinates of all of its intersections with grid lines are calculated in advance of the numerical evolution. In reference to the grid cell $C$ with top vertex $c$, we distinguish between two scenarios: Either $C$ is traversed by $\cal S$ (``particle cell'') or it is not (``vacuum cell''). We deal with each of these two scenarios separately below. 

%The form of the FD formula depends on the position of the field point relative to the worldline. We integrate the field equation over the cell bounded by the field points $X_{00}$, $X_{01}$, $X_{10}$ and $X_{11}$. Integrations over this cell will be notated by $\int_\diamondsuit$. If the particle's worldline does not cross within the integration cell then we apply a "vacuum" formula. However, if the worldline does cross through the cell then there are extra contributions which depend on trajectory of the particle through the cell. The four types of non-vacuum cells (`UU', `UV', `VU' and `VV') are shown in Figure \ref{ScatterNonVacCells}. We will describe the method to get the vacuum and non-vacuum cells separately.

%The scheme is constructed such that it is convergent to quadratic order (i.e. the global error scales as ${\cal O}(h^2)$). For this it is sufficient to use the field points $X_{01}$, $X_{10}$ and $X_{11}$. To do this we need two rays of initial data along constant $u=u_0$ and $v=v_0$. We choose this initial data to be zero for all points on the rays. This non-physical initial data produces initial bursts of radiation which decay rapidly. To ensure we have the physical solution we run the code for the appropriate amount of time such that the bursts have time to radiate away.

\subsection{Vacuum cells}
\label{app:FDSVac}

First we consider the simpler case where the particle's worldline does not cross the integration cell. Then a sufficiently accurate FD approximation for $X_{00}$ can be written based on the three value $X_{01}$, $X_{10}$ and $X_{11}$ alone: Integrating each of the two terms on the left-hand side of the RW equation (\ref{RWeq}) over the grid cell $C$, we have 
\begin{equation}
\int_C X_{,uv} \: du dv = X_{00} - X_{01} - X_{10} + X_{11}
\end{equation}
(exactly), and 
\begin{equation}
\int_C {\cal W}(r)X \: du dv = \frac{1}{2} h^2{\cal W}(r_c) \left( X_{01} + X_{10} \right) + O(h^4),
\end{equation}
where 
\begin{equation}
{\cal W}(r) := \frac{f}{4}\left(\frac{\lambda_1}{r^2}-\frac{6M}{r^3} \right),
\end{equation}
and $r_c$ is the value of $r$ at point $c$. 
The vacuum RW equation then gives
\begin{equation}
X_{00} = - X_{11} + (X_{01} + X_{10}) \left( 1 - \frac{h^2}{2} {\cal W}(r_c) \right)+O(h^4),
\label{FDSRWVac}
\end{equation}
which we use as our FD formula for vacuum points.

\begin{figure}[h!]
\centering
\includegraphics[width=\linewidth]{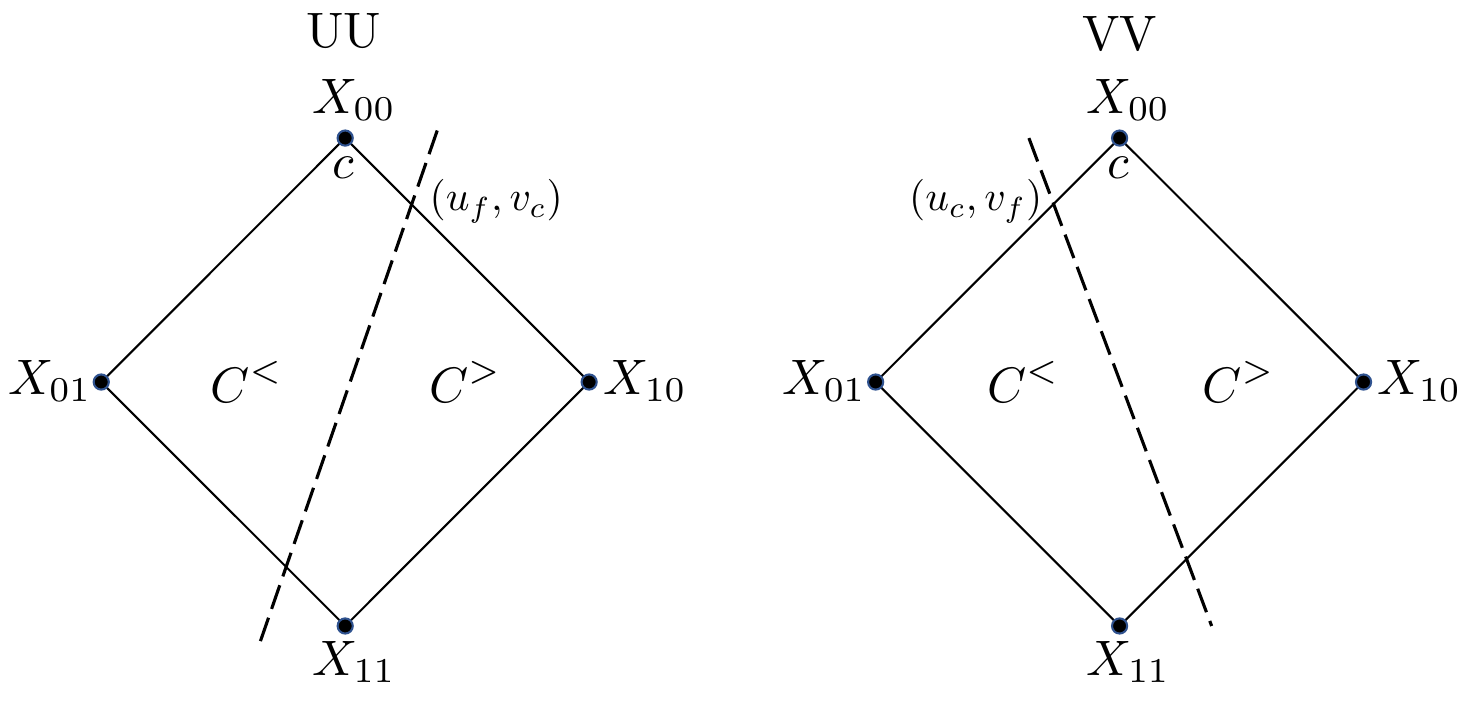} \\
\vspace{3mm}
\includegraphics[width=\linewidth]{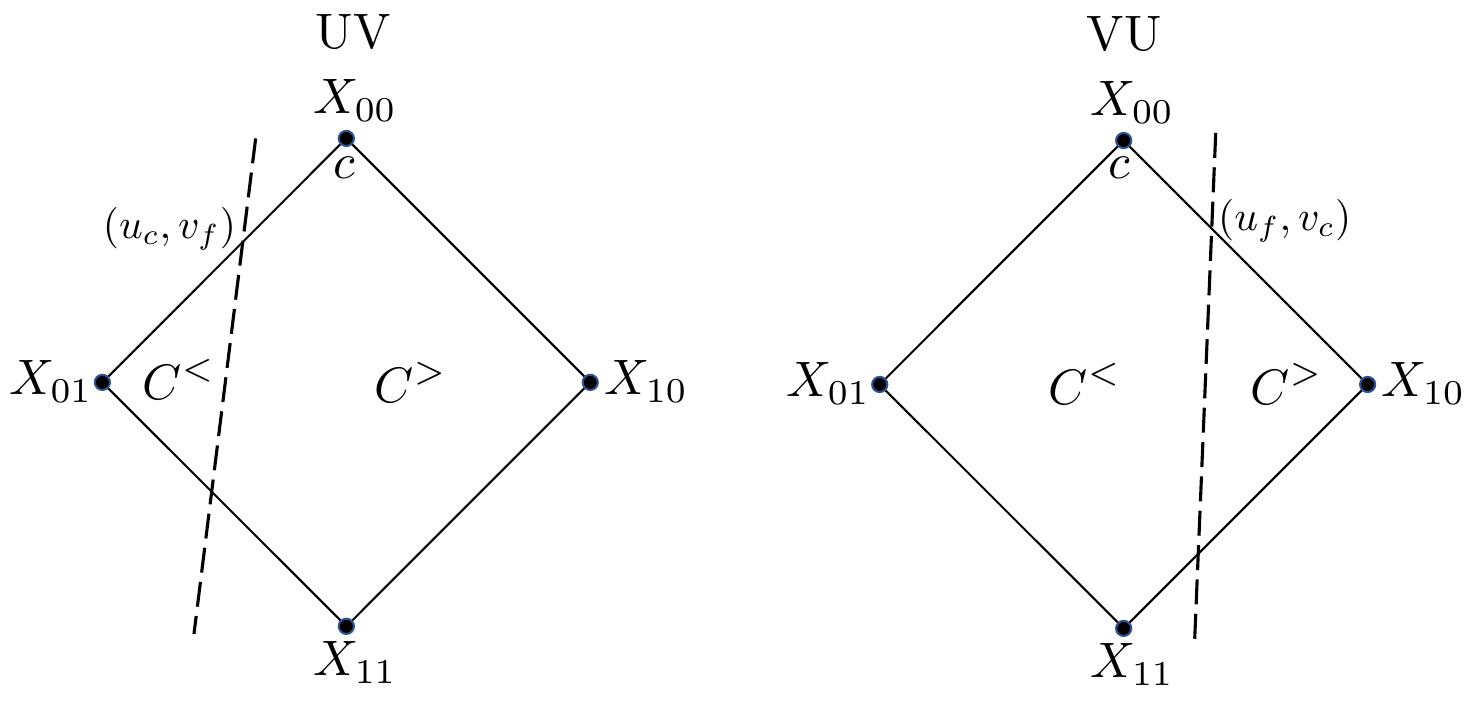}
\caption{A particle cell is traversed by the particle's worldline (dashed curve) in one of four possible ways: cases `UU', `VV', `UV' and `VU', illustrated here. A different variant of the FD formula applies in each case, as described in the text.   The apex of the cell is the point $c$ at $(u,v) = (u_c,v_c)$, and the particle exits the cell at $(u_f,v_c)$ or $(u_c,v_f)$, depending on the case.
%we indicate the coordinate values at which the worldline enters the cell and exits it. 
The vacuum portions of the cell left and right of the worldline are $C^<$ and $C^>$, respectively.}
\label{ScatterNonVacCells}
\end{figure}

\subsection{Particle cells}
\label{app:FDSNonVac}

The vacuum formula (\ref{FDSRWVac}) does not work for cells that are traversed by the worldline, due to the discontinuity in the field across $\cal S$. The worldline splits $C$ into two disjoint vacuum regions, $C^>$ and $C^<$, as shown in Fig.\  \ref{ScatterNonVacCells}, which displays the four possible scenarios. Since $X$ is smooth on each of the two vacuum regions, we can expand it piecewise in a Taylor series about point $c$, in the form 
\begin{equation}
X^{\gtrless} = \sum_{i+j=0}^N \frac{c_{ij}^{\gtrless}}{i! j!} \tilde u^i \tilde v^j + O(h^{N+1}),
\label{2DTaylorX}
\end{equation}
where $\tilde u := u-u_c$, $\tilde v := v-v_c$, and different expansion coefficients apply on each side of $\cal S$: $c_{ij}^{<}$ are used in $C^<$, and $c_{ij}^{>}$ are used in $C^>$. The idea now is to derive the values of $c_{ij}^{\gtrless}$ based on a sufficient number of data points $X^{\gtrless}_{nk}$, plus the analytically known jumps in $X$ and its derivatives on $\cal S$. We note that, since the total number of particle cells scales as $h^{-1}$, it is acceptable for our local FD scheme to have an error as great as $O(h^3)$ (but not greater) at each particle cell. 

To achieve such accuracy we take $N=2$ in Eq.\ (\ref{2DTaylorX}), leaving us with 12 coefficients $c_{ij}^{\gtrless}$ to determine. We use the 6 data points $\{X_{00},X_{01},X_{10},X_{11},X_{02},X_{20}\}$ to supply 6 constraints, and 6 additional constraints are obtained from the known jumps $\{[X],[X_{,u}],[X_{,v}],[X_{,uu}],[X_{,uv}],[X_{,vv}]\}$, imposed at the point where the worldline exits the cell $C$ [i.e., referring to Fig.\ \ref{ScatterNonVacCells}, either the point $(u_f,v_c)$ or the point $(u_c,v_f)$, depending on the case]. For example, in cases `UU' and `VU' we have the constraint 
\begin{equation}
\J{X}_{(u_f,v_c)} = \sum_{i=0}^2 \frac{(c_{i0}^>-c_{i0}^<)}{i!} \tilde u^i_f  + O(h^3),
\label{2DTaylorXJump}
\end{equation}
where $\tilde u_f = u_f-u_c$. Solving the 12 simultaneous equations for $c_{ij}^{\gtrless}$ and then substituting these coefficients back in (\ref{2DTaylorX}), gives an expression for $X^{\gtrless}$, accurate through $O(h^2)$ in the vicinity of point $c$ [i.e., with an error $O(h^3)$], in terms of the above 6 field points (which include the unknown $X_{00}$) and above 6 jumps. 

Considering first the principal part of the RW equation, we thus obtain, 
\begin{align}\label{FDS:Xvuterm}
X^{\gtrless}_{,uv} &= c^{\gtrless}_{11} +O(h)
\nonumber\\
&= h^{-2}\left(X_{00}-X_{01}-X_{10}+X_{11}+J_1^A\right) +O(h),
\end{align}
where $A\in\{UU,VV,UV,VU\}$ labels the case in question, with
\begin{align}
J_1^{UU}&=h[X_{,v}]+h(u_c-u_f)[X_{,uv}]-\frac{h^2}{2}\left(2[X_{,uv}]+[X_{,vv}]\right),
\nonumber\\
J_1^{VV}&=-h[X_{,u}]-h(v_c-v_f)[X_{,uv}]+\frac{h^2}{2}\left(2[X_{,uv}]+[X_{,uu}]\right),
\nonumber\\
J_1^{UV}&=-[X]+(h-v_c+v_f)[X_{,v}]-\frac{1}{2}(h-v_c+v_f)^2[X_{,vv}],
\nonumber\\
J_1^{VU}&=[X]-(h-u_c+u_f)[X_{,u}]+\frac{1}{2}(h-u_c+u_f)^2[X_{,uu}].
\end{align}
For the term ${\cal W}X$ of the RW equation we wish to obtain a FD approximation that does not involve $X_{00}$, and the form of (\ref{FDS:Xvuterm}) implies that we only require a leading-order, $O(h^0)$ approximation for this term. We choose to achieve this by taking $N=1$ in Eq.\ (\ref{2DTaylorX}), and then solving for the six coefficients $c^{\gtrless}_{ij}$ ($i+j\leq 1$) using the 3 data points $\{X_{01},X_{10},X_{11}\}$ and 3 jumps $\{[X],[X_{,u}],[X_{,v}]\}$, again evaluated at $(u_f,v_c)$ or $(u_c,v_f)$. This gives 
\begin{align}
X^{\gtrless} &= c^{\gtrless}_{00} +O(h)
\nonumber\\
& = X_{01} + X_{10}- X_{11} + J_2^A +O(h),
\end{align}
with 
\begin{align}
J_2^{UU}=0=J_2^{VV},
\quad\quad
J_2^{UV}=[X]=- J_2^{VU}.
\end{align}
Hence,
\begin{equation}\label{FDS:WXterm}
{\cal W}X^{\gtrless} = {\cal W}(r_c) (X_{01} + X_{10}- X_{11} + J_2^A) +O(h).
\end{equation}

Imposing finally the vacuum RW equation $X_{,uv}+{\cal W}X=0$, we obtain, using (\ref{FDS:Xvuterm}) and (\ref{FDS:WXterm}),
\begin{align}
X_{00} =& (X_{01} + X_{10}- X_{11}) \left( 1 - h^2{\cal W}(r_c) \right) - J_1^A
\nonumber\\
& - h^2{\cal W}(r_c)J_2^A + O(h^3),
\label{FDSparticle}
\end{align}
which is our FD formula for particle cells.

Note that our second-order-convergent FD scheme, consisting of Eq.\ (\ref{FDSRWVac}) for vacuum cells with Eq.\ (\ref{FDSparticle}) for particle cells, requires as input only the three field data points $X_{01}$, $X_{10}$ and $X_{11}$ (as well as the known jumps). This is convenient, as it means that at each characteristic evolution step we require data on a single previously calculated characteristic ray.

%%%%%%%%%%%%%%%%%%%%%%%%%%%%%%%%%%%%%%%%%%%%%%%%%%%%%%%%%%%%%%%%%%%
%%%%%%%%%%%%%%%%%%%%%%%%%%%%%%%%%%%%%%%%%%%%%%%%%%%%%%%%%%%%%%%%%%%
\bibliographystyle{unsrt}
\bibliography{newbiblio}
%%%%%%%%%%%%%%%%%%%%%%%%%%%%%%%%%%%%%%%%%%%%%%%%%%%%%%%%%%%%%%%%%%
%%%%%%%%%%%%%%%%%%%%%%%%%%%%%%%%%%%%%%%%%%%%%%%%%%%%%%%%%%%%%%%%%%%

\end{document}